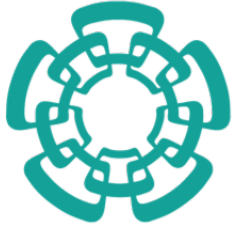

Center for Research and Advanced Studies of the
National Polytechnic Institute
**Cinvestav**

# Path integrals and deformation quantization: the fermionic case

Thesis submitted in partial fulfillment of the requirements
for the degree of
Master of Science in Physics

## Anuar Kafuri Zarzosa

Thesis Advisor:
Dr. Héctor Hugo García Compeán
Department of Physics

Mexico City
October 2025

October 13, 2025

*To my family,*
*my source of strength, endurance, resilience, focus, and serenity.*

# Acknowledgments

First, I want to thank my family, who have been fundamental to me throughout this marvelous process known as my master's degree programme. Special thanks to my parents, whose unconditional support was decisive for the development of my career. I also want to thank my aunts and uncles, whose love and presence were a great comfort. Special thanks go to my siblings, whose encouragement was always important.

I would love to deeply thank my girlfriend, whose love, support, and empathy have been invaluable to my performance throughout this process.

I want to thank my extraordinary thesis advisor, professor, and mentor, Dr. Hector Hugo García-Compeán, whose patience, dedication, support, and orientation were absolutely crucial throughout the development of this project.

I want to thank Cinvestav—from the security guards and cleaning staff to all my colleagues (many of whom are now good friends) and all my marvelous professors—who inspired me to always work diligently and with responsibility, wherever life takes me in the future.

Finally, I want to thank CONAHCYT (now SECIHTI) for granting me a full-time scholarship that allowed me to dedicate myself 100% to my master's degree programme.

Anuar Kafuri Zarzosa          Mexico City          October 13, 2025

# Abstract

This thesis addresses a fundamental problem in deformation quantization: the difficulty of calculating the star-exponential, the symbol of the evolution operator, due to convergence issues. Inspired by the formalism that connects the star-exponential with the quantum propagator for bosonic systems, this work develops the analogous extension for the fermionic case. A rigorous method, based on Grassmann variables and coherent states, is constructed to obtain a closed-form expression for the fermionic star-exponential from its associated propagator. As a primary application, a fermionic version of the Feynman-Kac formula is derived within this formalism, allowing for the calculation of the ground state energy directly in phase space. Finally, the method is validated by successfully applying it to the simple and driven harmonic oscillators, where it is demonstrated that a simplified ("naive") approach (with an ad-hoc "remediation") is a valid weak-coupling limit of the rigorous ("meticulous") formalism, thereby providing a new and powerful computational tool for the study of fermionic systems.

# Resumen

Esta tesis aborda un problema fundamental en la cuantización por deformación: la dificultad de calcular la exponencial estrella, símbolo del operador de evolución, debido a problemas de convergencia. Inspirado en el formalismo que conecta la exponencial estrella con el propagador cuántico para sistemas bosónicos, este trabajo desarrolla la extensión análoga para el caso fermiónico. Se construye un método riguroso, basado en variables de Grassmann y estados coherentes, para obtener una expresión en forma cerrada de la exponencial estrella fermiónica a partir de su propagador asociado. Como aplicación principal, se deriva una versión fermiónica de la fórmula de Feynman-Kac dentro de este formalismo, permitiendo calcular la energía del estado base directamente en el espacio fase. Finalmente, se valida el método aplicándolo exitosamente a los osciladores armónicos (simple y forzado), donde se demuestra que un enfoque simplificado ("ingenuo" con una "remediación" ad-hoc) es un límite de acoplamiento débil válido del formalismo riguroso ("meticuloso"), proveyendo así una nueva y poderosa herramienta computacional para el estudio de sistemas fermiónicos.

# Contents

# List of Figures

# List of Tables

# Chapter 1
# Introduction

Quantum Mechanics is, without a doubt, the most prolific, precise and predictive theory scientists have conceived. Its bizarre features have challenged the classical intuitions of every generation of physicists since the founding fathers' spark of genius around one hundred years ago[1]. The notions of space, energy and time were shaken deeply, with inexorable results that defy the continuous perception of matter (v.g., Heisenberg's uncertainty principle, Pauli's exclusion principle, inter alia). It is a rich and diverse theory that admits equivalent representations (v.g., Heisenberg's picture, Schrödinger's picture, Interaction picture) as well as formulations (v.g., Standard quantum mechanics, Path integral Quantum Mechanics, etc.). In the canonical and, in a sense, *classical* formulation built by Heisenberg, Schrödinger, Jordan, Born, Dirac, Pauli, et al., the fundamental property of the physical description is the **quantum state**, rather than a function of space assigned to a particle[1]. Another key notion takes part within the vast and, per se, interesting theory of Hilbert spaces. It is, of course, the concept of an *operator*. In fact, it can be stated without loss of generality that traditional quantum mechanics combines operators and states to yield observables, without considering the specific representation one is working with.

However, there is an *elephant in the room* that is subtly omitted in the introductory courses and goes unnoticed as one works further and further into the depths of the theory (in a *Shut up and calculate!* pragmatic philosophy [2]). Our perceptions and previous models are constrained by classical mechanics: we assume matter is continuous, completely filled, perfectly definable in time and space, and absolutely deterministic; however, these principles simply do not hold as one approaches the proper energy limits and the quantum effects start to take place. It is a reasonable perception, naturally based on Bohr's correspondence principle[2], that there should be a perfectly rigorous, well-defined, and self-consistent method that connects both theories (not only predictions) with a minimal set of extra assumptions. Let us analyze this way of thinking in more detail.

---
[1]This is self-evident from the fact that there is no single wavefunction for an electron, proton, etc.
[2]The behavior of quantum systems must converge to classical physics in the limit of large quantum numbers.

This thesis is organized as follows: Chapter 1 provides a brief introduction to quantization, establishing the origins and utility of deformation quantization, and includes a concise overview of the path integral formalism. In Chapter 2, the formalism for bosons is developed, focusing on the mathematical structure of deformation quantization and constructing the framework for the star product. Chapter 3 addresses the case for fermions, emphasizing the peculiar properties arising from the Grassmannian nature of the variables. In Chapter 4, a formula for the star-exponential is derived via an integral equation for the propagator, applicable to both bosons and fermions[3], with its use demonstrated through specific examples. In Chapter 5, as an application of the formalism, a Feynman-Kac formula is derived for both bosonic and fermionic systems, yielding results consistent with the existing literature. Finally, the conclusion presents an analysis of the state-of-the-art of the theory, alongside a discussion of its limitations and open problems.

## 1.1 From the Classical to the Quantum Realm

When building the elements of quantum theory, it is common procedure to state the following recipe: $A \to \hat{A}$ and $\{A, B\}_{\text{PB}} \to \frac{1}{i\hbar}[\hat{A}, \hat{B}]$, which essentially permutes a function of coordinates $A$ to an operator $\hat{A}$ and the classical phase-space *Poisson bracket* to the commutator, respectively. Formally they can be framed in mathematical terms as follows.

**Definition 1.1.1** (Poisson bracket in canonical form [3])**.** *Let $f$ and $g$ be two smooth functions on $\mathbb{R}^{2n}$, where an element of $\mathbb{R}^{2n}$ is thought of as a pair $(x, p)$, with $x \in \mathbb{R}^n$ representing the position of a particle and $p \in \mathbb{R}^n$ representing the momentum. Then the* Poisson bracket *of $f$ and $g$, denoted $\{f, g\}_{\text{PB}}$, is the function on $\mathbb{R}^{2n}$ given by*

$$\{f, g\}_{\text{PB}} = \sum_{j=1}^{n} \left( \frac{\partial f}{\partial x_j} \frac{\partial g}{\partial p_j} - \frac{\partial f}{\partial p_j} \frac{\partial g}{\partial x_j} \right).$$

### Principal Properties of the Poisson Bracket [4]

Let $f, g, h$ be functions on phase space (in the mathematical definition 1.1.1 it is $\mathbb{R}^{2n}$) and let $\alpha, \beta$ be scalars. The Poisson bracket $\{f, g\}$ is defined for any pair of such functions and satisfies the following fundamental properties that establish the structure of Hamiltonian mechanics.

---

[3]With its corresponding subtleties.

1. **Antisymmetry (Skew-Symmetry):** The bracket changes sign upon interchange of its arguments. A direct consequence is that the Poisson bracket of any function with itself vanishes.
$$\{f, g\} = -\{g, f\}. \quad \implies \quad \{f, f\} = 0.$$

2. **Bilinearity:** The bracket operation is linear in each of its two arguments.
$$\{\alpha f + \beta g, h\} = \alpha \{f, h\} + \beta \{g, h\}.$$
$$\{f, \alpha g + \beta h\} = \alpha \{f, g\} + \beta \{f, h\}.$$

3. **Leibniz's Rule (Product Rule):** The bracket acts as a derivative with respect to products.
$$\{fg, h\} = f\{g, h\} + \{f, h\}g,$$
and due to antisymmetry:
$$\{f, gh\} = g\{f, h\} + \{f, g\}h.$$

4. **Jacobi Identity:**
$$\{f, \{g, h\}\} + \{g, \{h, f\}\} + \{h, \{f, g\}\} = 0.$$

5. **Time Evolution of Observables:** The total time derivative of any observable $f$ is given by its Poisson bracket with the system's Hamiltonian, $H$, plus any explicit time dependence.
$$\frac{df}{dt} = \{f, H\} + \frac{\partial f}{\partial t}.$$
If an observable does not explicitly depend on time ($\partial f / \partial t = 0$), it is conserved if and only if its Poisson bracket with the Hamiltonian is zero.

**Definition 1.1.2** (Commutator [5]). *Let $\hat{A}$ and $\hat{B}$ be two operators acting on a Hilbert space $\mathbb{H}$. Then the* commutator *of $\hat{A}$ and $\hat{B}$, denoted $[\hat{A}, \hat{B}]$, is the operator defined by*
$$[\hat{A}, \hat{B}] = \hat{A}\hat{B} - \hat{B}\hat{A}.$$

## Principal Properties of Commutators [5]

1. **Antisymmetry (Skew-Symmetry):** The commutator changes sign upon interchange of its operator arguments. Consequently, any operator commutes with itself.
$$[\hat{F}, \hat{G}] = -[\hat{G}, \hat{F}]. \quad \implies \quad [\hat{F}, \hat{F}] = 0.$$

2. **Bilinearity:** The commutator is linear in each of its arguments. For any scalar constants $\alpha, \beta$:
$$[\alpha \hat{F} + \beta \hat{G}, \hat{H}] = \alpha [\hat{F}, \hat{H}] + \beta [\hat{G}, \hat{H}].$$
$$[\hat{F}, \alpha \hat{G} + \beta \hat{H}] = \alpha [\hat{F}, \hat{G}] + \beta [\hat{F}, \hat{H}].$$

3. **Leibniz's Rule (Product Rule):** The commutator satisfies the same product rule identity.

$$[\hat{F}\hat{G}, \hat{H}] = \hat{F}[\hat{G}, \hat{H}] + [\hat{F}, \hat{H}]\hat{G},$$

and due to antisymmetry:

$$[\hat{F}, \hat{G}\hat{H}] = \hat{G}[\hat{F}, \hat{H}] + [\hat{F}, \hat{G}]\hat{H}.$$

4. **Jacobi Identity:** This identity holds for commutators as well, establishing that the set of operators on a Hilbert space, equipped with the commutator, also forms a Lie algebra.

$$[\hat{F}, [\hat{G}, \hat{H}]] + [\hat{G}, [\hat{H}, \hat{F}]] + [\hat{H}, [\hat{F}, \hat{G}]] = 0.$$

5. **Time Evolution of Observables (Heisenberg Equation):** The total time derivative of any quantum observable $\hat{F}$ is given by its commutator with the system's Hamiltonian operator, $\hat{H}$, plus any explicit time dependence. This is the Heisenberg equation of motion.

$$\frac{d\hat{F}}{dt} = \frac{1}{i\hbar}[\hat{F}, \hat{H}] + \frac{\partial \hat{F}}{\partial t}.$$

If an observable does not explicitly depend on time ($\partial \hat{F}/\partial t = 0$), it is a conserved quantity (a constant of motion) if and only if it commutes with the Hamiltonian, $[\hat{F}, \hat{H}] = 0$.

In canonical quantum mechanics, such as the one elaborated in detail in Cohen-Tannoudji et al's book [5], the following postulates are established.

**Postulate 1.1.1** (First Postulate). *At a fixed time $t_0$, the state of an isolated physical system is defined by specifying a ket $|\psi(t_0)\rangle$ belonging to the state space $\mathcal{E}$.*

**Postulate 1.1.2** (Second Postulate). *Every measurable physical quantity $\mathbb{A}$ is described by an operator A acting in $\mathcal{E}$; this operator is called an* observable.

**Postulate 1.1.3** (Third Postulate). *The only possible result of the measurement of a physical quantity $\mathbb{A}$ is one of the eigenvalues of the corresponding observable A.*

**Postulate 1.1.4** (Fourth Postulate — Discrete Non-Degenerate Spectrum). *When the physical quantity $\mathbb{A}$ is measured on a system in the normalized state $|\psi\rangle$, the probability $\mathbb{P}(a_n)$ of obtaining the non-degenerate eigenvalue $a_n$ of the corresponding observable A is*

$$\mathbb{P}(a_n) = |\langle u_n|\psi\rangle|^2,$$

*where $|u_n\rangle$ is the normalized eigenvector of A associated with the eigenvalue $a_n$.*

**Postulate 1.1.5** (Fourth Postulate — Discrete Degenerate Spectrum). *When the physical quantity $\mathbb{A}$ is measured on a system in the normalized state $|\psi\rangle$, the probability $\mathbb{P}(a_n)$ of obtaining the eigenvalue $a_n$ of the corresponding observable A is*

$$\mathbb{P}(a_n) = \sum_{i=1}^{g_n} \left|\langle u_n^i | \psi \rangle\right|^2,$$

*where $g_n$ is the degree of degeneracy of $a_n$, and $\{|u_n^i\rangle\}_{i=1}^{g_n}$ is an orthonormal basis for the eigensubspace $\mathcal{E}_n$ associated with the eigenvalue $a_n$.*

**Postulate 1.1.6** (Fourth Postulate — Continuous Non-Degenerate Spectrum). *When the physical quantity $\mathbb{A}$ is measured on a system in the normalized state $|\psi\rangle$, the probability $d\mathbb{P}(\alpha)$ of obtaining a result in the interval $[\alpha, \alpha + d\alpha]$ is*

$$d\mathbb{P}(\alpha) = \left|\langle \nu_\alpha | \psi \rangle\right|^2 d\alpha,$$

*where $|\nu_\alpha\rangle$ is the eigenvector corresponding to the eigenvalue $\alpha$ of the observable A associated with $\mathbb{A}$.*

**Postulate 1.1.7** (Fifth Postulate). *If the measurement of the physical quantity $\mathbb{A}$ on the system in the state $|\psi\rangle$ gives the result $a_n$, then immediately after the measurement, the system is in the normalized projection:*

$$\frac{P_n |\psi\rangle}{\sqrt{\langle \psi | P_n | \psi \rangle}},$$

*where $P_n$ is the projection operator onto the eigensubspace associated with $a_n$.*

**Postulate 1.1.8** (Sixth Postulate). *The time evolution of the state vector $|\psi(t)\rangle$ is governed by the Schrödinger equation:*

$$i\hbar \frac{d}{dt} |\psi(t)\rangle = H(t) |\psi(t)\rangle,$$

*where $H(t)$ is the observable corresponding to the total energy of the system.*

Armed with these postulates, then, it is straightforward to establish the *quantization rules*, which can be stated in the following way:

> The observable $\hat{A}$ which describes a classically defined physical quantity $A$ is obtained by replacing, in the suitably symmetrized expression for $A$, the classical variables **r** and **p** with the operators $\hat{\mathbf{R}}$ and $\hat{\mathbf{P}}$, respectively [5].

Nonetheless, Cohen is direct and states the following: **"We shall see, however, that there exist quantum physical quantities that have no classical equivalent and which are therefore defined directly by the corresponding observables (e.g., particle spin)"** [5].

We can see that this quantization scheme is not self sufficient even in its own canonical scheme. The situation can be even more dramatic, as was shown by Groenewold [6] in 1946. It is illustrative to see his famous counterexample to Dirac's quantization scheme.

## Groenewold's Counterexample

Let us study now in detail the famous identity that Groenewold used to disprove the generality of Dirac's quantization scheme [7]. The classical identity is:

$$\{x^3, p^3\}_{PB} + \frac{1}{12}\{\{p^2, x^3\}_{PB}, \{x^2, p^3\}_{PB}\}_{PB} = 0. \tag{1.1.1}$$

### 1. Classical Derivation (Poisson Brackets)

We use the definition of the Poisson Bracket (PB) for two functions $A(x, p)$ and $B(x, p)$, defined in 1.1.1, and the fundamental relation $\{x, p\}_{PB} = 1$.

**Step 1: Calculate the first term** $\{x^3, p^3\}_{PB}$

Applying the definition directly:

$$\begin{aligned}\{x^3, p^3\}_{PB} &= \frac{\partial(x^3)}{\partial x}\frac{\partial(p^3)}{\partial p} - \frac{\partial(x^3)}{\partial p}\frac{\partial(p^3)}{\partial x}, \\ &= (3x^2)(3p^2) - (0)(0), \\ &= 9x^2 p^2.\end{aligned}$$

**Step 2: Calculate the inner Poisson Brackets**

$$\begin{aligned}\text{a) } \{p^2, x^3\}_{PB} &= \frac{\partial(p^2)}{\partial x}\frac{\partial(x^3)}{\partial p} - \frac{\partial(p^2)}{\partial p}\frac{\partial(x^3)}{\partial x}, \\ &= (0)(0) - (2p)(3x^2), \\ &= -6px^2.\end{aligned}$$

$$\begin{aligned}\text{b) } \{x^2, p^3\}_{PB} &= \frac{\partial(x^2)}{\partial x}\frac{\partial(p^3)}{\partial p} - \frac{\partial(x^2)}{\partial p}\frac{\partial(p^3)}{\partial x}, \\ &= (2x)(3p^2) - (0)(0), \\ &= 6xp^2.\end{aligned}$$

**Step 3: Calculate the outer Poisson Bracket**

Now we compute the PB of the two results from the previous step:

$$\{\{p^2, x^3\}_{PB}, \{x^2, p^3\}_{PB}\}_{PB} = \{-6px^2, 6xp^2\}_{PB},$$
$$= \frac{\partial(-6px^2)}{\partial x}\frac{\partial(6xp^2)}{\partial p} - \frac{\partial(-6px^2)}{\partial p}\frac{\partial(6xp^2)}{\partial x},$$
$$= (-12px)(12xp) - (-6x^2)(6p^2),$$
$$= -144x^2p^2 + 36x^2p^2,$$
$$= -108x^2p^2.$$

**Step 4: Assemble the complete expression**

We substitute the results from Steps 1 and 3 into the original equation:

$$9x^2p^2 + \frac{1}{12}(-108x^2p^2) = 9x^2p^2 - 9x^2p^2,$$
$$= 0.$$

As expected, the classical expression is identically zero.

## 2. Quantum Derivation (Promotion to Commutators)

We now apply Dirac's quantization "recipe", which promotes variables to operators and Poisson Brackets to commutators, according to the rule $\{A, B\}_{PB} \to \frac{1}{i\hbar}[\hat{A}, \hat{B}]$, with the fundamental canonical commutation relation: $[\hat{x}, \hat{p}] = i\hbar$. Some useful results are shown in the first section of the appendix, and a general commutator identity is proven [1] , facilitating some calculations.

**Step 1: Promote the first term**

We seek the quantum analogue of $\{x^3, p^3\}_{PB}$, which is $\frac{1}{i\hbar}[\hat{x}^3, \hat{p}^3]$.

$$[\hat{x}^3, \hat{p}^3] = \hat{x}^2[\hat{x}, \hat{p}^3] + [\hat{x}^2, \hat{p}^3]\hat{x}.$$
$$\text{where } [\hat{x}, \hat{p}^3] = 3i\hbar\hat{p}^2.$$
$$\text{and } [\hat{x}^2, \hat{p}^3] = \hat{x}[\hat{x}, \hat{p}^3] + [\hat{x}, \hat{p}^3]\hat{x} = 3i\hbar(\hat{x}\hat{p}^2 + \hat{p}^2\hat{x}).$$
Substituting back:
$$[\hat{x}^3, \hat{p}^3] = \hat{x}^2(3i\hbar\hat{p}^2) + 3i\hbar(\hat{x}\hat{p}^2 + \hat{p}^2\hat{x})\hat{x},$$
$$= 3i\hbar\left(\hat{x}^2\hat{p}^2 + \hat{x}\hat{p}^2\hat{x} + \hat{p}^2\hat{x}^2\right).$$

The quantum analogue of the first term is: $\frac{1}{i\hbar}[\hat{x}^3, \hat{p}^3] = 3(\hat{x}^2\hat{p}^2 + \hat{x}\hat{p}^2\hat{x} + \hat{p}^2\hat{x}^2)$.

**Step 2: Promote the inner terms of the second term**

a) Analogue of $\{p^2, x^3\}_{PB}$ :
$$\frac{1}{i\hbar}[\hat{p}^2, \hat{x}^3] = \frac{-1}{i\hbar}[\hat{x}^3, \hat{p}^2],$$
$$= \frac{-1}{i\hbar}[3i\hbar(\hat{x}^2\hat{p} + \hat{p}\hat{x}^2)],$$
$$= -3(\hat{x}^2\hat{p} + \hat{p}\hat{x}^2).$$

b) Analogue of $\{x^2, p^3\}_{PB}$ :
$$\frac{1}{i\hbar}[\hat{x}^2, \hat{p}^3] = \frac{1}{i\hbar}[3i\hbar(\hat{x}\hat{p}^2 + \hat{p}^2\hat{x})],$$
$$= 3(\hat{x}\hat{p}^2 + \hat{p}^2\hat{x}).$$

**Step 3: Promote the outer term and calculate the anomaly**

Now we promote the outermost PB, using the quantum operators we just found:

$$\frac{1}{12}\left(\frac{1}{i\hbar}\left[-3(\hat{x}^2\hat{p} + \hat{p}\hat{x}^2), 3(\hat{x}\hat{p}^2 + \hat{p}^2\hat{x})\right]\right) = \frac{-9}{12i\hbar}\left[\hat{x}^2\hat{p} + \hat{p}\hat{x}^2, \hat{x}\hat{p}^2 + \hat{p}^2\hat{x}\right].$$

The calculation of this commutator is notoriously extensive but can be step by step deduced using [1]. The final result, however, is the following:

$$\frac{-9}{12i\hbar}\left[\hat{x}^2\hat{p} + \hat{p}\hat{x}^2, \hat{x}\hat{p}^2 + \hat{p}^2\hat{x}\right]$$
$$= -3(x^2p^2 + xp^2x + p^2x^2 + \hbar^2).$$

**Step 4: Assemble the quantum expression and conclusion**

The sum of all the operator-dependent terms in the full quantum expression exactly cancels, just as in the classical case. However, the scalar anomaly from Step 3 remains. Therefore, the quantum analogue of the vanishing classical expression is not zero:

$$\left(\text{Quantum Analogue of } \{x^3, p^3\}\right) + \left(\text{Quantum Analogue of } \frac{1}{12}\{\{...\}, \{...\}\}\right) = -3\hbar^2.$$

It is now evident that Dirac's intuitive quantisation scheme is not correct for a general mapping from classical configurations to the quantum realm. Let us now specify and formalize the principal properties that such a mapping must fulfill.

## 1.2 Quantization

Any valid quantization scheme must not only be consistent with quantum mechanical principles (i.e. from postulate 1.1.1 to postulate 1.1.8) but also provide a distinct map $Q$. This map transforms classical observables—real functions $f$ on the phase space $\Gamma$—into self-adjoint operators $Q(f)$ on the quantum Hilbert space $\mathcal{H}$ [8]. To construct a fiducial correspondence map, $Q : f \rightarrow Q(f)$, we must impose the following properties:

(i) The map reproduces the fundamental observables:
$$Q(1) = \hat{I}, \quad Q(x) = \hat{X}, \quad \text{and} \quad Q(p) = \hat{P},$$
where $\hat{I}$ is the identity operator, and $\hat{X}$ and $\hat{P}$ are the usual position and momentum operators, with $\hat{P} = -i\hbar \frac{\partial}{\partial x}$. in the position representation.

(ii) The correspondence must be linear. That is, for any constants $\alpha, \beta \in \mathbb{R}$ and observables $f, g$:
$$Q(\alpha f + \beta g) = \alpha Q(f) + \beta Q(g).$$

(iii) The map must preserve the Lie algebra structure (Dirac's condition):
$$[Q(f), Q(g)] = i\hbar Q(\{f, g\}_{\text{PB}}).$$

(iv) For any smooth function $\varphi : \mathbb{R} \to \mathbb{R}$, the map should respect function composition (von Neumann's rule) :
$$Q(\varphi \circ f) = \varphi(Q(f)).$$

In general, according to Groenewold's no-go theorem [8][6] **there does not exist a linear map that takes the Poisson algebra into the Lie algebra of the corresponding operators**. This implies that the process of quantization has more to do with educated guesses (in a sense, it is an art) than with an objective methodological criterion that connects classical and quantum mechanics. For a beautiful discussion and historical background, see Todorov's *Quantization is a mystery* [9]. To solve this problem several methods have been discussed (v.g. geometric quantization, Kähler and HyperKähler manifolds quantization, phase space quantization, etc.); we will discuss here the most simple solution: it is conceivable to assume that condition (iii) is fulfilled only asymptotically in the limit $\hbar \to 0$. This approach is commonly denominated as **deformation quantization**.

## 1.3 Deformation Quantization

The origins of deformation quantization can be found in the attempt of formulating *phase space quantum mechanics*. The following paragraphs are synthesized from [10]. Readers are encouraged to explore the ideas further in this outstanding and concise book.

**Quantum Mechanics in Phase Space (QMPS)**

The formulation of quantum mechanics in phase space (QMPS) was born from a veridical paradox. Initial criticisms, famously advanced by figures like Niels Bohr, contended that the very notion of phase-space trajectories was **fundamentally incompatible** with the uncertainty principle [4]. This objection, traceable to the metaphysics of the Copenhagen interpretation [12], suggested that any attempt to assign **simultaneous** position and momentum values to a quantum particle was misguided. However, this perspective overlooks a deeper truth: QMPS is not a naive classical theory but a fully consistent and powerful reformulation of quantum mechanics, entirely equivalent to the standard Hilbert space and path integral approaches. The key insight, which took decades to fully crystallize, is that classical *c-number variables like position ($x$) and momentum ($p$) can indeed coexist with quantum rules*, provided that the **underlying mathematical algebra is deformed to enforce non-commutativity**.

This resolution has its roots in the early contributions of several independent pioneers in quantum theory. In 1927, Hermann Weyl introduced a seminal correspondence rule mapping phase-space functions to Weyl-ordered quantum operators, believing he had found the definitive quantization prescription [5]. Shortly after, in 1932, Eugene Wigner developed what is now known as the **Wigner function** to calculate quantum corrections to classical statistical mechanics. In doing so, he made the profound discovery that this quasi-probability distribution could take on **negative values**—a direct signature of quantum nonlocality [14] and interference. Around the same time, John von Neumann used the Weyl correspondence to prove the **uniqueness of Schrödinger's representation** [15] and, in his work, implicitly discovered the convolution rule for operator symbols, though he did not pursue its full implications. These foundational contributions provided the essential, although disconnected, pieces of a new quantum picture.

It was not until the 1940s, however, that these distinct elements were forged into a complete and autonomous theory through the independent, wartime breakthroughs of Hilbrand Groenewold and Joe Moyal. In 1946, Groenewold, having developed the theory autonomously during the war, introduced the modern star product ($\star$) as the mathematical mechanism needed to **reconcile classical variables with quantum non-commutativity**. His work culminated in the celebrated Groenewold-Van Hove theorem, which proved that **a naive map between Poisson brackets and commutators is impossible**; instead, quantum commutators represent a non-trivial deformation of their classical counterparts. Independently, in 1949, Joe Moyal arrived at an equivalent formalism using characteristic functions, linking the Wigner function to expectation values and formulating time evolution through what is now called the Moyal bracket—the explicit deformation of the Poisson bracket.

The elegance of this formalism rests upon a trio of powerful mathematical structures. The cornerstone is the star product ($\star$), a noncommutative binary operation that replaces standard multiplication. It deforms the algebra of phase-space functions such that they replicate the operator algebra of the Heisenberg picture, perfectly encoding the relation $x \star p - p \star x = i\hbar$. The second structure is the Wigner function $F(x, p)$, a quasi-probability density whose permitted negative values are not a flaw but a feature, embodying the uncertainty principle by occupying phase-space domains no larger than $(\hbar/2)^n$. Finally, the Moyal bracket, defined as the antisymmetrized star product $\{\{A, B\}\} \equiv (A \star B - B \star A)/(i\hbar)$, serves as the quantum generalization of the Poisson bracket, contracting to it in the classical limit as $\hbar \to 0$.

---

[4]Indeed Bohr claimed "it was obvious that such trajectories violated the uncertainty principle" [11]

[5]The Weyl-Wigner-Moyal (WWM) correspondence essentially establishes an isomorphism that connects the Heisenberg operator algebra with the corresponding algebra of operator symbols. Within this framework, the conventional product of operators is mapped to the associative and noncommutative Moyal $\star$ product [13].

Despite its internal consistency and elegance, the phase-space formulation faced significant resistance, most notably from Paul Dirac, who in a 1945 letter to Moyal dismissed the approach as "not neat" and claimed it "obviously" violated the uncertainty principle, seemingly overlooking the subtleties of Wigner's earlier work. This opposition, combined with the concurrent rise of Richard Feynman's powerful path integral formalism, left QMPS largely overshadowed for decades. Its vindication came in the 1950s and 1960s through the work of physicists like Takabayasi and Baker, who rigorously established its logical autonomy and formal equivalence to standard quantum mechanics. For a masterful account of this historical journey, from its controversial beginnings to its eventual vindication, the reader is encouraged to consult the detailed review in [10] for deeper insights. From that point, it gained traction for its unique advantages in specialized fields, solidifying its place as a valid and insightful quantum formalism.

Let us study now how the modern version of deformation quantization (dq) came into existence.

**First Era**

The foundational papers are [16] and [17], in which F. Bayen, M. Flato, C. Fronsdal, A. Lichnerowicz, and D. Sternheimer (hereafter, the BFFLS papers) set the stage for the formal study of deformation quantization. In the first BFFLS paper, they present a mathematical study of the differentiable deformations of the algebras associated with phase space. They were the first to recognize that deformations of the Lie algebra of $C^\infty$ functions generalize the Moyal bracket, while deformations of the associative algebra of $C^\infty$ functions give rise to noncommutative and associative algebras isomorphic to the operator algebras of quantum theory. In particular, by studying deformations invariant under any Lie algebra of distinguished observables, they generalized the usual quantization procedure based on Heisenberg's algebra [16]. On the other hand, the second paper shows via some examples that the spectra of physical observables can be determined by direct phase-space methods [17]. In this sense, both papers were crucial in establishing that deformation theory and quantization can be constructed as a complete and autonomous quantum theory.

Although these results were fundamental in constructing the theory and laying its foundations, they were limited to phase space functions. Moyal's product is in this sense a special case of another class of products, fact that took around 20 years to be rigorously proven.

**Second Era**

M. Kontsevich transformed the state-of-the-art in deformation quantization by constructing another star product in addition to Moyal's. More specifically, he proved that *every finite-dimensional Poisson manifold X admits a canonical deformation quantization* [18]. He also extended and finally proved what was then the most important open problem in the field: the formality conjecture, which connects the Lie superalgebra of polyvector fields on X with the Hochschild complex associated with the algebra of functions on X.

**Utility and problems**

It is a self-consistent, independent, and powerful formulation that makes it possible to quantize some intractable open problems (e.g., gauge theories which are both commutative [19] and non-commutative [20]) and offers an eclectic view and mechanism to merge the quantum and classical realms *asymptotically*.

Today, the significance of deformation quantization is undeniable. It provides a natural framework for studying noncommutative spacetimes, offers direct and intuitive access to semiclassical limits through $\hbar$-expansions, and is an indispensable tool in quantum optics and for modeling decoherence in quantum computing. It emerged from a series of independent breakthroughs to resolve an apparent paradox, and in doing so, gifted us a self-consistent and powerful perspective on the deep connection between the classical and quantum realms.

However, there is an important problem that remains open: the question of convergence of Star functions [8]. In fact, this renders the theoretically solid formalism useless because one cannot predict a measurable result—such as the spectrum of a given physical system—that could ultimately be tested in the laboratory. Such a test would be necessary to consolidate deformation quantization as a powerful formalism with explanatory power.

Besides this, there is a clever way to circumvent this issue through a result that connects star exponentials to propagators—the building blocks of Feynman's path integral formulation of quantum mechanics—as will be explained in the next section.

## 1.4 Path Integrals proposal

Feynman et al. [21] proposed an alternative formulation of canonical quantum mechanics in which the key ingredients are the following [22]:

- **Lagrangian**[6] **and action principle**

$$L(q, \dot{q}, t) = \tfrac{1}{2}m\dot{q}^2 - V(q),$$

---
[6]For simplicity the simplest model is assumed

$$S[q] = \int_{t_i}^{t_f} L\big(q(t), \dot{q}(t), t\big)\, dt.$$

The lagrangian encodes the dynamics of the system, while the action is a functional of the entire path q(t).

- **Propagator**

$$K(q_f, t_f; q_i, t_i) = \big\langle q_f \big| e^{-\frac{i}{\hbar}H(t_f - t_i)} \big| q_i \big\rangle,$$

$$= \int_{q(t_i)=q_i}^{q(t_f)=q_f} \mathcal{D}q \; \exp\!\Big(\tfrac{i}{\hbar} S[q]\Big).$$

$K(q_f, t_f; q_i, t_i)$ is the basic quantum amplitude that connects $(t_i, q_i) \to (t_f, q_f)$, while $\mathcal{D}q$ integrates over all paths q(t) with those endpoints, though it may not be rigorosuly defined as a *measure*.

- **Wick Rotation and Euclidean Action**

$$t \to -i\tau, \qquad S[q] \to i\, S_E[q], \qquad iS[q]/\hbar \to -S_E[q]/\hbar.$$

where

$$S_E = \int \left[ \tfrac{1}{2} m \left(\frac{dq}{d\tau}\right)^2 + V(q) \right] d\tau$$

It is a useful application of complex analysis to yield a dissipative action principle that helps construct the following item.

- **Partition function**

$$Z(\beta) = \operatorname{Tr} e^{-\beta H},$$

$$= \int_{q(0)=q(\beta\hbar)} \mathcal{D}q(\tau) \; \exp\!\Big(-\tfrac{1}{\hbar} S_E[q]\Big), \quad \beta = \tfrac{1}{k_B T}.$$

The partition function, Z, is the generating object for thermodynamic quantities in statistical mechanics. It's also extensively used in quantum field theory (QFT) and string theory (ST) to calculate the spectrum of a specific model.

- **Generating Functional**

$$Z[J] = \int \mathcal{D}q \; \exp\!\Big(\tfrac{i}{\hbar} S[q] + \tfrac{i}{\hbar}\! \int dt\, J(t)\, q(t)\Big).$$

If we introduce an external source J(t) to build Green's functions we arrive at the result below.

- **Correlation Function (n-points in this case)**

$$\langle q(t_1) \cdots q(t_n) \rangle = \frac{1}{Z[0]} \, (-i\hbar)^n \, \frac{\delta^n Z[J]}{\delta J(t_1) \cdots \delta J(t_n)} \bigg|_{J=0}.$$

These correlation functions encode all the observable dynamics.

The path integral formulation starts with the system's Lagrangian, from which the action functional S[q] is constructed via temporal integration. The system's quantum mechanical propagator (or transition amplitude) K is then determined by substituting this action into the real-time path integral, which functionally integrates over all trajectories with fixed endpoints. To analyze the system's thermal and statistical properties, a Wick rotation to imaginary time ($\tau$) is performed. This procedure maps the action S[q] to its Euclidean counterpart $S_E[q]$. The resulting Euclidean path integral, evaluated over all paths with periodic boundary conditions, yields the partition function Z($\beta$). Furthermore, by introducing an external source J(t), one defines the generating functional Z[J]. This functional is a powerful tool, as all n-point correlation functions of the system can be systematically derived by its repeated functional differentiation with respect to the source. Finally, a perturbative expansion of Z[J] in the theory's interaction terms gives rise to Feynman diagrams. The rules for constructing these diagrams are read directly from the quadratic (free) and higher-order (interaction) parts of the action.

The path integral formalism thereby **provides a unified and cohesive framework that encompasses transition amplitudes, partition functions, and correlation functions**. For a deeper discussion of these notions for bosons, fermions, and supersymmetric (SUSY) systems from both quantum mechanical and quantum field theory perspectives, see the extraordinary book by A. Das, *Field Theory: A Path Integral Approach* [22].

As we will see in the following chapters, there's an ingenious way to connect the propagator of a given Hamiltonian with its associated star exponential and, finally, to estimate its ground state energy via the Feynman-Kac formula. This allows the predictions to be compared with known academic results.

# Chapter 2
# WWM formalism for bosonic classical systems

It is now important to outline the basic mathematical properties that Deformation Quantization comprises. For a more detailed exposition consider reading the outstanding work ellaborated by A. Hirshfel: *Current Aspects of Deformation Quantization* [23] and G. Karaali:*Deformation Quantization - a brief summary* [24]. The first yet pretty old-fashioned notation can be found in [25].

## 2.1 The Mathematics of Deformation Quantization

**The Moyal ⋆ Product**

The archetypal example of a star product, also referred to as a *formal deformation* of the algebra (C.4), is the Moyal ∗-product, which applies to the simple case of a flat manifold $M = \mathbb{R}^d$ equipped with a constant Poisson structure [24]. This structure is defined by the constant bivector $\alpha$:

$$\alpha = \sum_{i,j} \frac{1}{2} \alpha^{ij} \frac{\partial}{\partial x_i} \wedge \frac{\partial}{\partial x_j}. \tag{2.1.1}$$

where $\alpha^{ij} = -\alpha^{ji} \in \mathbb{R}$, $x_i$ are the coordinates on $\mathbb{R}^d$. For any two functions $f$ and $g$, this bivector induces the familiar Poisson bracket:

$$\{f, g\} = \sum_{i,j} \alpha^{ij} \frac{\partial f}{\partial x_i} \frac{\partial g}{\partial x_j}. \tag{2.1.2}$$

The Moyal $*$-product then arises by exponentiating this Poisson operator to deform the standard pointwise multiplication of functions:

$$f * g = \left( \exp(-i\frac{\hbar}{2} \sum_{i,j} \alpha^{ij} \frac{\partial}{\partial x_i} \wedge \frac{\partial}{\partial x_j}) \right)(f, g). \tag{2.1.3}$$

In canonical coordinates it is expressed as follows:

$$(f *_M g)(q, p) = f(q, p) \exp\left( \frac{i\hbar}{2} (\overleftarrow{\partial}_q \overrightarrow{\partial}_p - \overleftarrow{\partial}_p \overrightarrow{\partial}_q) \right) g(q, p), \tag{2.1.4}$$

where $\overleftarrow{\partial}_i$ operates on $f$ and $\overrightarrow{\partial}_j$ on $g$, and is also equivalent to the following shift formula:

$$(f *_M g)(q, p) = f\left( q + \frac{i\hbar}{2} \overrightarrow{\partial}_p, p - \frac{i\hbar}{2} \overrightarrow{\partial}_q \right) g(q, p), \tag{2.1.5}$$

or as a Fourier integral:

$$(f *_M g)(q, p) = \frac{1}{\hbar^2 \pi^2} \int dq_1 dq_2 dp_1 dp_2 f(q_1, p_1) g(q_2, p_2)$$
$$\times \exp\left[ \frac{2}{i\hbar} (q_2(p_1 - p) + p_2(q - q_1) + (pq_1 - qp_1)) \right]. \tag{2.1.6}$$

The integral representation of Star products will be crucial for our comparison with the Path integral formalism, and it is interesting to see the following geometric interpretation [23]: Denote points in phase space by vectors, for example in 2 dimensions

$$\vec{r} = \begin{pmatrix} q \\ p \end{pmatrix}, \quad \vec{r}_1 = \begin{pmatrix} q_1 \\ p_1 \end{pmatrix}, \quad \vec{r}_2 = \begin{pmatrix} q_2 \\ p_2 \end{pmatrix}. \tag{2.1.7}$$

Now consider the triangle in phase space shown below.

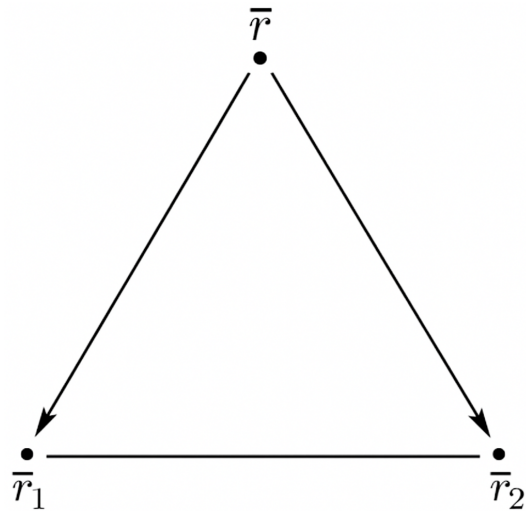

**Figure 2.1:** Phase-space triangle

Its area [1] is given by:

$$A(\vec{r}, \vec{r}_1, \vec{r}_2) = \frac{1}{2}(\vec{r} - \vec{r}_1) \wedge (\vec{r} - \vec{r}_2) = \frac{1}{2}[p(q_2 - q_1) + p_1(q - q_2) + p_2(q_1 - q)], \quad (2.1.8)$$

which is proportional to the exponent in the previous formula. Therefore

$$(f * g)(\vec{r}) = \int d\vec{r}_1 \wedge d\vec{r}_2 f(\vec{r}_1) g(\vec{r}_2) \exp\left[\frac{4i}{\hbar} A(\vec{r}, \vec{r}_1, \vec{r}_2)\right]. \quad (2.1.9)$$

This construction represents the first known non-trivial example of a deformation quantization.

The star product is fundamentally *nonlocal* in all its representations. This means that to calculate the product of two functions at a single point $(q, p)$, you need more than just the values of the functions at that specific point. Instead, the calculation requires *global* information about the functions, which is equivalent to knowing either all of their derivatives at that point or their values across the entire phase space. The underlying idea can be generalized from flat space to any Poisson manifold that possesses a flat, torsion-free connection [24].

### Elementary cohomology and Kontsevich's star product

From Moyal's star product one can deduce that another possibility could emerge if one restricts the exponential to the first functional term i.e.

$$e^{aA+bB} \to e^{aA} \quad \forall a, b \in \mathbb{C} \quad (2.1.10)$$

which yields the following star product

$$f *_S g = f \, e^{i\hbar \overleftarrow{\partial}_q \overrightarrow{\partial}_p} g. \quad \text{(Standard Star Product)}$$

While the following definitions are not the most rigorous, they are sufficient for our purposes. For greater mathematical precision, the reader is referred to [24]; for a formal deduction of the general result, consider [18].

Two star products, $*$ and $*'$, are considered c-equivalent if they're related by an invertible operator, $T$. This operator is a formal power series in $\hbar$, where the $T_n$ are differential operators:

$$T = \sum_{n=0}^{\infty} \hbar^n T_n. \quad (2.1.11)$$

The equivalence condition requires that $T$ acts as a homomorphism between the two deformed algebras:

$$T(f *' g) = Tf * Tg. \quad (2.1.12)$$

---

[1]symplectic volume.

A key example is the relationship between the Moyal ($*_M$) and standard ($*_S$) star products, which are c-equivalent. The specific transition operator relating them is:

$$T = \exp\left[-\frac{i\hbar}{2}\vec{\partial}_q\vec{\partial}_p\right]. \tag{2.1.13}$$

It's crucial to understand that c-equivalence—a mathematical property rooted in cohomology—does not in itself imply physical equivalence. For a more precise introduction to the concept, see the appendix.

We can expand Moyal's star product in a formal series and obtain the following:

$$f * g = fg + \frac{i\hbar}{2}\alpha^{ij}(\partial_i f)(\partial_j g) + \frac{1}{2!}\left(\frac{i\hbar}{2}\right)^2 \alpha^{ij}\alpha^{kl}(\partial_{ik}f)(\partial_{jl}g) + \cdots . \tag{2.1.14}$$

Kontsevich introduced a powerful graphical calculus to represent the star product as an infinite series of diagrams. In this visual language, the functions $f$ and $g$ are vertices (dots), and differential operators $\partial_i$ are arrows that terminate on these vertices. The bivector $\alpha^{ij}$ is represented by another type of vertex from which pairs of arrows originate. For instance, the first-order term $\alpha^{ij}(\partial_i f)(\partial_j g)$ corresponds to a simple diagram with two arrows pointing to $f$ and $g$:

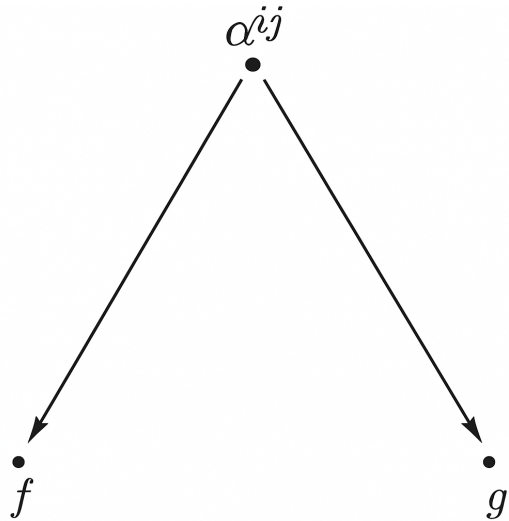

**Figure 2.2:** Vertex diagram

For the simple Moyal product on a flat space, where $\alpha^{ij}$ is constant, the complete series is the sum of all such possible diagrams, which conveniently exponentiates:

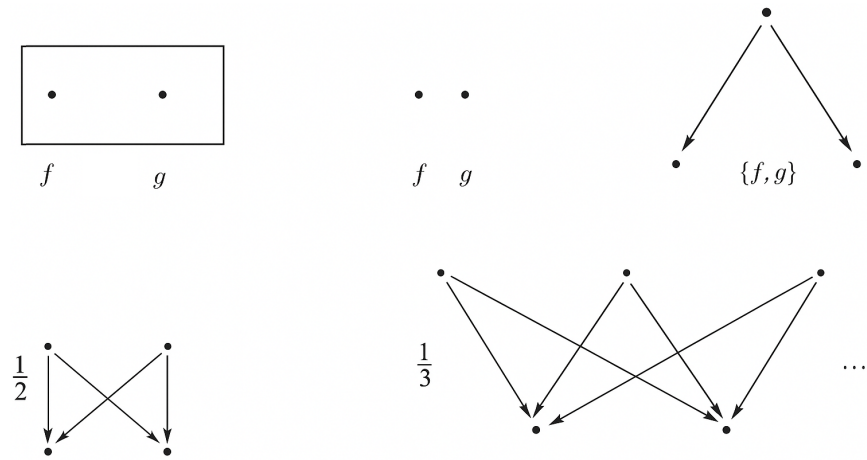

**Figure 2.3:** Series expansion for the Moyal product

Kontsevich's breakthrough was generalizing this to any Poisson manifold by also allowing the derivative-arrows to terminate on the $\alpha$ vertices themselves, accounting for the fact that $\alpha^{ij}(x)$ is no longer constant

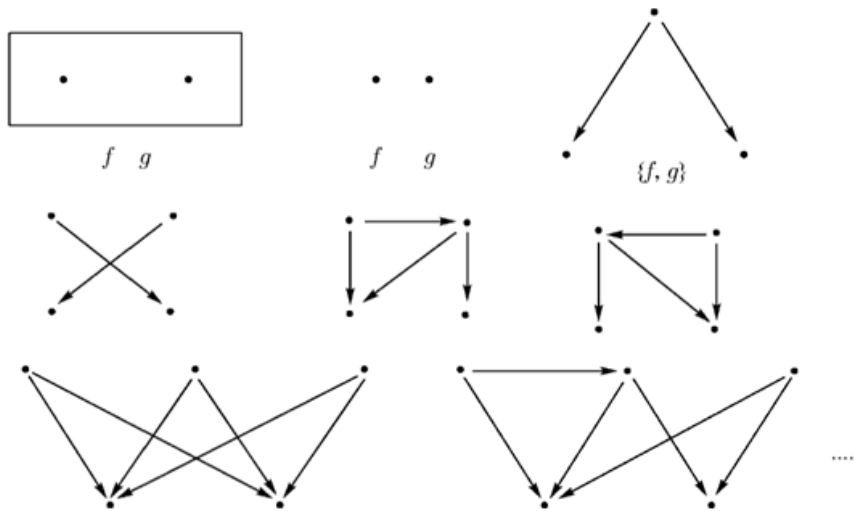

**Figure 2.4:** Series expansion for the Kontsevich product

The full Kontsevich star product is then a weighted sum over all "admissible" graphs, with a rule that excludes loop diagrams:

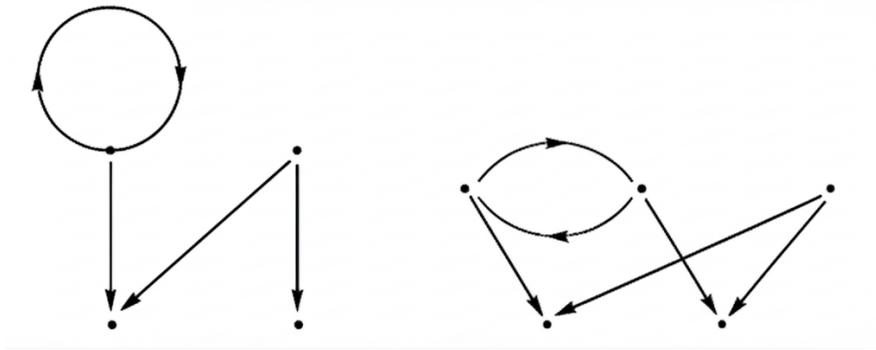

**Figure 2.5:** Loop graphs

Kontsevich also provided a prescription to calculate the weight of each graph via specific angular integrals.

The formula for the first few terms in the resulting series for the Kontsevich product is:

$$f * g = fg + \frac{i\hbar}{2}\alpha^{ij}(\partial_i f)(\partial_j g) + \left(\frac{i\hbar}{2}\right)^2 \left\{\frac{1}{2}\alpha^{ij}\alpha^{kl}(\partial_{ik}f)(\partial_{jl}g) \right. $$
$$\left. + \frac{1}{3}\alpha^{im}[(\partial_m \alpha^{jk})(\partial_{ij}f)(\partial_k g) + (\partial_{ij}g)(\partial_k f)]\right\} + \cdots \quad . \quad (2.1.15)$$

Although there are some subtleties in considering its exponentiation compared to the Moyal case, when the Poisson structure is nondegenerate (e.g., when it is symplectic), the Poisson coefficients are constants in the Darboux coordinate system, and the Kontsevich product reduces to the Moyal product.

This expanding field of study incorporates approaches from both pure mathematics and theoretical physics, but its essential point is that **by choosing a star product, one chooses a quantization scheme**. We will now introduce the notation used throughout this work, consistent with that of [8] for the description of bosons and [13] for the case of fermions.

## 2.2 Deformation Quantization and star products

Let us start by considering Weyl's quantization map [8]. It provides a precise rule for connecting the classical and quantum worlds. It takes a classical observable—a square-integrable function $f(\mathbf{x}, \mathbf{p}) \in L^2(\mathbb{R}^{2n})$ on the phase space $\Gamma = \mathbb{R}^{2n}$—and associates it with a corresponding quantum observable. This quantum observable is an operator acting on the Hilbert space $\mathcal{H} = L^2(\mathbb{R}^n)$ as follows:

$$\mathcal{Q}_W(f) = \frac{1}{(2\pi)^n}\int_{\mathbb{R}^{2n}} \widetilde{f}(\mathbf{a},\mathbf{b})e^{i(\mathbf{a}\cdot\widehat{\mathbf{X}}+\mathbf{b}\cdot\widehat{\mathbf{P}})}d\mathbf{a}\,d\mathbf{b}, \qquad (2.2.1)$$

where $\tilde{f}(\mathbf{a}, \mathbf{b})$ denotes the Fourier transform on $\mathbb{R}^{2n}$, and the components of the operators $\hat{\mathbf{X}}$, $\hat{\mathbf{P}}$ satisfy the canonical commutation relations $[\hat{X}_i, \hat{P}_j] = i\hbar\delta_{ij}$, with $i, j = 1, \ldots, n$. For the rigorous mathematical construction of Weyl's quantization see Yufan Ge's work *The Weyl Quantization: a brief introduction* [3]. By explicitly writing the Fourier transform of a classical observable, $\tilde{f}(a, b)$, as

$$\tilde{f}(\mathbf{a}, \mathbf{b}) = \frac{1}{(2\pi)^n} \int_{\mathbb{R}^{2n}} f(\mathbf{u}, \mathbf{v}) \, e^{-i(\mathbf{u}\cdot\mathbf{a} + \mathbf{v}\cdot\mathbf{b})} \, d\mathbf{u} \, d\mathbf{v}, \tag{2.2.2}$$

the Weyl quantization map $Q_W(f)$ can be expressed in the following integral form:

$$Q_W(f) = \frac{1}{(2\pi\hbar)^{2n}} \int_{\mathbb{R}^{2n}} f(\mathbf{x}, \mathbf{p}) \, \hat{\Omega}(\mathbf{x}, \mathbf{p}) \, d\mathbf{x} \, d\mathbf{p}. \tag{2.2.3}$$

Here $\hat{\Omega}(\mathbf{x}, \mathbf{p})$ is the Weyl-Stratonovich operator, which is defined by the integral:

$$\hat{\Omega}(\mathbf{x}, \mathbf{p}) = \left(\frac{\hbar}{2\pi}\right)^n \int_{\mathbb{R}^{2n}} e^{i\left(\mathbf{a}\cdot(\hat{\mathbf{X}} - \mathbf{x}) + \mathbf{b}\cdot(\hat{\mathbf{P}} - \mathbf{p})\right)} \, d\mathbf{a} \, d\mathbf{b}. \quad \text{(Weyl-Stratonovich Quantizer)}$$

It admits a well defined trace [26] and satisfies the following properties:

$$\operatorname{tr}\left(\hat{\Omega}(\mathbf{x}, \mathbf{p})\right) = 1. \quad \text{(Normalization)}$$

$$\operatorname{tr}\left(\hat{\Omega}(\mathbf{x}, \mathbf{p}) \, \hat{\Omega}(\mathbf{x}', \mathbf{p}')\right) = 2\pi\hbar \, \delta(\mathbf{x} - \mathbf{x}') \, \delta(\mathbf{p} - \mathbf{p}'). \quad \text{(Orthogonality)}$$

$$\hat{\Omega}(\mathbf{x}, \mathbf{p})^\dagger = \hat{\Omega}(\mathbf{x}, \mathbf{p}). \quad \text{(Self-Adjointness)}$$

By applying the Weyl quantization map on any $\psi \in L^2(R^n)$, we obtain

$$Q_W(f)\psi(\mathbf{x}) = \frac{1}{(2\pi\hbar)^n} \int_{\mathbb{R}^{2n}} f\left(\frac{\mathbf{x} + \mathbf{y}}{2}, \mathbf{p}\right) e^{-\frac{i}{\hbar}(\mathbf{y}-\mathbf{x})\cdot\mathbf{p}} \psi(\mathbf{y}) \, d\mathbf{p} \, d\mathbf{y},$$

$$= \int_{\mathbb{R}^n} \kappa_f(\mathbf{x}, \mathbf{y}) \, \psi(\mathbf{y}) \, d\mathbf{y}. \tag{2.2.4}$$

where $\kappa_f(\mathbf{x}, \mathbf{y}) \in L^2(\mathbb{R}^{2n})$ is the integral kernel characterizing the Hilbert–Schmidt (HS) operator $Q_W(f) := \hat{F} \in \text{HS}(L^2(\mathbb{R}^n))$ [26].

The inverse Weyl quantization map can be found by multiplying equation (2.2.3) by the operator $\hat{\Omega}(\mathbf{x}, \mathbf{p})$ and then taking the trace of both sides. This procedure yields:

$$Q_W^{-1}(\hat{F})(\mathbf{x}, \mathbf{p}) = \operatorname{tr}\left(\hat{\Omega}(\mathbf{x}, \mathbf{p}) \, \hat{F}\right),$$

$$= \hbar^n \int_{\mathbb{R}^n} \kappa_f\left(\mathbf{x} - \tfrac{\hbar}{2}\mathbf{y}, \mathbf{x} + \tfrac{\hbar}{2}\mathbf{y}\right) e^{i\mathbf{y}\cdot\mathbf{p}} \, d\mathbf{y}. \tag{2.2.5}$$

where $\hat{F} \in \text{HS}(L^2(\mathbb{R}^n))$ is a Hilbert–Schmidt operator. Weyl's inversion formula defines a symbol f from its quantization $\hat{F}$ [26].

Using the Weyl map and its inverse, we can determine the Wigner function associated with a given density operator $\hat{\rho}$. This function is the phase-space equivalent of the density operator acting on the Hilbert space $L^2(\mathbb{R}^n)$.

Let $\hat{\rho}$ be the density operator associated with the quantum state $\psi \in L^2(R^n)$ that is:

$$\hat{\rho} = \hat{\rho}^{\dagger} \quad . \qquad \text{(self-adjoint)}$$

$$\hat{\rho} \geq 0 \quad . \qquad \text{(positive semi-definite)}$$

$$\text{tr}(\hat{\rho}) = 1 \quad . \qquad \text{(normalized)}$$

written as an integral operator as follows:

$$\hat{\rho}\varphi(x) = \psi(x) \int_{\mathbb{R}^n} \overline{\psi(y)}\, \varphi(y)\, dy, \tag{2.2.6}$$

and equivalently in Dirac's notation:

$$\hat{\rho} = |\psi\rangle\langle\psi|. \tag{2.2.7}$$

By using (2.2.5) one can deduce its corresponding symbol:

$$\rho(\boldsymbol{x}, \boldsymbol{p}) = \frac{1}{(2\pi\hbar)^n} \int_{\mathbb{R}^n} \psi\left(\boldsymbol{x} + \frac{\boldsymbol{y}}{2}\right) \overline{\psi}\left(\boldsymbol{x} - \frac{\boldsymbol{y}}{2}\right) e^{-\frac{i}{\hbar}\boldsymbol{y}\cdot\boldsymbol{p}} d\boldsymbol{y}. \tag{2.2.8}$$

This is the famous Wigner function. In direct analogy to statistical mechanics, it characterizes quantum states using a quasi-probability distribution [2] on phase space. Additionally, the Wigner function can be used to calculate the expectation value of an operator by integrating its corresponding symbol over phase space, as follows:

$$\langle \psi, \hat{A}\psi \rangle = \int_{\mathbb{R}^{2n}} \rho(\mathbf{x}, \mathbf{p})\, A(\mathbf{x}, \mathbf{p})\, d\mathbf{x}\, d\mathbf{p}, \tag{2.2.9}$$

where $A = Q_W^{-1}(\hat{A})$ is the symbol associated with the operator $\hat{A} \in \text{HS}(L^2(\mathbb{R}^n))$, obtained via the inverse Weyl quantization map.

### Star Products

Since the Weyl map $Q_W : L^2(\mathbb{R}^{2n}) \to \mathcal{HS}(L^2(\mathbb{R}^n))$ is bijective and the product of Hilbert-Schmidt operators is closed [8], it implies that there is a unique function in $L^2(\mathbb{R}^{2n})$, denoted by $f \star g$, such that:

$$Q_W(f)\, Q_W(g) = Q_W(f \star g). \tag{2.2.10}$$

---

[2] The Wigner function can take negative values, distinguishing it from a true probability distribution

Moyal's star product is characterized by using the inverse Weyl map[3] as:

$$(f \star g)(\mathbf{x}, \mathbf{p}) = Q_W^{-1}\left(Q_W(f)\, Q_W(g)\right),$$
$$= \mathrm{tr}\left(\hat{\Omega}(\mathbf{x}, \mathbf{p})\, \hat{F}\, \hat{G}\right). \tag{2.2.11}$$

Using (2.2.3) and (2.2.5) yields the integral representation of the star product:

$$(f \star g)(\mathbf{x}, \mathbf{p}) = \frac{1}{(\pi\hbar)^2} \int_{\mathbb{R}^4} f(\mathbf{a}, \mathbf{b}) g(\mathbf{c}, \mathbf{d}) e^{-\frac{2i}{\hbar}(\mathbf{p}\cdot(\mathbf{a}-\mathbf{c})+\mathbf{x}\cdot(\mathbf{d}-\mathbf{b})+(\mathbf{c}\cdot\mathbf{b}-\mathbf{a}\cdot\mathbf{d}))}\, d\mathbf{a}\, d\mathbf{b}\, d\mathbf{c}\, d\mathbf{d}\ . \tag{2.2.12}$$

It is straightforward to prove that it satisfies:

$$f \star (g \star h) = (f \star g) \star h, \quad \text{(Associativity)} \tag{2.2.13}$$

$$f \star g \neq g \star f. \quad \text{(Noncommutativity)} \tag{2.2.14}$$

The former arises directly from the integral representation performing a detailed calculation, while the latter comes from the Baker-Campbell-Hausdorff (BCH) identity:

$$e^{\hat{A}} e^{\hat{B}} = \exp\left(\hat{A} + \hat{B} + \frac{1}{2}[\hat{A}, \hat{B}] + \frac{1}{12}[\hat{A}, [\hat{A}, \hat{B}]] - \frac{1}{12}[\hat{B}, [\hat{A}, \hat{B}]] + \cdots\right). \tag{2.2.15}$$

By Taylor expanding functions $f, g \in C^\infty(\mathbb{R}^{2n})$ inside the integral representation, one obtains the Moyal star product in its differential representation:

$$(f \star g)(\mathbf{x}, \mathbf{p}) = \sum_{m,n=0}^{\infty} \left(\frac{i\hbar}{2}\right)^{m+n} \frac{(-1)^m}{m!\, n!} \left(\partial_\mathbf{p}^m \partial_\mathbf{x}^n f\right)\left(\partial_\mathbf{p}^n \partial_\mathbf{x}^m g\right),$$
$$= f(x, p) \exp\left[\frac{i\hbar}{2}\left(\overleftarrow{\partial}_\mathbf{x} \cdot \overrightarrow{\partial}_\mathbf{p} - \overleftarrow{\partial}_\mathbf{p} \cdot \overrightarrow{\partial}_\mathbf{x}\right)\right] g(x, p). \tag{2.2.16}$$

where the differential operators $\overleftarrow{\partial}_\mathbf{x}, \overleftarrow{\partial}_\mathbf{p}$ act to the left (on $f$), and $\overrightarrow{\partial}_\mathbf{x}, \overrightarrow{\partial}_\mathbf{p}$ act to the right (on $g$). In parallel with the standard Hilbert space formalism, **spectral properties in deformation quantization are determined by solving star-eigenvalue equations**. For the specific case of a stationary Wigner function, the structure of the star product leads to the following expression:

$$H(\mathbf{x}, \mathbf{p}) \star \rho(\mathbf{x}, \mathbf{p}) = H\left(\mathbf{x} + \frac{i\hbar}{2}\overrightarrow{\partial}_\mathbf{p},\ \mathbf{p} - \frac{i\hbar}{2}\overrightarrow{\partial}_\mathbf{x}\right) \rho(\mathbf{x}, \mathbf{p}),$$
$$= E\, \rho(\mathbf{x}, \mathbf{p}). \tag{2.2.17}$$

where $H(\mathbf{x}, \mathbf{p})$ is the classical Hamiltonian function and $E$ corresponds to the energy eigenvalues of the quantum Hamiltonian operator, $\hat{H}\psi = E\psi$, in the Schrödinger representation.

---

[3]Also known as Wigner map

**Evolution of the system**

The time evolution of a system's density operator, $\hat{\rho}$, in the Schrödinger picture is governed by the von Neumann equation

$$\frac{\partial \hat{\rho}}{\partial t} = -\frac{i}{\hbar}[\hat{H}, \hat{\rho}]. \tag{2.2.18}$$

For a time-independent Hamiltonian, $\hat{H}$, the formal solution to this equation is given by:

$$\hat{\rho}(t) = \hat{U}(t)\hat{\rho}(0)\hat{U}^\dagger(t). \tag{2.2.19}$$

where $\hat{U}(t) = e^{-i\hat{H}t/\hbar}$ is the unitary time evolution operator.

Within the deformation quantization framework, the Weyl correspondence translates the von Neumann equation into its phase-space equivalent, the Moyal equation of motion for the Wigner function:

$$\begin{aligned}\frac{\partial \rho(\mathbf{x}, \mathbf{p})}{\partial t} &= -\frac{i}{\hbar}\left(H \star \rho - \rho \star H\right), \\ &= \{\rho, H\}_M.\end{aligned} \tag{2.2.20}$$

Its solution is obtained formally in terms of the **star exponential**:

$$\text{Exp}_\star\left(-\frac{i}{\hbar}tH\right) \equiv \sum_{n=0}^{\infty} \frac{1}{n!}\left(-\frac{i}{\hbar}t\right)^n H^{\star n}, \tag{2.2.21}$$

with $H^{\star n} = H \star H \cdots \star H$ ($n$ factors) as,

$$\rho(t) = Exp_\star\left(-\frac{i}{\hbar}tH\right) \star \rho \star Exp_\star\left(\frac{i}{\hbar}tH\right). \tag{2.2.22}$$

where $\rho(t)$ denotes the Wigner function at time $t$.

A key property of the star exponential is that, when treated as a distribution, it admits a Fourier-Dirichlet expansion:

$$Exp_\star\left(-\frac{i}{\hbar}tH\right) = \sum_{n=0}^{\infty} e^{-\frac{i}{\hbar}tE_n} \rho_n. \tag{2.2.23}$$

where $E_n$ denotes the eigenvalues of the Hamiltonian operator $\hat{H}$. The functions $\rho_n$ are the phase-space symbols obtained by applying the inverse Weyl map (2.2.5) to the projection operator $\hat{P}_n = |n\rangle\langle n|$, which is associated with the normalized energy eigenstate $|n\rangle$. These symbols are defined such that:

$$H \star \rho_n = E_n \rho_n, \tag{2.2.24}$$

$$\rho_m \star \rho_n = \delta_{mn} \rho_n. \tag{2.2.25}$$

To determine the star exponential, $\text{Exp}_\star(-itH/\hbar)$, for a given Hamiltonian $H(x,p)$, one must solve the following star-differential equation:

$$H \star \text{Exp} \star \left(-\frac{i}{\hbar}tH\right) = i\hbar \frac{d}{dt}\text{Exp} \star \left(-\frac{i}{\hbar}tH\right). \quad (2.2.26)$$

As proven in [16], a solution to this equation, denoted $\varphi(t, H)$, can be constructed only if its power series in $t$ converges in the sense of distributions. This means that $\varphi(t, H)$ must be a continuous linear functional acting on the space of smooth functions with compact support, $C_c^\infty(\mathbb{R}^{2n})$.

The investigation of convergence for star products presents major obstacles due to the highly nontrivial analytic structure of the formal series appearing in (2.2.26). Furthermore, as mentioned in [27], the quest for convergence remains one of the most important open problems in deformation quantization. To overcome the above-mentioned difficulties, we will investigate the properties of the star exponential and its connection to the path integral formalism.

# Chapter 3
# WWM formalism for fermionic classical systems

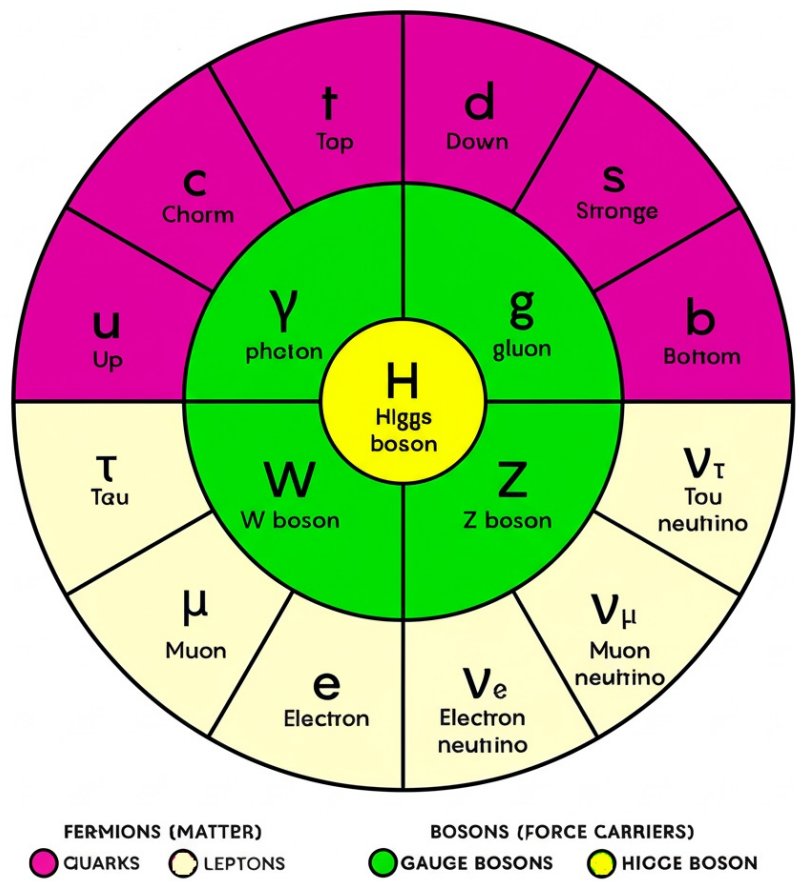

**Figure 3.1:** Standard model of elementary particles

The bosonic discussion we had before implicitly assumed some properties of matter that we did not make explicit. It is now necessary to outline in a clear and introductory style the fundamental differences concerning the nature of distinct particles. According to the standard model of particle physics, the other type of particles besides bosons are fermions. As can be seen in (3.1), this bi-modal description of the constituents of matter gives rise to all the other ones: leptons, quarks, scalar bosons, gauge bosons, etc. Besides the standard model, there is a powerful result in quantum field theory, called the *Spin-Statistics theorem*, that ensures the existence of both types of particles. The reader is encouraged to visit the paper *Spin-Statistics Theorem in Path Integral Formulation* [28] by K. Fujikawa for a precise discussion of it and a nonstandard proof based on path integrals, which is very adequate for our purposes in this work. Let us see some fundamental differences that characterize fermions in opposition to bosons.

1. **Spin** According to the strong conditions imposed by the Spin-Statistics theorem, bosons have integer spin numbers (e.g., 0, 1, ...) whereas fermions have half-integer spin numbers (e.g., 1/2, 3/2, ...).
2. **Statistics** Statistics in this context can be understood as the proper description of the distribution that a large collection of identical particles acquire among the distinct energy levels (eigenvalues).
   Bosons obey the Bose-Einstein distribution:

   **Average occupation number of bosons in a quantum state with energy $\epsilon$**

   $$\langle n(\epsilon) \rangle = \frac{1}{e^{(\epsilon-\mu)/k_B T} - 1}, \qquad (3.0.1)$$

   where:

   $\epsilon$ : energy of the quantum state,
   $\mu$ : chemical potential ,
   $k_B$ : Boltzmann constant,
   $T$ : absolute temperature in kelvins.

   While fermions do the analogous with Fermi-Dirac statistics:

   **Average occupation number of fermions in a quantum state with energy $\epsilon$**

   $$\langle n(\epsilon) \rangle = \frac{1}{e^{(\epsilon-\mu)/k_B T} + 1}. \qquad (3.0.2)$$

3. **Pauli exclusion principle** No two identical fermions can "occupy" the same quantum state simultaneously; in other words, the configuration of quantum numbers within a specific state is unique. This key property will be the cornerstone of the mathematical description we will employ to address and extend the results we arrived at for the case of bosons.

**Table 3.1:** Comparison between Bose-Einstein and Fermi-Dirac Distributions

| Feature | Bose-Einstein | Fermi-Dirac |
| --- | --- | --- |
| Particle Type | Bosons (integer spin) | Fermions (half-integer spin) |
| Denominator sign | $-1$ | $+1$ |
| Max occupancy | Unlimited | 1 (Pauli exclusion) |
| Examples | Photons, Higgs Boson | Electrons, Neutrons |

As we have seen, a proper physical description of fermions must account for the three properties summarized in Table (3.1). Fulfilling this requirement necessitates the use of a different kind of mathematical object, which in turn introduces subtleties to the standard tools of mathematical analysis—such as differentiation, integration, and approximation methods, *inter alia*. We are, of course, referring to **Grassmann variables**.

### 3.0.1 Grassmann variables

For an in-depth discussion of this topic within the context of field theory, see the marvelous book by A. Das [22], from which the following results are derived.

To properly describe the physics of fermions, it is necessary to introduce a different class of mathematical objects known as **Grassmann variables**. These are defined by their fundamental property of being completely **anticommuting**. For a set of Grassmann variables $\theta_i$, this means:

$$\theta_i \theta_j + \theta_j \theta_i = 0, \quad i,j = 1, 2, \ldots, n. \tag{3.0.3}$$

A crucial and immediate consequence of this rule is that Grassmann variables are **nilpotent**: the square of any single variable is identically zero:

$$\theta_i^2 = 0. \tag{3.0.4}$$

This property dramatically simplifies functions of Grassmann variables. For instance, any function of a single variable $\theta$ can be expressed *exactly* by a simple linear Taylor expansion:

$$f(\theta) = a + b\theta. \tag{3.0.5}$$

**Grassmann Calculus: Differentiation**

The anticommuting nature of these variables requires a careful definition of differentiation. One must specify the direction of the derivative, leading to distinct **left** and **right** derivatives which yield different results, for example:

$$\text{(Right Derivative):} \quad \frac{\partial}{\partial \theta_i}(\theta_j \theta_k) = \theta_j \left(\frac{\partial \theta_k}{\partial \theta_i}\right) - \left(\frac{\partial \theta_j}{\partial \theta_i}\right) \theta_k$$
$$= \delta_{ik}\theta_j - \delta_{ij}\theta_k. \tag{3.0.6}$$
$$\text{(Left Derivative):} \quad \frac{\partial}{\partial \theta_i}(\theta_j \theta_k) = \left(\frac{\partial \theta_j}{\partial \theta_i}\right) \theta_k - \theta_j \left(\frac{\partial \theta_k}{\partial \theta_i}\right),$$
$$= \delta_{ij}\theta_k - \delta_{ik}\theta_j. \tag{3.0.7}$$

For all subsequent discussions, we will adopt the convention of using **left derivatives**. These derivatives form an algebra with properties analogous to the variables themselves: they also anticommute and are nilpotent.

$$\frac{\partial}{\partial \theta_i}\frac{\partial}{\partial \theta_j} + \frac{\partial}{\partial \theta_j}\frac{\partial}{\partial \theta_i} = 0. \tag{3.0.8}$$

$$\left(\frac{\partial}{\partial \theta_i}\right)^2 = 0. \tag{3.0.9}$$

Furthermore, the fundamental relation between a derivative and a coordinate is given not by a commutator, but by an **anticommutator**:

$$\left\{\frac{\partial}{\partial \theta_i}, \theta_j\right\} = \frac{\partial}{\partial \theta_i}\theta_j + \theta_j\frac{\partial}{\partial \theta_i},$$
$$= \delta_{ij}. \tag{3.0.10}$$

**Grassmann Calculus: Integration**

Integration over Grassmann variables **(Berezin integration)** is defined by properties analogous to ordinary calculus, namely that the integral of a total derivative vanishes and the derivative of an integral is zero.

$$ID = 0, \qquad DI = 0. \tag{3.0.11}$$

Since the Grassmann derivative is nilpotent, it already satisfies these conditions. This leads to the remarkable and powerful conclusion that for Grassmann variables, **integration is identical to differentiation**:

$$I = D. \tag{3.0.12}$$

Thus, for a function of a single Grassmann variable, the integral is defined as:

$$\int d\theta \, f(\theta) = \frac{\partial f(\theta)}{\partial \theta}. \tag{3.0.13}$$

This definition immediately yields the **fundamental integration rules**:

$$\int d\theta = 0, \qquad \int d\theta \, \theta = 1. \tag{3.0.14}$$

This is a profound departure from the calculus of ordinary commuting variables. An immediate consequence is an inverted rule for the change of variables. If $\theta' = a\theta$, then:

$$\int d\theta\, f(\theta) = a \int d\theta'\, f\left(\frac{\theta'}{a}\right). \tag{3.0.15}$$

This generalizes to multiple variables, where the Jacobian appears in the numerator. For a linear transformation $\theta'_i = a_{ij}\theta_j$:

$$\int \prod_{i=1}^n d\theta_i f(\theta_i) = (\det a_{ij}) \int \prod_{i=1}^n d\theta'_i f\left((a^{-1})_{ij}\theta'_j\right). \tag{3.0.16}$$

### Delta Function and the Gaussian Integral

The unique properties of Grassmann calculus allow for a simple definition of the **delta function**:

$$\delta(\theta) = \theta. \tag{3.0.17}$$

One can verify that this satisfies the required normalization and shifting properties:

$$\int d\theta\, \delta(\theta) = \int d\theta\, \theta = 1. \tag{3.0.18}$$

$$\int d\theta\, \delta(\theta) f(\theta) = \int d\theta\, \theta(a + b\theta) = a = f(0). \tag{3.0.19}$$

Its integral representation is also straightforward:

$$\int d\zeta\, e^{i\zeta\theta} = i\theta = i\delta(\theta). \tag{3.0.20}$$

The most important result for physical applications is the **Gaussian integral**. For two sets of independent Grassmann variables and sources, the integral is:

$$I = \int \prod_{i,j} d\theta^*_i\, d\theta_j\, e^{-(\theta^*_i M_{ij} \theta_j + c^*_i \theta_i + \theta^*_i c_i)}. \tag{3.0.21}$$

By performing a suitable change of variables, one arrives at the result:

$$I = N(\det M_{ij}) e^{c^*_i M^{-1}_{ij} c_j}. \tag{3.0.22}$$

The critical feature here is the **positive power of the determinant** ($\det M_{ij}$) in the numerator. This is the exact opposite of the result for an integral over ordinary (bosonic) variables, where the determinant would appear in the denominator.

We will now establish the foundational results for the fermionic case. This section will state the sign and integral conventions used throughout this work, as well as the key equations needed for the subsequent chapters on star exponentials and the Feynman-Kac formula. For a more in-depth discussion, the reader is referred to the paper *Weyl-Wigner-Moyal Formalism for Fermi Classical Systems* by I. Galaviz, H. Garcia-Compean, M. Przanowski, and F.J. Turrubiates [13].

### 3.0.2 Preliminaries

Let $\psi = (\psi_1, \ldots, \psi_n)$ and $\pi = (\pi_1, \ldots, \pi_n)$ be complex Grassmann coordinates on the phase space $\Gamma_F^{2n}$. The momenta $\pi_j$ are given by

$$\pi_j = i\psi_j^*, \quad j = 1, \ldots, n. \tag{3.0.23}$$

Canonical quantisation rules establish:

$$\{\widehat{\psi}_j, \widehat{\pi}_k\}_+ = i\hbar \delta_{jk}, \tag{3.0.24}$$

$$\{\widehat{\psi}_j, \widehat{\psi}_k\}_+ = 0 = \{\widehat{\pi}_j, \widehat{\pi}_k\}_+, \tag{3.0.25}$$

$$\{\widehat{\psi}_j, \widehat{\psi}_k^*\}_+ = \hbar \delta_{jk}. \tag{3.0.26}$$

The proper creation and annihilation operators can be built as:

$$\widehat{b}_j := \frac{\widehat{\psi}_j}{\sqrt{\hbar}}, \quad \widehat{b}_j^* := \frac{\widehat{\psi}_j^*}{\sqrt{\hbar}}. \tag{3.0.27}$$

The vacuum state is given by the ket $|0\rangle$ or bra $\langle 0|$ such that:

$$\widehat{b}_j|0\rangle = 0, \quad \langle 0|\widehat{b}_j^* = 0, \quad \forall j, \tag{3.0.28}$$

satisfying the normalization condition $\langle 0|0\rangle = 1$. The basis of all states can be constructed from excitations of the vacuum state $|0\rangle$ and it is given by

$$|j, k, l, \ldots\rangle := \widehat{b}_j^* \widehat{b}_k^* \widehat{b}_l^* \cdots |0\rangle. \tag{3.0.29}$$

Consider that if $l \notin \{j, k, \ldots\}$ one gets

$$\widehat{b}_l|j, k, \ldots\rangle = 0, \tag{3.0.30}$$

$$\widehat{b}_l^*|j, k, \ldots\rangle = |l, j, k, \ldots\rangle. \tag{3.0.31}$$

Moreover

$$\widehat{b}_l|l, j, k, \ldots\rangle = |j, k, \ldots\rangle, \tag{3.0.32}$$

$$\widehat{b}_l^*|l, j, k, \ldots\rangle = 0. \tag{3.0.33}$$

Analogously the dual basis is defined as follows:

$$\langle j, k, l, \ldots | := \langle 0| \cdots \widehat{b}_l \widehat{b}_k \widehat{b}_j. \tag{3.0.34}$$

It is straightforward to verify that

$$\langle j_1, k_1, l_1, \ldots | j_2, k_2, l_2, \ldots\rangle = \begin{cases} 0 & \text{if } \{j_1, k_1, l_1, \ldots\} \neq \{j_2, k_2, l_2, \ldots\} \\ 1 & \text{if } j_1 = j_2, k_1 = k_2, l_1 = l_2, \ldots \end{cases}. \tag{3.0.35}$$

433### 3.0.3 Coherent states for fermions

We look for the state $|\psi_1,\ldots,\psi_n\rangle \equiv |\psi\rangle$ that satisfies the following (eigenvalue) condition[29]:

$$\widehat{\psi}_j|\psi\rangle = \psi_j|\psi\rangle \quad \forall j. \tag{3.0.36}$$

It is straightforward to see that $|\psi\rangle$ has the following form

$$|\psi\rangle = \exp\left\{-\frac{i}{\hbar}\sum_{j=1}^n \widehat{\pi}_j \psi_j\right\}|0\rangle. \tag{3.0.37}$$

Indeed

$$\widehat{\psi}_k|\psi\rangle = \widehat{\psi}_k \exp\left\{-\frac{i}{\hbar}\widehat{\pi}_k\psi_k\right\} \cdot \exp\left\{-\frac{i}{\hbar}\sum_{j\neq k}\widehat{\pi}_j\psi_j\right\}|0\rangle,$$

$$= \psi_k \exp\left\{-\frac{i}{\hbar}\sum_{j\neq k}\widehat{\pi}_j\psi_j\right\}|0\rangle,$$

$$= \psi_k|\psi\rangle. \tag{3.0.38}$$

Where we have used $\widehat{\psi}_k|0\rangle = 0$. and

$$\widehat{\psi}_k \exp\left\{-\frac{i}{\hbar}\sum_{j\neq k}\widehat{\pi}_j\psi_j\right\} = \exp\left\{-\frac{i}{\hbar}\sum_{j\neq k}\widehat{\pi}_j\psi_j\right\}\widehat{\psi}_k, \tag{3.0.39}$$

$$\psi_k \exp\left\{-\frac{i}{\hbar}\sum_{j\neq k}\widehat{\pi}_j\psi_j\right\} = \psi_k \exp\left\{-\frac{i}{\hbar}\sum_{j=1}^n \widehat{\pi}_j\psi_j\right\}. \tag{3.0.40}$$

Combining the previous results it is not difficult to get the crucial translation formula:

$$\exp\left\{-\frac{i}{\hbar}\sum_{j=1}^n \widehat{\pi}_j\xi_j\right\}|\psi\rangle = |\psi + \xi\rangle. \tag{3.0.41}$$

Now let us find the corresponding bra. We seek a solution of the form:

$$\langle\psi|\widehat{\psi}_j = \langle\psi|\psi_j \quad \forall j. \tag{3.0.42}$$

It is a bit cumbersome but one can deduce it is [1]

$$\langle\psi| := \langle 0|\widehat{\psi}_1 \cdots \widehat{\psi}_n \exp\left\{-\frac{i}{\hbar}\sum_{j=1}^n \psi_j\widehat{\pi}_j\right\},$$

$$= \langle 0|\left(\prod_{J=1}^n \widehat{\psi}_J\right)\exp\left\{\frac{i}{\hbar}\sum_{J=1}^n \widehat{\pi}_J\psi_J\right\}. \tag{3.0.43}$$

---

[1] The ordering of $\widehat{\psi}_j$ can be arbitrarily chosen but then must be fixed. We choose the ordering $\widehat{\psi}_1\cdots\widehat{\psi}_n$ for simplicity.

One can immediately see that contrary to the bosonic case we have, $\langle\psi| \neq (|\psi\rangle)^*$. We can also see that the ket $|\psi\rangle$ commutes with all Grassmann numbers $\eta$, i.e., $\eta|\psi\rangle = |\psi\rangle\eta$. But for the bra defined for $\langle\psi|$ one arrives at:

$$\eta\langle\psi| = (-1)^{e_\eta \cdot n}\langle\psi|\eta, \tag{3.0.44}$$

where $e_\eta = 1$ for odd Grassmann numbers or $e_\eta = 0$ for even Grassmann numbers. The inner product $\langle\psi'|\psi\rangle$ satisfies:

$$\langle\psi'|\psi\rangle = \langle 0|\widehat{\psi}_1 \cdots \widehat{\psi}_n \exp\left\{\frac{i}{\hbar}\sum_{J=1}^{n}\widehat{\pi}_J\left(\psi'_j - \psi_j\right)\right\}|0\rangle,$$

$$= \prod_{J=1}^{n}\left(\psi_j - \psi'_j\right),$$

$$=: \delta(\psi - \psi'). \tag{3.0.45}$$

In the case of the $\pi$ basis one can find the following analogous results:

$$|\pi\rangle = \exp\left\{-\frac{i}{\hbar}\sum_{J=1}^{n}\widehat{\psi}_j\pi_j\right\}\prod_{J=1}^{n}\widehat{\pi}_j|0\rangle, \tag{3.0.46}$$

$$\langle\pi| = \langle 0|\exp\left\{-\frac{i}{\hbar}\sum_{J=1}^{n}\pi_j\widehat{\psi}_j\right\} = \langle 0|\exp\left\{\frac{i}{\hbar}\sum_{J=1}^{n}\widehat{\psi}_j\pi_j\right\}. \tag{3.0.47}$$

Self-evidently $\langle\pi| \neq (|\pi\rangle)^*$. The inner product $\langle\pi'|\pi\rangle$ yields:

$$\langle\pi'|\pi\rangle = \prod_{J=1}^{n}\left(\pi'_j - \pi_j\right),$$

$$= \delta(\pi' - \pi). \tag{3.0.48}$$

Mixing kets and bras of $\psi$ and $\pi$ one finds:

$$\langle\pi|\psi\rangle = \exp\left\{-\frac{i}{\hbar}\sum_{J=1}^{n}\pi_J\psi_j\right\}, \tag{3.0.49}$$

$$\langle\psi|\pi\rangle = (i^n\hbar)^n \exp\left\{\frac{i}{\hbar}\sum_{j=1}^{n}\pi_j\psi_j\right\}. \tag{3.0.50}$$

We will adhere to the integral conventions from Weinberg's book [30]:

$$\int \psi_j d\psi_j = -\int d\psi_j \psi_j = 1, \quad \int \pi_j d\pi_j = -\int d\pi_j \pi_j = 1 \quad \forall j. \tag{3.0.51}$$

which yield:

$$\int |\psi\rangle \mathcal{D}\psi \langle\psi| = 1, \tag{3.0.52}$$

where $\mathcal{D}\psi = d\psi_n \cdots d\psi_1$. Analogously one finds:

$$\int |\pi\rangle (-1)^n \mathcal{D}\pi \langle \pi| = 1. \qquad (3.0.53)$$

It is now important to deduce one cornerstone of the formalism: the Stratonovich-Weyl quantiser.

### 3.0.4 Stratonovich-Weyl quantiser

Let us start considering Weyl's map ($f = f(\pi, \psi) \in \Gamma^{2n}$ *a classical observable*):

$$Q_W(f) = \int \widetilde{f}(\lambda, \mu) \exp\left\{ i \sum_{j=1}^n (\hat{\pi}_j \lambda_j + \hat{\psi}_j \mu_j) \right\} \prod d\lambda d\mu, \qquad (3.0.54)$$

where $\prod d\lambda d\mu := d\lambda_1 d\mu_1 \cdots d\lambda_n d\mu_n$, $\widetilde{f}(\lambda, \mu)$ is the fourier transform of f:

$$\widetilde{f}(\lambda, \mu) := \int f(\pi, \psi) \exp\left\{ -i \sum_{j=1}^n (\pi_j \lambda_j + \psi_j \mu_j) \right\} \prod d\pi d\psi, \qquad (3.0.55)$$

and $\prod d\pi d\psi := d\pi_1 d\psi_1 \cdots d\pi_n d\psi_n$. Using this explicit expansion of the transform and considering f to be a smooth function on $\Gamma_F^{2n}$ we arrive at:

$$Q_W(f) = \int f(\pi, \psi) \hat{\Omega}(\pi, \psi) \prod d\pi d\psi, \qquad (3.0.56)$$

where the operator $\hat{\Omega}(\pi, \psi)$ corresponds to the Stratonovich-Weyl operator/quantiser, given by:

$$\hat{\Omega}(\pi, \psi) = \int \exp\left\{ i \sum_{j=1}^n [(\hat{\pi}_j - \pi_j)\lambda_j + (\hat{\psi}_j - \psi_j)\mu_j] \right\} \prod d\lambda d\mu. \qquad (3.0.57)$$

Using the translation (3.0.41) and the normalization equations (3.0.52) we find after some calculations the following equivalent expression:

$$\hat{\Omega}(\pi, \psi) = i^n \int D\lambda \exp\left\{ -i \sum_{j=1}^n \pi_j \lambda_j \right\} \left| \psi - \frac{\hbar\lambda}{2} \right\rangle \left\langle \psi + \frac{\hbar\lambda}{2} \right|. \qquad (3.0.58)$$

In the $\pi$−basis it takes the form:

$$\hat{\Omega}(\pi, \psi) = (-1)^n \int \hat{\Omega}(\pi, \psi) |\pi'\rangle \mathcal{D}\pi' \langle \pi'|,$$

$$= (-i)^n \int \mathcal{D}\mu \exp\left\{ -i \sum_{j=1}^n \psi_j \mu_j \right\} \left| \pi - \frac{\hbar\mu}{2} \right\rangle \left\langle \pi + \frac{\hbar\mu}{2} \right|, \qquad (3.0.59)$$

where we have used (3.0.46) and (3.0.53).

This operator admits a well-defined trace:

$$\text{tr}\{\widehat{A}\} := c \int \mathcal{D}\psi \langle\psi|\widehat{A}|\psi\rangle, \tag{3.0.60}$$

where $c$ is to be determined from the condition

$$\text{tr}\{\widehat{\Omega}(\pi, \psi)\} = 1. \tag{3.0.61}$$

It yields the following equation

$$\text{tr}\{\widehat{A}\} = (i\hbar)^{-n} \int \mathcal{D}\psi \langle\psi|\widehat{A}|\psi\rangle,$$
$$= (i\hbar)^{-n} \int \mathcal{D}\pi \langle\pi|\widehat{A}|\pi\rangle. \tag{3.0.62}$$

which satisfies the following properties:

$$\text{tr}\{\widehat{\Omega}(\pi', \psi')\widehat{\Omega}(\pi'', \psi'')\} = (\psi'_1 - \psi''_1)(\pi'_1 - \pi''_1) \cdots (\psi'_n - \psi''_n)(\pi'_n - \pi''_n),$$
$$= \delta(\psi' - \psi'', \pi' - \pi''). \tag{3.0.63}$$

$$\text{tr}\{\widehat{\Omega}(\pi, \psi)\} = 1. \tag{3.0.64}$$

$$\text{tr}\{\widehat{A}\widehat{B}\} = (-1)^{\hat{e}_A \hat{e}_B} \text{tr}\{\widehat{B}\widehat{A}\}. \tag{3.0.65}$$

where $\hat{e}_{A(B)}$ are the Grassmann parities of $\hat{A}(\hat{B})$.

Consider now Weyl's inverse quantization map:

$$Q_W^{-1}(f(\pi, \psi)) = tr\{\hat{f}\widehat{\Omega}(\pi, \psi)\}. \tag{3.0.66}$$

It defines a symbol f from its quantisation $\hat{f}$. We will restrict our analysis to the case of integrable functions for simplicity. Employing (3.0.52) and (3.0.65) one arrives at the following relation:

$$f(\pi, \psi) = \text{tr}\{\hat{f}\widehat{\Omega}(\pi, \psi)\},$$
$$= \text{tr}\{\widehat{\Omega}(\pi, \psi)\hat{f}\}. \tag{3.0.67}$$

Using Weyl's quantisation map and its inverse we can calculate the Wigner function corresponding to the density operator acting on the Hilbert space:

$$\rho_W(\pi, \psi) = \text{tr}\{\hat{\rho}\widehat{\Omega}(\pi, \psi)\},$$
$$= (i\hbar)^{-n} \int \mathcal{D}\psi' \langle\psi'|\hat{\rho}\widehat{\Omega}(\pi, \psi)|\psi'\rangle,$$
$$= \hbar^{-n} \int \mathcal{D}\lambda \exp\left\{-i\sum_{j=1}^{n} \pi_j \lambda_j\right\} \left\langle \psi + \frac{\hbar\lambda}{2} \middle| \hat{\rho} \middle| \psi - \frac{\hbar\lambda}{2} \right\rangle. \tag{3.0.68}$$

It is a *quasi probability* distribution function, which can be used to determine the expectation values of operators by integrating their associated symbols over the grassmann phase space:

$$\langle \widehat{A} \rangle = \int \rho(\pi, \psi) A(\pi, \psi) \prod d\pi d\psi, \qquad (3.0.69)$$

where $A = Q_W^{-1}(\widehat{A})$. is obtained using Weyl's inverse quantisation map.

Now let us explore the star products for the case of fermions.

### 3.0.5 Star products

Using the bijectivity of Weyl's map and the closure of Hilbert-Schmidt operators we find that there is a unique function on $\Gamma_F^{2n}$, $f \star g$ :

$$Q_W(f)Q_W(g) = Q_W(f \star g). \qquad (3.0.70)$$

It is the called the **Moyal star product**:

$$(f \star g)(\pi, \psi) = Q_W^{-1}(Q_W(f)Q_W(g)),$$
$$= \text{tr}\left\{\widehat{f}\widehat{g}\widehat{\Omega}(\pi, \psi)\right\}. \qquad (3.0.71)$$

We need the following identity. The details of its deduction are provided in [13].

$$\text{tr}\{\widehat{\Omega}(\pi', \psi')\widehat{\Omega}(\pi'', \psi'')\widehat{\Omega}(\pi, \psi)\} = \left(\frac{i\hbar}{2}\right)^{2n} \exp\left\{-\frac{2i}{\hbar}(\pi'(\psi'' - \psi) + \pi''(\psi - \psi') + \pi(\psi' - \psi''))\right\}. \qquad (3.0.72)$$

Combining it with (3.0.56) inside the trace we find:

$$(f \star g)(\pi, \psi) = \left(\frac{i\hbar}{2}\right)^{2n} \int f(\pi', \psi') g(\pi'', \psi'')$$
$$\times \exp\left\{-\frac{2i}{\hbar}[\pi'(\psi'' - \psi) + \pi''(\psi - \psi') + \pi(\psi' - \psi'')]\right\} \mathcal{D}\pi' \mathcal{D}\psi' \mathcal{D}\pi'' \mathcal{D}\psi''. \qquad (3.0.73)$$

By changing variables: $\Psi'' = \psi' - \psi$, $\Pi'' = \pi' - \pi$, $\Psi''' = \psi'' - \psi$, $\Pi''' = \pi'' - \pi$, the Moyal product takes the form

$$(f \star g)(\pi, \psi) = \left(\frac{i\hbar}{2}\right)^{2n} \int f(\Pi' + \pi, \Psi'' + \psi) g(\Pi'' + \pi, \Psi'' + \psi)$$
$$\times \exp\left\{-\frac{2i}{\hbar}[\Pi''\Psi'' - \Pi''\Psi']\right\} \mathcal{D}\Pi' \mathcal{D}\Psi'' \mathcal{D}\Pi'' \mathcal{D}\Psi'', \qquad (3.0.74)$$

which is an integral representation of the star product.

Expanding $f$ and $g$ into a Taylor series and after some manipulations it yields:

$$(f \star g)(\pi, \psi) = f(\pi, \psi) \exp\left\{\frac{i\hbar}{2}\hat{\mathcal{P}}_F\right\} g(\pi, \psi), \qquad (3.0.75)$$

where

$$\hat{\mathcal{P}}_F = \frac{\overleftarrow{\partial}}{\partial \pi}\frac{\overrightarrow{\partial}}{\partial \psi} + \frac{\overleftarrow{\partial}}{\partial \psi}\frac{\overrightarrow{\partial}}{\partial \pi}, \qquad (3.0.76)$$

with $\overrightarrow{\partial}$ and $\overleftarrow{\partial}$ denoting the right and left derivatives, respectively.

With these subtleties established, the previous bosonic analysis for the Moyal bracket, as well as the star-genvalue equations holds. Let us explore in the following chapters its connection with propagators and with the Feynman-Kac formula.

# Chapter 4
# Star exponentials from propagators

For clarity, we'll treat the bosonic and fermionic cases independently in separate sections

## 4.1 Star exponential for bosons

Full details and a thorough discussion can be found in *Star Exponentials from Propagators and Path Integrals* by J. Berra–Montiel, H. Garcia-Compean, and A. Molgado [8].

We begin by considering a quantum particle moving under the influence of a potential $V(x)$. The system's Hamiltonian is given by:

$$\hat{H} = \frac{\hat{P}^2}{2m} + V(\hat{X}). \tag{4.1.1}$$

In the coordinate representation, the corresponding Cauchy initial value problem is the time-dependent Schrödinger equation:

$$i\hbar \frac{\partial \psi}{\partial t} = -\frac{\hbar^2}{2m} \frac{\partial^2 \psi}{\partial x^2} + V(x)\psi, \tag{4.1.2}$$

subject to an initial condition of the form:

$$\psi(x, t)|_{t=t_0} = \psi(x_0, t_0). \tag{4.1.3}$$

Under general conditions on the potential $V(x)$ (such as local integrability and boundedness), this Cauchy problem admits a fundamental solution, denoted by $K(x, t; x_0, t_0)$. This solution satisfies the Schrödinger equation (4.1.2) and meets the initial condition [26]:

$$\lim_{t \to t_0} K(x, t; x_0, t_0) = \delta(x - x_0). \tag{4.1.4}$$

Consequently, the solution to the Cauchy problem can be written as an integral equation:

$$\psi(x, t) = \int_{\mathbb{R}} K(x, t; x_0, t_0)\psi(x_0, t_0)dx_0. \tag{4.1.5}$$

This fundamental solution $K(x, t; x_0, t_0)$ is called the **quantum mechanical propagator** or transition amplitude. In Dirac's terminology, the propagator is written as:

$$K(x, t; x_0, t_0) = \langle x, t | x_0, t_0 \rangle, \tag{4.1.6}$$
$$= \langle x | e^{-\frac{i}{\hbar}\hat{H}(t-t_0)} | x_0 \rangle. \tag{4.1.7}$$

If the spectrum of the Hamiltonian is known, the propagator has a spectral decomposition:

$$K(x, t; x_0, t_0) = \sum_{n=0}^{\infty} e^{-\frac{i}{\hbar}E_n(t-t_0)} \psi_n(x)\psi_n^*(x_0), \tag{4.1.8}$$

where $\{\psi_n(x)\}_{n=0}^{\infty}$ is a complete set of eigenfunctions of $\hat{H}$ with corresponding eigenvalues $E_n$.

Our next task is to express the quantum propagator, $K(x_f, t_f; x_0, t_0)$, as an integral transform of the star exponential. This is achieved by finding the phase-space symbol associated with the time evolution operator via the inverse Weyl quantization map.

Since the star product defines a homomorphism between classical observables ($C^{\infty}(\mathbb{R}^2)$) and operators on the Hilbert space ($L^2(\mathbb{R})$), the symbol corresponding to the formal evolution operator,

$$e^{-\frac{i}{\hbar}t\hat{H}} = \hat{I} - \frac{it}{\hbar}\hat{H} + \frac{1}{2!}\left(-\frac{it}{\hbar}\right)^2 \hat{H}^2 + \cdots, \tag{4.1.9}$$

is given by the star exponential [16, 10]:

$$\text{Exp}_\star\left(-\frac{i}{\hbar}tH\right) = 1 - \frac{it}{\hbar}H + \frac{1}{2!}\left(-\frac{it}{\hbar}\right)^2 H \star H + \cdots. \tag{4.1.10}$$

To connect this to the propagator, we consider a non-diagonal density operator $\hat{\rho}_{f,0} = |x_0\rangle\langle x_f|$, where $|x_0\rangle$ and $|x_f\rangle$ are eigenstates of the position operator $\hat{X}$. Using the Weyl inversion formula (2.2.5), the corresponding non-diagonal Wigner function is:

$$\rho_{f,0}(x, p) := Q_W^{-1}(|x_0\rangle\langle x_f|),$$
$$= \frac{1}{2\pi\hbar} e^{\frac{i}{\hbar}(x_f-x_0)p} \delta\left(x - \frac{x_f + x_0}{2}\right). \tag{4.1.11}$$

Using the integral properties of the Wigner function (2.2.9), we can now express the propagator as an integral over phase space involving the Wigner function and the star exponential:

$$K(x_f, t_f; x_0, t_0) = \int_{\mathbb{R}^2} \rho_{f,0}(x, p) \, \text{Exp}_\star\left(-\frac{i}{\hbar}(t_f - t_0)H(x, p)\right) dx\, dp,$$
$$= \frac{1}{2\pi\hbar} \int_{\mathbb{R}} e^{\frac{i}{\hbar}(x_f-x_0)p} \, \text{Exp}_\star\left(-\frac{i}{\hbar}(t_f - t_0)H\left(\frac{x_f + x_0}{2}, p\right)\right) dp. \tag{4.1.12}$$

This expression shows that the propagator is the Fourier transform of the star exponential. To make this relationship more explicit, we introduce the new variables $q = (x_f + x_0)/2.$ and $q' = (x_f - x_0)/2.$, and set $t_f = t, t_0 = 0$. By taking the inverse Fourier transform of (4.1.12), we can solve for the star exponential:

$$\mathrm{Exp}_\star\left(-\frac{i}{\hbar}tH(q,p)\right) = 2\int_\mathbb{R} e^{-\frac{i}{\hbar}2q'p} K(q+q',t;q-q',0)dq'. \tag{4.1.13}$$

Equation (4.1.13) provides an advantageous method for calculating star exponentials, as it relies on the known propagator rather than the convergence of the formal series, which, as mentioned previously, can be difficult to establish.

Let us now turn to some key examples.

### 4.1.1 Bosonic harmonic oscillator's star exponential

Consider the harmonic oscillator in one dimension:

$$H_{\mathrm{ho}}(q,p) = \frac{p^2}{2m} + \frac{m\omega^2 q^2}{2}. \tag{4.1.14}$$

For the specific instance of the quantum harmonic oscillator, the propagator $K_{ho}$ is known exactly [31]:

$$K_{ho}(x_f, t_f; x_0, t_0) = \sqrt{\frac{m\omega}{2\pi i\hbar \sin(\omega T)}} \exp\left[\frac{im\omega}{2\hbar \sin(\omega T)}\left((x_f^2 + x_0^2)\cos(\omega T) - 2x_f x_0\right)\right], \tag{4.1.15}$$

where $T = t_f - t_0$.

Using the general relationship established in (4.1.13), we can insert this known propagator to directly calculate the star exponential for the harmonic oscillator Hamiltonian, $H_{ho}$. This calculation yields the closed-form expression:

$$\mathrm{Exp}\star\left(-\frac{i}{\hbar}tH_{ho}(q,p)\right) = \left(\cos\left(\frac{\omega t}{2}\right)\right)^{-1} \exp\left[\frac{2H_{ho}(q,p)}{i\omega\hbar}\tan\left(\frac{\omega t}{2}\right)\right]. \tag{4.1.16}$$

A key feature of this star exponential is that, under certain conditions, it admits a Fourier-Dirichlet expansion that reveals the system's spectral content [17]:

$$\left(\cos\left(\frac{\omega t}{2}\right)\right)^{-1} \exp\left[\frac{2H_{ho}(q,p)}{i\omega\hbar}\tan\left(\frac{\omega t}{2}\right)\right] = \sum_{n=0}^\infty \rho_n(q,p)e^{-it\omega(n+1/2)}. \tag{4.1.17}$$

where the functions $\rho_n(q,p)$ are the diagonal Wigner functions corresponding to the energy eigenvectors with eigenvalues $E_n = \hbar\omega(n + 1/2)$. Their explicit form involves Laguerre polynomials, $L_n$:

$$\rho_n(q,p) = \frac{(-1)^n}{\pi\hbar} e^{-\frac{2H_{ho}(q,p)}{\omega\hbar}} L_n\left(\frac{4H(q,p)}{\omega\hbar}\right). \tag{4.1.18}$$

### 4.1.2   Bosonic general quadratic Lagrangian's star exponential

The next example considers systems defined by a general one-dimensional quadratic Lagrangian of the form:

$$L(q(t), \dot{q}(t), t) = \frac{m}{2}\dot{q}^2 - \frac{c(t)}{2}q^2 + f(t)q, \tag{4.1.19}$$

where $c(t)$ and $f(t)$ are continuous functions.

The propagator $K$ for such a system can be written as a path integral over all trajectories connecting the initial and final points, weighted by the exponential of the classical action $S$ [21]:

$$K(q_f, t_f; q_0, t_0) = \int \mathcal{D}q(t) e^{\frac{i}{\hbar}S(q(t))}, \tag{4.1.20}$$

where the classical action $S(q(t))$ is the time integral of the Lagrangian:

$$S(q(t)) = \int_{t_0}^{t_f} L(q(t), \dot{q}(t), t) dt. \tag{4.1.21}$$

It's important to note that the path integral measure, $\mathcal{D}q(t)$, is not a standard Lebesgue measure, as **one does not exist for an infinite-dimensional space**. It is formally understood as a limit over discretized paths [31].

For quadratic systems, the path integral can be evaluated exactly, and the propagator takes the well-known form [32]:

$$K(q_f, t_f; q_0, t_0) = \sqrt{\frac{m}{2\pi i \hbar \varphi(t_f)}} e^{\frac{i}{\hbar} S_c(q_f, t_f; q_0, t_0)}, \tag{4.1.22}$$

where $S_c$ is the action evaluated along the classical path $q_c(t)$ satisfying the boundary conditions. The function $\varphi(t)$ in the prefactor is determined by solving the differential equation:

$$m\frac{d^2\varphi}{dt^2} + c(t)\varphi(t) = 0, \tag{4.1.23}$$

subject to the conditions $\varphi(t_0) = 0$ and $\dot{\varphi}(t_0) = 1$. The phase of the prefactor is related to the Maslov index $\nu$ [33]:

$$\frac{1}{\sqrt{\varphi(t_f)}} = e^{-i\pi\nu/2} \frac{1}{\sqrt{|\varphi(t_f)|}}. \tag{4.1.24}$$

With the exact propagator known, we can use the formula (4.1.13) to find the corresponding star exponential:

$$\mathrm{Exp}_\star\left(-\frac{i}{\hbar}tH(q,p)\right) = \sqrt{\frac{2m}{\pi i \hbar \varphi(t)}} \int_{\mathbb{R}} e^{-\frac{i}{\hbar}2q'p} e^{\frac{i}{\hbar}S_c(q+q',t;q-q',0)} dq'. \tag{4.1.25}$$

#### An Important Clarification

Because the Lagrangian (4.1.19) is explicitly time-dependent, the resulting star exponential (4.1.25) is not the symbol of the simple evolution operator $e^{-it\hat{H}/\hbar}$. Instead, it corresponds to the symbol of the time-ordered evolution operator, which is typically expressed as a Dyson series [31].

## 4.2 Star exponential for fermions

### 4.2.1 Naive approach for the star exponential

Let us start by considering the canonical quantum mechanics state:

$$\hat{\theta} = \hat{\theta}(\psi, t). \tag{4.2.1}$$

Its evolution satisfies Dirac's equation

$$H\theta = \pi\dot{\psi}\theta(\psi, \pi),$$
$$= E\theta, \tag{4.2.2}$$

with initial conditions

$$\theta(\psi, t)|_{t=t_0} = \theta(\psi_0, t_0). \tag{4.2.3}$$

The propagator must also satisfy the initial condition

$$\lim_{t \to t_0} K(\psi, t, \psi_0, t_0) = \delta(\psi - \psi_0). \tag{4.2.4}$$

where the limit is taken in the distributional sense

$$\lim_{t \to t_0} \int K(\psi, t, \psi_0, t_0)\theta(\psi)d\psi = \theta(\psi_0). \tag{4.2.5}$$

In Dirac's terminology it can be expressed as:

$$K(\psi, t, \psi_0, t_0) = <\psi, t|\psi_0, t_0>,$$
$$= <\psi|e^{-\frac{i}{\hbar}\hat{H}(t-t_0)}|\psi_0>,$$
$$= \sum_{n=0}^{\infty} e^{-\frac{i}{\hbar}E_n(t-t_0)}\theta_n(\psi)\bar{\theta}_n(\psi_0), \tag{4.2.6}$$

where the last equality is satisfied if the spectrum of the Hamiltonian is known.

**Non diagonal density operator**

Let $\hat{\rho}_{f,o} = |\psi_0\rangle \langle \psi_f|$ be a non diagonal density operator, with $|\psi_0\rangle$ and $|\psi_f\rangle$ being eigen vectors associated to

$$\hat{\psi}: \quad \hat{\psi}_j |\psi\rangle = \psi_j |\psi\rangle \quad \forall j, \tag{4.2.7}$$

with

$$|\psi\rangle = \exp\{-\frac{i}{\hbar} \sum_{j=i}^{n} \hat{\pi}_j \psi_j\} |0\rangle, \tag{4.2.8}$$

and

$$\langle \psi| = \langle 0| (\prod_{j=1}^{n} \hat{\psi}_j) \exp\{\frac{i}{\hbar} \sum_{j=1}^{n} \hat{\pi}_j \psi_j\}. \tag{4.2.9}$$

Using Weyl's inverse quantisation map:

$$\rho_{f,o}(\psi, \pi) = Q_W^{-1}(|\psi_0\rangle \langle \psi_f|),$$
$$= tr\{|\psi_0\rangle \langle \psi_f| \hat{\Omega}(\pi, \psi)\},$$
$$= (i\hbar)^{-n} \int D\psi' <\psi'|\psi_0><\psi_f|\hat{\Omega}(\pi,\psi)|\psi'>,$$
$$= \quad \ldots \quad ,$$
$$= (\hbar)^{-n} \int D\lambda \delta(\psi - \frac{\hbar\lambda}{2} - \psi_f)\delta(\psi_0 - \psi - \frac{\hbar\lambda}{2}) \exp\{-i\sum_{j=1}^{n} \pi_j \lambda_j\}. \tag{4.2.10}$$

This defines a system of equations:

$$\psi - \frac{\hbar\lambda}{2} - \psi_f = 0, \tag{4.2.11}$$

$$\psi_0 - \psi - \frac{\hbar\lambda}{2} = 0. \tag{4.2.12}$$

Solving it yields:

$$\psi = \frac{\psi_0 + \psi_f}{2}, \tag{4.2.13}$$

$$\lambda = \frac{\psi_0 - \psi_f}{\hbar}. \tag{4.2.14}$$

This eliminates the integral sign and one of the deltas, yielding as a final result:

$$\rho_{f,o} = (\frac{1}{\hbar})^n \exp\{\frac{i}{\hbar} \sum_{j=1}^{n} \pi_j (\psi_f^j - \psi_0^j)\} \delta\left(\psi - \frac{(\psi_f + \psi_o)}{2}\right). \tag{4.2.15}$$

Now let us calculate a proper expression of the propagator using Dirac's terminology and the integral property of $<\hat{A}>$:

$$K(\psi_f, t_f, \psi_0, t_0) = \langle \psi_f | \exp\{-\frac{i}{\hbar}\hat{H}(t_f - t_0)\} | \psi_0 \rangle,$$

$$= \int \rho_{f,o}(\psi, \pi) \exp_\star\{-\frac{i}{\hbar}(t_f - t_0)H(\psi, \pi)\} D\psi D\pi,$$

$$= \frac{1}{\hbar^n} \int \exp\left\{\frac{i}{\hbar}\sum_{j=1}^n \pi_j(\psi_f^j - \psi_0^j)\right\} \delta\left(\psi - \frac{(\psi_f + \psi_0)}{2}\right) \exp_\star\{-\frac{i}{\hbar}(t_f - t_0)H(\psi, \pi)\} D\psi D\pi,$$

$$= \frac{1}{\hbar^n} \int \exp\left\{\frac{i}{\hbar}\sum_{j=1}^n \pi_j(\psi_f^j - \psi_0^j)\right\} \exp_\star\left\{-\frac{i}{\hbar}(t_f - t_0)H(\frac{\psi_f + \psi_0}{2}, \pi)\right\} D\pi.$$

$$(4.2.16)$$

This shows the propagator can be formulated as the Fourier transform of a star exponential! If we introduce the following variables

$$\Psi_j = \frac{\psi_f^j + \psi_o^j}{2}, \quad (4.2.17)$$

$$\Psi'_j = \frac{\psi_f^j - \psi_o^j}{2}, \quad (4.2.18)$$

$$t_f = t, \quad (4.2.19)$$

$$t_0 = 0, \quad (4.2.20)$$

the propagator takes the form:

$$K(\Psi + \Psi', t, \Psi - \Psi', 0) = \frac{1}{\hbar^n} \int \exp\{\frac{i}{\hbar}\sum_j^n 2\pi_j \Psi'_j\} \exp_\star\{-\frac{i}{\hbar}tH(\Psi, \pi)\} D\pi. \quad (4.2.21)$$

Computing the inverse transform we get to the final formula of the **fermionic star exponential**:

$$\exp_\star\{-\frac{i}{\hbar}tH(\Psi, \Pi)\} = \int \exp\{-\frac{i}{\hbar}2\sum_j^n \pi_j \Psi'_j\} K(\Psi + \Psi', t, \Psi - \Psi', 0) D\Psi'. \quad (4.2.22)$$

This formula will be crucial in determining the star exponential of a physical system if its propagator admits a closed form.

However, there are some problems concerning the previous development. First of all, using only the $\psi$ representation introduces a problem concerning the **Grassmann parity**. In the Grassmann formalism, it is important to notice that under the integral sign, $d\psi$ is not a 1-form but also a Grassmann object with an associated parity. Specifically, $d\psi$ is Grassmann-odd. The action of an even object times an odd one yields an odd object. This is a problem because, as can be seen in 4.1.13, the expression is odd. This raises the question of how an even object (the Hamiltonian), which also yields another even object (the Propagator), can finally coalesce into an odd object. This appears to be a fundamental objection to the alleged formula. There is another procedure that preserves the parity properly and also has a natural physical interpretation for the fermionic propagator. Let us explore it in detail.

### 4.2.2   Meticulous approach for the star exponential

Notice first the following ([13]):

$$\langle \pi | = \langle \psi^* |, \tag{4.2.23}$$
$$= (|\psi\rangle)^*. \tag{4.2.24}$$

which implies:

$$K(\psi_f, t; \psi_0, 0) \to K(\pi_f, t; \psi_0, 0), \tag{4.2.25}$$
$$= \langle \pi_f | e^{-\frac{i}{\hbar} t \hat{H}} | \psi_0 \rangle. \tag{4.2.26}$$

$$\rho_{f,o}(\psi, \pi) = Q_W^{-1}(|\psi_0\rangle \langle \psi_f|) \to \rho_{f,o}(\psi, \pi) = Q_W^{-1}(|\psi_0\rangle \langle \pi_f|). \tag{4.2.27}$$

Let us do again the calculation step by step, using some results available in [13].

$$\rho_{f,o}(\pi, \psi) = \hbar^{-n} \int D\lambda \exp\{-i \sum_{j=1}^n \pi_j \lambda_j\} \langle \psi + \frac{\hbar \lambda}{2} | \hat{\rho} | \psi - \frac{\hbar \lambda}{2} \rangle,$$

$$= \hbar^{-n} \int D\lambda \exp\{-i \sum_{j=1}^n \pi_j \lambda_j\} \langle \psi + \frac{\hbar \lambda}{2} | | \psi_0 \rangle \langle \pi_f | | \psi - \frac{\hbar \lambda}{2} \rangle,$$

$$= \hbar^{-n} \int D\lambda \exp\{-i \sum_j \pi_j \lambda_j\} \delta(\psi_0 - (\psi + \frac{\hbar \lambda}{2})) \exp\{-\frac{i}{\hbar} \sum_j \pi_j^f(\psi_j - \frac{\hbar \lambda_j}{2})\}. \tag{4.2.28}$$

Now using the formula for grassmann variables:

$$\delta(f(\theta)) = \sum_i \delta(\theta - \theta_i) |f'(\theta_i)| \tag{4.2.29}$$

where $\theta_i$ stands for the zero's inside the delta symbol, we find:

$$\delta(\psi_0 - (\psi + \frac{\hbar \lambda}{2})) = (\frac{\hbar}{2})^n \delta(\lambda - \frac{2}{\hbar}(\psi_0 - \psi)). \tag{4.2.30}$$

$\Longrightarrow$

$$\rho_{f,o}(\pi, \psi) = 2^{-n} \int D\lambda \exp\{-i \sum_j \pi_j \lambda_j\} \delta\left(\lambda - \frac{2}{\hbar}(\psi_0 - \psi)\right) \exp\left\{-\frac{i}{\hbar} \sum_j \pi_j^f(\psi_j - \frac{\hbar \lambda_j}{2})\right\},$$

$$= 2^{-n} \exp\left\{-\frac{i}{\hbar} \sum_j \pi_j 2(\psi_j^0 - \psi_j)\right\} \exp\left\{-\frac{i}{\hbar} \sum_j \pi_j^f(2\psi_j - \psi_j^0)\right\},$$

$$= \ldots,$$

$$= 2^{-n} \exp\left\{i\frac{2}{\hbar} \sum_j (\pi_j - \pi_j^f)(\psi_j - \psi_j^0)\right\} \exp\left\{-\frac{i}{\hbar} \sum_j \pi_j^f \psi_j^0\right\}. \tag{4.2.31}$$

Consider now the new variables:

$$\chi_j := \psi_j - \psi_0 \quad , \quad \eta_j := \pi_j - \pi_j^f. \tag{4.2.32}$$

$\Longrightarrow$

$$D\psi D\pi = D\chi D\eta. \tag{4.2.33}$$

$$A \cdot B := \sum_{j=1}^{n} A_j B_j. \tag{4.2.34}$$

$\Longrightarrow$

$$\rho_{f,0} = 2^{-n} \exp\left\{\frac{i2}{\hbar}\eta \cdot \chi\right\} \exp\left\{-\frac{i}{\hbar}\pi_j^f \cdot \psi_j^0\right\}. \tag{4.2.35}$$

For the propagator then we find

$$K(\pi_f, t; \psi_0, 0) = \langle \pi_f | \exp\left\{-\frac{i}{\hbar} t \hat{H}\right\} | \psi_0 \rangle,$$

$$= \int \rho_{f,0}(\psi, \pi) Exp_\star \left\{-\frac{it}{\hbar} H(\psi, \pi)\right\} D\psi D\pi,$$

$$= 2^{-n} \exp\left\{-\frac{i}{\hbar}\pi_j^f \cdot \psi_j^0\right\} \int D\chi D\eta \exp\left\{\frac{i2}{\hbar}\eta \cdot \chi\right\} Exp_\star \left\{-\frac{it}{\hbar} H(\chi + \psi^0, \eta + \pi^f)\right\}.$$
$$\tag{4.2.36}$$

We must solve for the star exponential inside the integral.

Consider firstly the following variables (built in straight analogy to the bosonic case):

$$\Psi = \frac{\psi_f + \psi_0}{2}, \tag{4.2.37}$$

$$\Psi' = \frac{\psi_f - \psi_0}{2}, \tag{4.2.38}$$

$$\Pi = \frac{\pi_f + \pi_0}{2}, \tag{4.2.39}$$

$$\Pi' = \frac{\pi_f - \pi_0}{2}, \tag{4.2.40}$$

$\Longrightarrow$

$$K(\pi_f, t; \psi_0, 0) \to K(\Pi + \Pi', t; \Psi - \Psi'). \tag{4.2.41}$$

After a lot of cumbersome calculations the following transformation was established.

$$C \exp\left\{\frac{i}{\hbar}\Pi\cdot\Psi\right\}\int D\Psi' D\Pi' \exp\left\{-\frac{i2}{\hbar}\Pi'\cdot\Psi'\right\} K(\Pi+\Pi', t; \Psi-\Psi') := \tilde{F}(\Psi, \Pi) \quad (4.2.42)$$

where C is a constant that ensures the normalization of the result. Let us verify it step by step.

$$\begin{aligned}
\tilde{F}(\Psi, \Pi) &= C\exp\left\{\frac{i}{\hbar}\Pi\cdot\Psi\right\}\int D\Psi' D\Pi' \exp\left\{-\frac{i2}{\hbar}\Pi'\cdot\Psi'\right\} K(\Pi+\Pi', t; \Psi-\Psi'), \\
&= C\exp\left\{\frac{i}{\hbar}\Pi\cdot\Psi\right\}\int D\Psi' D\Pi' \exp\left\{-\frac{i2}{\hbar}\Pi'\cdot\Psi'\right\} 2^{-n} \exp\left\{-\frac{i}{\hbar}\pi_j^f\cdot\psi_j^0\right\} \\
&\quad\times \int D\chi D\eta \exp\left\{\frac{i2}{\hbar}\eta\cdot\chi\right\} \exp_\star\left\{-\frac{it}{\hbar}H(\chi+\psi^0, \eta+\pi^f)\right\}, \\
&= C2^{-n}\exp\left\{\frac{i}{\hbar}\Pi\cdot\Psi\right\}\int D\chi D\eta \int D\Psi' D\Pi' \exp\left\{\frac{i2}{\hbar}(\eta\cdot\chi-\Pi'\cdot\Psi')\right\} \\
&\quad\times \exp\left\{-\frac{i}{\hbar}(\Pi+\Pi')(\Psi-\Psi')\right\} \exp_\star\left\{-\frac{it}{\hbar}H(\chi+\Psi-\Psi', \eta+\Pi+\Pi')\right\}, \\
&= C2^{-n}\int D\chi D\eta \int D\Psi' D\Pi' \exp\{S(\Pi, \Pi', \Psi, \Psi', \eta, \chi)\} \\
&\quad\times \exp_\star\left\{-\frac{it}{\hbar}H(\chi+\Psi-\Psi', \eta+\Pi+\Pi')\right\}. \quad (4.2.43)
\end{aligned}$$

$\Longleftrightarrow$

$$\begin{aligned}
S(\Pi, \Pi', \Psi, \Psi', \eta, \chi) &:= \frac{i}{\hbar}[\Pi\cdot\Psi - 2\Pi'\cdot\Psi' + 2\eta\cdot\chi - (\Pi+\Pi')(\Psi-\Psi')], \\
&= \frac{i}{\hbar}[-\Pi'\cdot\Psi' + 2\eta\cdot\chi + \Pi\cdot\Psi' - \Pi'\cdot\Psi]. \quad (4.2.44)
\end{aligned}$$

Consider now the following shifts:

$$\chi = \chi' + \Psi'. \quad (4.2.45)$$

$$\eta = \eta' - \Pi'. \quad (4.2.46)$$

$\Longrightarrow$

$$Exp_\star\left\{-\frac{it}{\hbar}H(\chi+\Psi-\Psi', \eta+\Pi+\Pi')\right\} \to Exp_\star\left\{-\frac{it}{\hbar}H(\chi'+\Psi, \eta'+\Pi)\right\}. \quad (4.2.47)$$

$$\begin{aligned}
\eta\cdot\chi &= (\eta'-\Pi')\cdot(\chi'+\Psi'), \\
&= \eta'\cdot\chi' + \eta'\cdot\Psi' - \Pi'\cdot\chi' - \Pi'\cdot\Psi'. \quad (4.2.48)
\end{aligned}$$

$$S = \frac{i}{\hbar}[-\Pi' \cdot \Psi' + 2\eta \cdot \chi + \Pi \cdot \Psi' - \Pi' \cdot \Psi],$$

$$= \frac{i}{\hbar}[2\eta' \cdot \chi' + 2\eta' \cdot \Psi' - 2\Pi' \cdot \chi' - 3\Pi' \cdot \Psi' + \Pi \cdot \Psi' - \Pi' \cdot \Psi],$$

$$= \frac{i}{\hbar}[2\eta' \cdot \chi' - \Psi' \cdot (2\eta' + \Pi - 3\Pi') - \Pi' \cdot (\Psi + 2\chi')]. \tag{4.2.49}$$

$$D\chi D\eta = D(\chi' + \Psi')D(\eta' - \Pi'),$$
$$= D\chi' D\eta' - D\chi' D\Pi' + D\Psi' D\eta' - D\Psi' D\Pi'. \tag{4.2.50}$$

$\implies$

$$D\chi D\eta D\Psi' D\Pi' = (D\chi' D\eta' - D\chi' D\Pi' + D\Psi' D\eta' - D\Psi' D\Pi')D\Psi' D\Pi',$$
$$= D\chi' D\eta' D\Psi' D\Pi'. \tag{4.2.51}$$

$\implies$

$$\tilde{F}(\Psi, \Pi) = 2^{-n} C \int D\chi' D\eta' \exp\left\{\frac{i2}{\hbar}\eta' \cdot \chi'\right\} Exp_\star \left\{-\frac{it}{\hbar}H(\chi' + \Psi, \eta' + \Pi)\right\} \tilde{I},$$
$$\tag{4.2.52}$$

where

$$\tilde{I} := \int D\Psi' D\Pi' \exp\{\frac{i}{\hbar}[-3\Pi' \cdot \Psi' - \Psi' \cdot (2\eta' + \Pi) - \Pi' \cdot (\Psi + 2\chi')]\}. \tag{4.2.53}$$

Using the following formula, proven in the appendix section,

$$J = \int Du\, Dv\, \exp(v^T M u + u^T a + v^T b),$$
$$= \det(M) \exp(-a^T M^{-1} b). \tag{4.2.54}$$

we find:

$$M = \frac{i}{\hbar}(-3)I, \tag{4.2.55}$$

$$a = -\frac{i}{\hbar}(2\eta' + \Pi), \tag{4.2.56}$$

$$b = -\frac{i}{\hbar}(\Psi + 2\chi'), \tag{4.2.57}$$

$\implies$

$$M^{-1} = \frac{\hbar i}{3}I, \tag{4.2.58}$$

$$det(M) = \left(-\frac{3i}{\hbar}\right)^n, \tag{4.2.59}$$

$$-a^T M^{-1} b = \frac{i}{3\hbar}(2\eta' + \Pi) \cdot (\Psi + 2\chi'), \tag{4.2.60}$$

yielding finally:

$$\tilde{I} = \left(-\frac{3i}{\hbar}\right)^n \exp\left\{\frac{i}{3\hbar}(2\eta' + \Pi) \cdot (\Psi + 2\chi')\right\}, \tag{4.2.61}$$

$\Longrightarrow$

$$\tilde{F}(\Psi, \Pi) = 2^{-n} C \left(-\frac{3i}{\hbar}\right)^n \int D\chi' D\eta' e^{\bar{S}} Exp_\star \left\{-\frac{it}{\hbar} H(\chi' + \Psi, \eta' + \Pi)\right\}, \tag{4.2.62}$$

where

$$\bar{S} = \frac{i}{\hbar}\left[\frac{10}{3}\eta' \cdot \chi' + \frac{2}{3}\eta' \cdot \Psi + \frac{1}{3}\Pi \cdot \Psi + \frac{2}{3}\Pi \cdot \chi'\right]. \tag{4.2.63}$$

Expanding the exponential up to second order given its Grassmann nature and using the **top-form saturation (projection) principle of Grassmann integrals** ($\int d\theta f(\omega) = 0 \iff w \neq \theta$) one finds:

$$\begin{aligned}
\tilde{F}(\Psi, \Pi) &= C\left(-\frac{3i}{2\hbar}\right)^n \int D\chi' D\eta' \left(1 + \frac{i}{\hbar}[\frac{10}{3}\eta' \cdot \chi' + \frac{2}{3}\eta' \cdot \Psi + \frac{1}{3}\Pi \cdot \Psi + \frac{2}{3}\Pi \cdot \chi']\right) \\
&\quad \times \left(Exp_\star\left(-\frac{it}{\hbar} H(\chi' + \Psi, \eta' + \Pi)\right)\right), \\
&= C\left(-\frac{3i}{2\hbar}\right)^n \int D\chi' D\eta' \eta' \chi' Exp_\star\left(-\frac{it}{\hbar} H(\chi' + \Psi, \eta' + \Pi)\right) \left(i\frac{10}{3\hbar}\right), \\
&= C\left(-\frac{3i}{2\hbar}\right)^n \left(i\frac{10}{3\hbar}\right) \int D\chi' D\eta' \delta(\chi') \delta(\eta') Exp_\star\left(-\frac{it}{\hbar} H(\chi' + \Psi, \eta' + \Pi)\right), \\
&= C\left(-\frac{3i}{2\hbar}\right)^n \left(i\frac{10}{3\hbar}\right) Exp_\star\left(-\frac{it}{\hbar} H(\Psi, \Pi)\right), \\
&= Exp_\star\left(-\frac{it}{\hbar} H(\Psi, \Pi)\right).
\end{aligned} \tag{4.2.64}$$

$\iff$

$$C = \left(\frac{2^{n-1} \hbar^{n+1}}{5 \cdot 3^{n-1}}\right) i^{3-3n}. \tag{4.2.65}$$

$\therefore$

$$\begin{aligned}
Exp_\star\left\{-\frac{it}{\hbar} H(\Psi, \Pi)\right\} &= i^{3-3n}\left(\frac{2^{n-1}\hbar^{n+1}}{5 \cdot 3^{n-1}}\right) \exp\left\{\frac{i}{\hbar}\Pi \cdot \Psi\right\} \\
&\quad \times \int D\Psi' D\Pi' \exp\left\{-\frac{i2}{\hbar}\Pi' \cdot \Psi'\right\} K(\Pi + \Pi', t; \Psi - \Psi').
\end{aligned} \tag{4.2.66}$$

Let us now see how it can be applied in the most elementary configurations in physics: oscillators.

### 4.2.3  Harmonic oscillator's star exponential

The following discussion is adapted from the book by A. Das [22]. For full details, the reader may consult that work as well as other key texts such as [30, 5].

As discussed earlier, a fundamental property of fermions is **Pauli's exclusion principle**. In the language of Grassmann variables, this principle corresponds to their **nilpotent** character, i.e., that $\theta^2 = 0$ for any Grassmann variable $\theta$. This is *a priori* a problem, given that the Hamiltonian dynamics of oscillators is generally encoded in a quadratic term. Let us start the discussion by considering first the Lagrangian formalism.

Consider $\psi$ and $\bar{\psi}$ are two independent Grassmann variables which are functions of time. The following Lagrangian can represent, via its equations of motion, the oscillator behavior:

$$L_1 = \frac{i}{2}(\bar{\psi}\dot{\psi} - \dot{\bar{\psi}}\psi) - \frac{w}{2}[\bar{\psi}, \psi], \tag{4.2.67}$$

where w can be understood as the frequency of oscillation. It is a matter of algebra to show that the following Lagrangian is equivalent:

$$L_2 = i\bar{\psi}\dot{\psi} - \frac{w}{2}[\bar{\psi}, \psi]. \tag{4.2.68}$$

For simplicity let us consider $L_1$. It is important to start by calculating the canonical conjugate momenta:

$$\Pi_\psi = \frac{\partial L_1}{\partial \dot{\psi}},$$
$$= -\frac{i}{2}\psi. \tag{4.2.69}$$

$$\Pi_{\bar{\psi}} = \frac{\partial L_1}{\partial \dot{\bar{\psi}}},$$
$$= -\frac{i}{2}\bar{\psi}. \tag{4.2.70}$$

We can note that it is a **constrained** system and therefore one should use Dirac's formalism to quantize it. Let us calculate its Hamiltonian via the Legendre transform:

$$H_f = \phantom{}^{"}\dot{q}p - L^{"},$$
$$= \dot{\psi}\Pi_\psi + \dot{\bar{\psi}}\Pi_{\bar{\psi}} - L,$$
$$= \ldots ,$$
$$= \frac{w}{2}[\bar{\psi}, \psi]. \tag{4.2.71}$$

Under our convention (3.0.23) these independent Grassmann variables are complex *a fortiori*. This means that we can identify the maps:

$$\psi \quad \rightarrow \quad \hat{\psi} \quad , \quad \bar{\psi} \quad \rightarrow \quad i\hat{\pi}. \tag{4.2.72}$$

This implies the Hamiltonian has the form:

$$H_f = \frac{iw}{2}[\hat{\pi}, \hat{\psi}]. \tag{4.2.73}$$

Now using (3.0.24) we have:

$$\begin{aligned} H_f &= \frac{iw}{2}[\hat{\pi}, \hat{\psi}], \\ &= \frac{iw}{2}(\hat{\pi}\hat{\psi} - \hat{\psi}\hat{\pi}), \\ &= \frac{iw}{2}(\hat{\pi}\hat{\psi} + (\hat{\pi}\hat{\psi} - i\hbar)), \\ &= w(i\hat{\pi}\hat{\psi} + \frac{\hbar}{2}). \end{aligned} \tag{4.2.74}$$

Let us calculate the evolution operator:

$$\begin{aligned} U(t, t_0) &= \exp\{-iH(t - t_0)\}, \\ &= \exp\{w\hat{\pi}\hat{\psi}(t - t_0) - \frac{iw\hbar}{2}(t - t_0)\}, \\ &= \exp\{w\hat{\psi}\pi(t - t_0)\}\exp\{-\frac{iw\hbar}{2}(t - t_0)\}, \\ &= \exp\{w\hat{\psi}\pi(t - t_0)\}[phase]. \end{aligned} \tag{4.2.75}$$

As it can be seen in the appendix, the Hamiltonian of a system is invariant under any global phase shift. We will adjust this term in the end of the calculation of the propagator.

### 4.2.4 Naive approach for the fermionic harmonic oscillator

Consider now the propagator in Dirac's notation:

$$K(\psi_f, t; \psi_0, 0) = \langle \psi_f | U(t, 0) | \psi_i \rangle. \tag{4.2.76}$$

There are several ways to tackle this calculation, including:

1. **Method 1:** Start by using the normalization identity (3.0.52) inside the braket. Then employ the delta function identity (3.0.45) to eliminate the integral sign and deduce the final result.
2. **Method 2:** Use the basis $\langle \pi |$, $| \psi \rangle$ which is more natural (it avoids the infinite product of $\hat{\psi}'s, \hat{\pi}'s$). Then get the associated eigenvalues and complete the calculation.
3. **Method 3:** Use the known method of path integrals for the simple system $\hat{H} = w\hat{\psi}\hat{\psi}$ and finally adjust the phase factor in the result.

Method 1 is cumbersome. Method 2 is a lot more difficult, because it enters the complex field of *phase-space* quantum fields, which require more boundary conditions to be imposed. Method 3 was the one used in this work. The final result is the following:

$$K(\psi_f, t_f; \psi_0, 0) = \exp\left\{\frac{iwt}{2}\right\} \exp\{\psi_f \exp\{-iwt\}\psi_i\}. \tag{4.2.77}$$

With this result let us now calculate its associated star exponential.

$$\exp_\star\{-\frac{i}{\hbar}tH(\Psi, \Pi)\} = \int \exp\{-\frac{i}{\hbar}2\sum_j^n \pi_j \Psi'_j\} K(\Psi + \Psi', t, \Psi - \Psi', 0) D\Psi',$$

$$\stackrel{1D}{=} \int \exp\{-\frac{i}{\hbar}2\pi\Psi'\} K(\Psi + \Psi', t, \Psi - \Psi', 0) d\Psi', \tag{4.2.78}$$

where:

$$\Psi = \frac{\psi_f + \psi_0}{2}, \tag{4.2.79}$$

$$\Psi' = \frac{\psi_f - \psi_0}{2}. \tag{4.2.80}$$

This implies the following form for the propagator:

$$K(\psi_f, t_f; \psi_0, 0) = \exp\{\frac{iwt}{2}\} \exp\{\psi_f \exp\{-iwt\}\psi_i\},$$

$$= \dots ,$$

$$= \exp\{\frac{iwt}{2}\} \exp\{2\Psi' \exp\{-iwt\}\Psi\}. \tag{4.2.81}$$

$\Longrightarrow$

$$\exp_\star\{-itH(\Psi, \pi)\} = \int \exp\{\frac{iwt}{2}\} \exp\{-\frac{i}{\hbar}2\pi\Psi'\} \exp\{2\Psi' \exp\{-iwt\}\Psi\} d\Psi',$$

$$= \int \exp\{\frac{iwt}{2}\}(1 - \frac{2i}{\hbar}\pi\Psi')(1 + 2\Psi' \exp\{-iwt\}\Psi) d\Psi',$$

$$= \int \exp\{\frac{iwt}{2}\}(1 + 2\Psi' \exp\{-iwt\}\Psi - 2\frac{i}{\hbar}\pi\Psi') d\Psi',$$

$$= \exp\{\frac{iwt}{2}\}(-2\frac{i}{\hbar}\pi + 2\exp\{-iwt\}\Psi). \tag{4.2.82}$$

As was stated since the development of the formalism, this result violates the Grassmann parity preservation. Let us see how it can be better understood within the so called meticulous approach.

### 4.2.5 Meticulous approach for the fermionic harmonic oscillator

As can be seen with a step by step deduction in the appendix, or otherwise by working in a 2d basis (in analogy to the fermionic oscillator example in [13]), the propagator K satisfies the following formula:

$$K(\pi_f, t; \psi_0, 0) = \exp\left\{i\frac{wt}{2}\right\} \exp\{\pi_f e^{-iwt}\psi_0\},$$

$$= \exp\left\{i\frac{wt}{2}\right\} \exp\{(\Pi + \Pi')e^{-iwt}(\Psi - \Psi')\},$$

$$= \ldots,$$

$$= \exp\left\{i\frac{wt}{2}\right\} + [\Pi\Psi - \Pi\Psi' + \Pi'\Psi - \Pi'\Psi'] \exp\left\{-i\frac{wt}{2}\right\}. \quad (4.2.83)$$

$$e^{-i\frac{2}{\hbar}\Pi'\cdot\Psi'} \stackrel{1D}{=} 1 - i\frac{2}{\hbar}\Pi'\Psi'. \quad (4.2.84)$$

$\Longrightarrow$

$$e^{-i\frac{2}{\hbar}\Pi'\Psi'} K = \exp\left\{i\frac{wt}{2}\right\} + [\Pi\Psi - \Pi\Psi' + \Pi'\Psi - \Pi'\Psi'] \exp\left\{-i\frac{wt}{2}\right\}$$

$$- i\frac{2}{\hbar}\Pi'\Psi' \exp\left\{i\frac{wt}{2}\right\} - i\frac{2}{\hbar}\Pi'\Psi'\Pi\Psi \exp\left\{-i\frac{wt}{2}\right\}. \quad (4.2.85)$$

$\Longrightarrow$

$$\int D\Psi' D\Pi' \exp\left\{-i\frac{2}{\hbar}\Pi'\Psi'\right\} K = -\exp\left\{-i\frac{wt}{2}\right\} - i\frac{2}{\hbar}\Pi\Psi \exp\left\{-i\frac{wt}{2}\right\} - i\frac{2}{\hbar}\exp\left\{i\frac{wt}{2}\right\}. \quad (4.2.86)$$

$\Longrightarrow$

$$C \exp\left\{\frac{i}{\hbar}\Pi\Psi\right\} \int D\Psi' D\Pi' \exp\left\{-i\frac{2}{\hbar}\Pi'\Psi'\right\} K = Exp_\star\left\{-\frac{it}{\hbar}(\Psi, \Pi)\right\},$$

$$= -\left(\frac{\hbar^2}{5}\right)\left[\exp\left\{\frac{i}{\hbar}\Pi\Psi\right\}\left(\exp\left\{-\frac{iwt}{2}\right\} + i\frac{2}{\hbar}\exp\left\{i\frac{wt}{2}\right\}\right) + i\frac{2}{\hbar}\Pi\Psi \exp\left\{-\frac{iwt}{2}\right\}\right]. \quad (4.2.87)$$

This result is not in contradiction with the preservation of Grassmann parity.
Let us now consider the case of the driven harmonic oscillator.

### 4.2.6 Fermionic driven harmonic oscillator

Consider a "driven" fermionic model of the style: $\hat{H} = \omega\hat{\pi}\hat{\psi} + \alpha\hat{\psi}$. We can notice that $\omega$ can be interpreted as a frequency while $\alpha$ is the coupling force term, grassmann in nature also. First notice that it is not Hermitian! The complex conjugate of the second one must be included:

$$\hat{H} = \omega\hat{\pi}\hat{\psi} + \alpha\hat{\psi} + \alpha^*\hat{\pi}. \tag{4.2.88}$$

Consider the following results in the basis: $\{|0\rangle, |1\rangle\}$:

$$\hat{\psi}|1\rangle = |0\rangle, \tag{4.2.89}$$

$$\hat{\psi}|0\rangle = 0, \tag{4.2.90}$$

$$\hat{\pi}|1\rangle = 0, \tag{4.2.91}$$

$$\hat{\pi}|0\rangle = |1\rangle, \tag{4.2.92}$$

and with the number operator $\hat{N} = \hat{\pi}\hat{\psi}$:

$$\hat{N}|1\rangle = |1\rangle, \tag{4.2.93}$$

$$\hat{N}|0\rangle = 0. \tag{4.2.94}$$

We find their matrix representation:

$$(\hat{\psi}) : \begin{pmatrix} \langle 0|\hat{\psi}|0\rangle & \langle 0|\hat{\psi}|1\rangle \\ \langle 1|\hat{\psi}|0\rangle & \langle 1|\hat{\psi}|1\rangle \end{pmatrix},$$

$$= \begin{pmatrix} 0 & 1 \\ 0 & 0 \end{pmatrix}. \tag{4.2.95}$$

$$(\hat{\pi}) : \begin{pmatrix} \langle 0|\hat{\pi}|0\rangle & \langle 0|\hat{\pi}|1\rangle \\ \langle 1|\hat{\pi}|0\rangle & \langle 1|\hat{\pi}|1\rangle \end{pmatrix},$$

$$= \begin{pmatrix} 0 & 0 \\ 1 & 0 \end{pmatrix}. \tag{4.2.96}$$

$$(\hat{N}) : \begin{pmatrix} \langle 0|\hat{N}|0\rangle & \langle 0|\hat{N}|1\rangle \\ \langle 1|\hat{N}|0\rangle & \langle 1|\hat{N}|1\rangle \end{pmatrix},$$

$$= \begin{pmatrix} 0 & 0 \\ 0 & 1 \end{pmatrix}. \tag{4.2.97}$$

$\implies$

$$(\hat{H}) = \omega \begin{pmatrix} 0 & 0 \\ 0 & 1 \end{pmatrix} + \alpha \begin{pmatrix} 0 & 1 \\ 0 & 0 \end{pmatrix} + \alpha^* \begin{pmatrix} 0 & 0 \\ 1 & 0 \end{pmatrix},$$

$$= \begin{pmatrix} 0 & \alpha \\ \alpha^* & \omega \end{pmatrix}. \tag{4.2.98}$$

Let us diagonalize it and find its ground state energy. Its characteristic equation arises from:

$$det \begin{bmatrix} -\lambda & \alpha \\ \alpha^* & \omega - \lambda \end{bmatrix} = -\lambda(\omega - \lambda) - |\alpha|^2,$$

$$= 0. \qquad (4.2.99)$$

$$\iff \quad \lambda^2 - \lambda\omega - |\alpha|^2 = 0. \qquad (4.2.100)$$

Now using the quadratic formula:

$$\lambda = \frac{\omega \pm \sqrt{\omega^2 + 4|\alpha|^2}}{2},$$

$$= \frac{\omega}{2} \pm \frac{1}{2}\sqrt{\omega^2 + 4|\alpha|^2}. \qquad (4.2.101)$$

If we calculate the following commutator

$$[\hat{H}, \hat{N}] = \omega(\hat{\pi}\hat{\psi})\hat{\pi}\hat{\psi} + \alpha\hat{\psi}\hat{\pi}\hat{\psi} + \cancel{\alpha^*\hat{\pi}\hat{\pi}\hat{\psi}}^{\,0},$$

$$= \omega(\hat{\pi}(i\hbar - \hat{\pi}\hat{\psi})\hat{\psi}) + \alpha(i\hbar - \hat{\pi}\hat{\psi})\hat{\psi},$$

$$= i\hbar(\omega\hat{\pi}\hat{\psi} + \alpha\hat{\psi}),$$

$$\neq 0. \qquad (4.2.102)$$

we deduce that the linear terms mix states with different particle numbers, requiring a transformation to a new basis where the Hamiltonian is diagonal. The energy eigenvalues obtained from this diagonalization correspond to the energies of the new quasi-particle [1] excitations!

Let us now proceed with its propagator.

### Propagator

Let us start by using Heiseberg's equations to understand the evolution of the operators $\{\hat{\psi}, \hat{\pi}\}$.

### Heisenberg's equation of motion

Let $\hat{A}$ be an operator. It's time evolution is dictated by Heisenberg's equation ($\hbar = 1$):

$$i\frac{d\hat{A}}{dt} = [\hat{A}, \hat{H}]. \qquad (4.2.103)$$

Consider the following identity:

$$[A, B] = AB - BA,$$
$$= AB + BA - 2BA,$$
$$= \{A, B\} - 2BA. \qquad (4.2.104)$$

---

[1] Quasi-particle: collective excitations in a material, treating them as if they were individual particles.

$$[A, BC] = B[A, C] + [A, B]C. \tag{4.2.105}$$

and recall that ($\hbar = 1$)

$$\{\hat{\psi}, \hat{\pi}\} = i. \tag{4.2.106}$$

Being $\alpha, \alpha^*$ Grassmann variables this implies:

$$[\hat{\psi}, \hat{H}] = [\hat{\psi}, \omega\hat{\pi}\hat{\psi} + \alpha\hat{\psi} + \alpha^*\hat{\pi}],$$

$$= \omega[\hat{\psi}, \hat{\pi}\hat{\psi}] + \underbrace{[\hat{\psi}, \alpha\hat{\psi}]}_{0}\overset{\hat{\psi}^2 = 0}{+} [\hat{\psi}, \alpha^*\hat{\pi}],$$

$$= \omega(\hat{\psi}\hat{\pi}\hat{\psi} - \hat{\pi}\underbrace{\hat{\psi}^2}_{0}) + \hat{\psi}\alpha^*\hat{\pi} - \alpha^*\hat{\pi}\hat{\psi},$$

$$= \omega((i - \hat{\pi}\hat{\psi})\hat{\psi}) - \alpha^*\hat{\psi}\hat{\pi} - \alpha^*\hat{\pi}\hat{\psi},$$

$$= i\omega\hat{\psi} - \alpha^*\{\hat{\psi}, \hat{\pi}\},$$

$$= i(\omega\hat{\psi} - \alpha^*). \tag{4.2.107}$$

Analogously one finds

$$[\hat{\pi}, \hat{H}] = \quad \ldots \quad ,$$

$$= i(-\omega\hat{\pi} - \alpha). \tag{4.2.108}$$

This yields the following system of equations:

$$\dot{\hat{\psi}} = \omega\hat{\psi} - \alpha^* \quad , \quad \dot{\hat{\pi}} = -\omega\hat{\pi} - \alpha. \tag{4.2.109}$$

Recall that an equation of the form

$$\frac{dy}{dx} + P(x)y = Q(x), \tag{4.2.110}$$

has an integrating factor

$$I.F = e^{\int P(x)dx}, \tag{4.2.111}$$

and solution:

$$y = \frac{1}{I.F}\left[\int I.F * Q(x)dx + C\right]. \tag{4.2.112}$$

In this case

$$I.F = e^{\int \omega dt} = e^{\omega t}, \tag{4.2.113}$$

$$\int I.F * Q(x)dx + c = \int e^{\omega t}\alpha^* dt + c$$

$$= \frac{\alpha^*}{\omega}e^{\omega t} + c, \tag{4.2.114}$$

$\Longrightarrow$

$$y = \frac{\alpha^*}{\omega} + ce^{-\omega t}. \tag{4.2.115}$$

Now, if we choose the parameters to be purely imaginary (a choice permitted by the general freedom in the constants $\omega, \alpha, \alpha^*, c$) and consider the initial state at time $t = 0$, we arrive at:

$$\hat{\psi}(t) = e^{-i\omega t}\hat{\psi}(0) + \frac{\alpha^*}{\omega}(1 - e^{-i\omega t}). \tag{4.2.116}$$

For the $\hat{\pi}$ case consider doing the following changes:

$$\hat{\psi} \to \hat{\pi}, \quad \omega \to -\omega, \quad \alpha \to \alpha^*, \tag{4.2.117}$$

which implies

$$\hat{\pi}(t) = e^{i\omega t}\hat{\pi}(0) + \frac{\alpha}{\omega}(e^{i\omega t} - 1). \tag{4.2.118}$$

These solutions will be used in the next steps.

As is known for the bosonic case [34], modulo some constants and variations due to the Grassmann nature of the variables, the solution admits the following general form:

$$K = Ne^{iS_{cl}}, \tag{4.2.119}$$

where $S_{cl}$ is the classical solution. We can split it into 4 terms:

1. **Homogeneous propagator**

   In the absence of sources ($\alpha = \alpha^* = 0$) we recover the harmonic oscillator result without the phase shift $-\frac{w}{2}$:

   $$T_1 = \bar{\psi}_f e^{-i\omega t}\psi_i. \tag{4.2.120}$$

2. **Linear source couplings. Case 1**

   It concerns the interaction of the sources with the initial and final states

   $$T_2 : \quad \bar{\psi}_f * (\psi(t)_{inhomogeneous}) = \bar{\psi}_f \frac{\alpha^*}{\omega}(1 - e^{-i\omega t}), \tag{4.2.121}$$
   $$= -\frac{\alpha^*}{\omega}(1 - e^{-i\omega t})\bar{\psi}. \tag{4.2.122}$$

3. **Linear source couplings. Case 2**

   It is of the form $\xi(t) * \psi_i$, where $\xi(t)$ comes from integrating the source term for $\hat{\pi}$ against the time-reversed evolution operator for $\hat{\pi}$ (it is reversed given the - sign in the equations of motion)

   $$D(t' - t) = e^{i\omega(t' - t)}, \tag{4.2.123}$$

$\Longrightarrow$

$$\xi(t) = -i \int_0^t \alpha D(t'-t)dt',$$
$$= -i\alpha \int_0^t e^{i\omega(t'-t)}dt',$$
$$= -i\alpha e^{-i\omega t} \int_0^t e^{i\omega t'}dt',$$
$$= -i\alpha e^{-i\omega t} [\frac{e^{i\omega t'}}{i\omega}]_0^t,$$
$$= -i\alpha e^{-i\omega t} [\frac{e^{i\omega t}-1}{i\omega}],$$
$$= -\frac{\alpha}{\omega}(1 - e^{-i\omega t}). \tag{4.2.124}$$

$\Longrightarrow$

$$\tau_3: \quad -\frac{\alpha}{\omega}(1 - e^{-i\omega t})\psi_i. \tag{4.2.125}$$

This is equivalent to calculating a term analogous to case 1 but for the $\hat{\pi}$ case and time reversed:

$$\tau_3: \quad (\pi(-t)_{inhomogeneous}) * \psi_i. \tag{4.2.126}$$

4. **Constant term calculations**

   This term arises from the classical action for the source interaction, $\Phi(t) = iS_{\text{interaction}}$, which requires the use of the generating functional method. For further details, the reader is referred to Chapter 9 ("Functional Methods") in the standard text by Peskin and Schroeder [35], as well as Chapter 5 ("Path Integrals for Fermions") in the book by A. Das [22]. The function has the form:

$$\Phi(t) = -\frac{i}{2}\int_0^t dt' \int_0^{t'} dt'' \bar{\eta}(t')D(t'-t'')\eta(t''), \tag{4.2.127}$$

where

$$\eta(t) = -i\alpha^*, \quad \bar{\eta}(t) = -i\alpha, \tag{4.2.128}$$

which emerges from (ec. 5.81 [35]):

$$Z[0] = \exp\{-\frac{i}{2}\int\int dt_1 dt_2 \bar{\theta}(t_1)G_F(t_1 - t_2)\theta(t_2)\}. \tag{4.2.129}$$

$\Longrightarrow$

$$\Phi(t) = -\int_0^t dt' \int_0^{t'} dt''(-i\alpha)e^{i\omega(t'-t'')}(-i\alpha^*) \quad + c,$$

$$= |\alpha|^2 \int_0^t dt' e^{i\omega t'} \int_0^{t'} dt'' e^{-i\omega t''} \quad + c,$$

$$= \ldots ,$$

$$= \frac{i|\alpha|^2}{\omega}(t' - \frac{1}{i\omega}e^{i\omega t'})|_0^t + c,$$

$$= \frac{i|\alpha|^2 t}{\omega} + \frac{|\alpha|^2}{\omega^2}(1 - e^{-i\omega t}) + c. \quad (4.2.130)$$

Fixing the constants properly we find (it is a matter of changing the global (1/2) factor):

$$\Phi(t) = -\frac{|\alpha|^2}{2\omega}(-i\omega t + e^{-i\omega t} - 1),$$

$$:= -f(\alpha, t, \omega). \quad (4.2.131)$$

Collecting all the terms one finally finds:

$$K(\pi_f, t; \psi_0, 0) = \exp\{\pi_f e^{-i\omega t}\psi_0 - \frac{1}{\omega}(1 - e^{-i\omega t})(\alpha^*\pi_f + \alpha\psi_0) - \frac{|\alpha|^2}{2\omega}(-i\omega t + e^{-i\omega t} - 1)\}. \quad (4.2.132)$$

It is straightforward to see that when we turn off the drivers ($\alpha = \alpha^* = 0$) and doing ($\omega \to w$) we recover the harmonic oscillator term minus the global phase shift ($\omega/2$).

Let us calculate now its star exponential.

### 4.2.7    Naive approach for the fermionic driven harmonic oscillator

(Provided mapping $\bar{\psi}_f \to \psi_f$):

$$\exp_\star\{-\frac{i}{\hbar}tH(\Psi, \Pi)\} = \int \exp\left\{-\frac{i}{\hbar}2\sum_j^n \pi_j \Psi'_j\right\} K(\Psi + \Psi', t, \Psi - \Psi', 0)D\Psi',$$

$$\stackrel{1D}{=} \int \exp\left\{-\frac{i}{\hbar}2\pi\Psi'\right\} K(\Psi + \Psi', t, \Psi - \Psi', 0)d\Psi'. \quad (4.2.133)$$

where:

$$\Psi = \frac{\psi_f + \psi_0}{2}, \quad (4.2.134)$$

$$\Psi' = \frac{\psi_f - \psi_0}{2}. \quad (4.2.135)$$

This implies the following form for the propagator:

$$\exp\{\psi_f e^{-i\omega t}\psi_0 - \frac{1}{\omega}(1 - e^{-i\omega t})(\alpha^*\psi_f + \alpha\psi_0) + f(\alpha, t, \omega)\}$$

$$= \exp\{(\Psi + \Psi')e^{-i\omega t}(\Psi - \Psi') - \frac{1}{\omega}(1 - e^{-i\omega t})(\alpha^*(\Psi + \Psi') + \alpha(\Psi - \Psi')) - f(\alpha, t, \omega)\},$$

$$= \exp\{2\Psi' e^{-i\omega t}\Psi - \frac{1}{\omega}(1 - e^{-i\omega t})((\alpha^* + \alpha)\Psi + (\alpha^* - \alpha)\Psi') - f(\alpha, t, \omega)\}.$$

(4.2.136)

Let us develop the integrand:

$$e^{-\frac{i}{\hbar}2\pi\Psi'}K(\Psi + \Psi', t, \Psi - \Psi', 0)$$

$$= (1 - \frac{i}{\hbar}2\pi\Psi')(1 + 2\Psi' e^{-i\omega t}\Psi - \frac{1}{\omega}(1 - e^{-i\omega t})((\alpha^* + \alpha)\Psi + (\alpha^* - \alpha)\Psi') - f(\alpha, t, \omega)),$$

$$= \ldots \; ,$$

$$= 2\Psi' e^{-i\omega t}\Psi - \frac{1}{\omega}(1 - e^{-i\omega t})(\alpha - \alpha^*)\Psi' - \frac{i}{\hbar}2\pi\Psi' + \frac{i}{\hbar\omega}2\pi\Psi'(1 - e^{-i\omega t})(\alpha^* + \alpha)\Psi +,$$

$$- \frac{i}{\hbar}2\pi\Psi' f(\alpha, \omega, t) + \quad \text{terms} \quad \text{without} \quad \Psi'.$$

(4.2.137)

Under the integral sign one finds:

$$\exp_\star\left\{-\frac{i}{\hbar}(tH(\Psi, \pi))\right\} = \int e^{-\frac{i}{\hbar}2\pi\Psi'}K(\Psi + \Psi', t, \Psi - \Psi', 0)d\Psi',$$

$$= 2e^{-i\omega t}\Psi - \frac{1}{\omega}(1 - e^{-i\omega t})(\alpha - \alpha^*) - \frac{i}{\hbar}2\pi + \frac{i}{\hbar\omega}2\pi(1 - e^{-i\omega t})(\alpha^* + \alpha)\Psi - \frac{i}{\hbar}2\pi f(\alpha, \omega, t).$$

(4.2.138)

It can be useful, given the purpose of using later the Feynman-Kac formula, to separate it in its trivial and non-trivial parts:

$$\exp_\star\{-\frac{i}{\hbar}(tH(\Psi, \pi))\} = (2e^{-i\omega t})\Psi - \frac{i2}{\hbar}(1 + f(\alpha, \omega, t))\pi - \frac{i2}{\hbar\omega}(1 - e^{-i\omega t})(\alpha + \alpha^*)\pi\Psi +$$

$$- \frac{1}{\omega}(1 - e^{-i\omega t})(\alpha - \alpha^*).$$

(4.2.139)

where

$$f(\alpha, t, \omega) = \frac{|\alpha|^2}{2\omega}(-i\omega t + e^{-i\omega t} - 1). \tag{4.2.140}$$

It is straightforward to verify that turning the drivers off ($\alpha = \alpha^* = 0.$) yields the same result as the harmonic oscillator (without the global phase shift). As was evident since the undriven example this result does not preserve Grassmann parity. Let us now consider the meticulous case.

### 4.2.8 Meticulous approach for the fermionic driven harmonic oscillator

Expanding the propagator one finds:

$$K = \exp\{\pi_f e^{-i\omega t}\psi_0 - \frac{1}{\omega}(1 - e^{-i\omega t})(\alpha^*\pi_f + \alpha\psi_0) - \frac{|\alpha|^2}{2\omega}(-i\omega t + e^{-i\omega t} - 1)\},$$

$$= 1 + \pi_f e^{-i\omega t}\psi_0 - \frac{1}{\omega}(1 - e^{-i\omega t})(\alpha^*\pi_f + \alpha\psi_0) - \frac{|\alpha|^2}{2\omega}(-i\omega t + e^{-i\omega t} - 1),$$

$$= 1 + (\Pi + \Pi')e^{-i\omega t}(\Psi - \Psi') + \frac{1}{\omega}(1 - e^{-i\omega t})((\Pi + \Pi')\alpha^* + (\Psi - \Psi')\alpha) - \frac{|\alpha|^2}{2\omega}(-i\omega t + e^{-i\omega t} - 1),$$

$$= 1 - \frac{|\alpha|^2}{2\omega}(-i\omega t + e^{-i\omega t} - 1) + \Pi\Psi\exp\{-i\omega t\} - \Pi\Psi'\exp\{-i\omega t\} + \Pi'\Psi\exp\{-i\omega t\}$$

$$-\Pi'\Psi'\exp\{-i\omega t\} + (\Pi\alpha^* + \Pi'\alpha^* + \Psi\alpha - \Psi'\alpha)\frac{1}{\omega}(1 - \exp\{-i\omega t\}). \tag{4.2.141}$$

$\Longrightarrow$

$$\exp\left\{-\frac{i}{\hbar}2\Pi'\Psi'\right\}K = \left(1 - i\frac{2}{\hbar}\Pi'\Psi'\right)\left(1 - \frac{|\alpha|^2}{2\omega}(-i\omega t + e^{-i\omega t} - 1) + \Pi\Psi\exp\{-i\omega t\}\right.$$

$$- \Pi\Psi'\exp\{-i\omega t\} + \Pi'\Psi\exp\{-i\omega t\} - \Pi'\Psi'\exp\{-i\omega t\}$$

$$\left. + (\Pi\alpha^* + \Pi'\alpha^* + \Psi\alpha - \Psi'\alpha)\frac{1}{\omega}(1 - \exp\{-i\omega t\})\right), \tag{4.2.142}$$

$$= 1 - \frac{|\alpha|^2}{2\omega}(-i\omega t + e^{-i\omega t} - 1) + \Pi\Psi\exp\{-i\omega t\} - \Pi\Psi'\exp\{-i\omega t\}$$

$$+ \Pi'\Psi\exp\{-i\omega t\} - \Pi'\Psi'\exp\{-i\omega t\}$$

$$+ (\Pi\alpha^* + \Pi'\alpha^* + \Psi\alpha - \Psi'\alpha)\frac{1}{\omega}(1 - \exp\{-i\omega t\})$$

$$- i\frac{2}{\hbar}\left[\Pi'\Psi'\left(1 - \frac{|\alpha|^2}{2\omega}(-i\omega t + e^{-i\omega t} - 1)\right)\right.$$

$$+ \Pi'\Psi'\Pi\Psi\exp\{-i\omega t\}$$

$$\left. + (\Pi'\Psi'\Pi\alpha^* + \Pi'\Psi'\Psi\alpha)\frac{1}{\omega}(1 - \exp\{-i\omega t\})\right]. \tag{4.2.143}$$

$\Longrightarrow$

$$\int D\Psi' D\Pi' \exp\{-\frac{i}{\hbar}2\Pi'\Psi'\}K = -\exp\{-i\omega t\} - i\frac{2}{\hbar}[1 - \frac{|\alpha|^2}{2\omega}(-i\omega t + e^{-i\omega t} - 1) + \Pi\Psi\exp\{-i\omega t\}$$

$$+ (\Pi\alpha^* + \Psi\alpha)\frac{1}{\omega}(1 - \exp\{-i\omega t\})]. \tag{4.2.144}$$

$\Longrightarrow$

$$
\begin{aligned}
C \exp&\left\{\frac{i}{\hbar}\Pi\Psi\right\} \int D\Psi' D\Pi' \exp\left\{-\frac{i2}{\hbar}\Pi'\Psi'\right\} K \\
&= \mathrm{Exp}_\star\left(-\frac{it}{\hbar}H(\Psi,\Pi)\right), \\
=C&\left\{-\exp\{-i\omega t\}\right. \\
&-i\frac{2}{\hbar}\left[1-\frac{|\alpha|^2}{2\omega}(-i\omega t + \exp\{-i\omega t\} - 1) + \Pi\Psi \exp\{-i\omega t\}\right. \\
&\qquad \left.+(\Pi\alpha^* + \Psi\alpha)\frac{1}{\omega}(1-\exp\{-i\omega t\})\right] \\
&\left.+\frac{i}{\hbar}\left(-\Pi\Psi\exp\{-i\omega t\} - i\frac{2}{\hbar}\Pi\Psi\left(1-\frac{|\alpha|^2}{2\omega}(-i\omega t + \exp\{-i\omega t\} - 1)\right)\right)\right\}, \\
=-&\left(\frac{\hbar^2}{5}\right)\left[\exp\left\{\frac{i}{\hbar}\Pi\Psi\right\}\left(\exp\{-i\omega t\} + i\frac{2}{\hbar}(1-\frac{|\alpha|^2}{2\omega}(-i\omega t + \exp\{-i\omega t\} - 1))\right)\right. \\
&\left.+i\frac{2}{\hbar}\left(\Pi\Psi\exp\{-i\omega t\} + (\Pi\alpha^* + \Psi\alpha)\frac{(1-\exp\{-i\omega t\})}{\omega}\right)\right]. \qquad (4.2.145)
\end{aligned}
$$

It is also straightforward to see that when we turn off the drivers ($\alpha = \alpha^* = 0$) we recover the harmonic oscillator term minus the global phase shift ($\omega/2$).

Let us now work an application of the previous results in the Feynman-Kac formula.

# Chapter 5
# Feynman-Kac's formula from star exponentials

The Feynman-Kac formula provides a powerful link between *the asymptotic behavior of a system's propagator and its ground state energy* [36]. Given the established connection between propagators and star exponentials—detailed in the previous chapter and in works such as [37, 8], inter alia—we will now establish an analogous Feynman-Kac formula directly within the framework of deformation quantization. For a comprehensive treatment of this topic, the reader is referred to the paper *The Feynman-Kac formula in deformation quantization* by J. Berra–Montiel, H.H. Garcia-Compean, and A. Molgado [38].

Analogous to the procedure for star exponentials, we'll now address each case in turn.

## 5.1 Feynman-Kac formula for bosons

P. Sharan demonstrated that Feynman's path integral is precisely the Fourier transform of the Moyal star exponential [37]. This foundational result was later extended to field theory [39], and relatively recent work (2020) [34] has established how to determine the star exponential directly from the quantum mechanical propagator. Let us start now with the procedure.

The Wigner function $\rho_n(x, p)$ is normalized such that its integral over all of phase space is equal to one [1]:

$$\int_{\mathbb{R}^2} \rho_n(x, p) dx dp = 1. \qquad \text{(Normalized Wigner Function)}$$

---
[1]modulo a global factor ($\frac{1}{2\pi\hbar}$)

Let us integrate the Fourier-Dirichlet expansion (2.2.23) over the entire phase space. Using the normalization property (Normalized Wigner Function ) of the Wigner functions, we find:

$$\frac{1}{2\pi\hbar} \int_{\mathbb{R}^2} \text{Exp}_\star \left(-\frac{i}{\hbar}tH\right) dxdp = \sum_{n=0}^{\infty} e^{-\frac{i}{\hbar}tE_n}. \tag{5.1.1}$$

Next, we perform an analytic continuation of the time variable via a Wick rotation, $\tau = it$. In the limit $\tau \to \infty$, the summation on the right-hand side becomes dominated by the ground state term ($E_0$), as the contributions from all higher energy states decay exponentially faster. This yields the asymptotic behavior:

$$\lim_{\tau \to \infty} \frac{e^{\frac{\tau}{\hbar}E_0}}{2\pi\hbar} \int_{\mathbb{R}^2} \text{Exp}_\star \left(-\frac{\tau}{\hbar}H\right) dxdp = 1, \tag{5.1.2}$$

where $E_0$ is the non-degenerate ground state energy (If the ground state is degenerate, the limit equals the degree of degeneracy).

From this expression, the ground state energy can be extracted by solving for $E_0$. This gives the deformation quantization version of the Feynman-Kac formula:

$$E_0 = -\lim_{\tau \to \infty} \frac{\hbar}{\tau} \ln\left[\frac{1}{2\pi\hbar} \int_{\mathbb{R}^2} \text{Exp}_\star \left(-\frac{\tau}{\hbar}H\right) dxdp\right]. \tag{5.1.3}$$

This outstanding result provides a method for determining ground state energy levels using neither operators nor Green's functions. Let's now illustrate this approach with some key examples.

### 5.1.1 Feynman-Kac formula for the bosonic harmonic oscillator

Let us analyze the case of a quantum harmonic oscillator in one dimension, with the Hamiltonian:

$$H_{ho}(x,p) = \frac{p^2}{2m} + \frac{m\omega^2 x^2}{2}. \tag{5.1.4}$$

For this instance, the Wick-rotated star exponential is [8, 17]:

$$\text{Exp}_\star \left(-\frac{\tau}{\hbar}H_{ho}(x,p)\right) = \frac{1}{\cosh(\frac{\omega\tau}{2})} \exp\left[\frac{2H_{ho}(x,p)}{\omega\hbar} \tanh\left(\frac{\omega\tau}{2}\right)\right]. \tag{5.1.5}$$

Integrating this expression over phase space yields:

$$\frac{1}{2\pi\hbar} \int_{\mathbb{R}^2} \text{Exp}_\star \left(-\frac{\tau}{\hbar}H_{ho}(x,p)\right) dxdp = \frac{1}{2}\text{csch}\left(\frac{\omega\tau}{2}\right). \tag{5.1.6}$$

By rewriting the hyperbolic cosecant as $\frac{1}{2}\text{csch}\left(\frac{\omega\tau}{2}\right) = \frac{e^{\omega\tau/2}}{e^{\omega\tau}-1}$, the Feynman-Kac formula for the ground state energy becomes:

$$E_0 = -\lim_{\tau \to \infty} \frac{\hbar}{\tau} \ln\left[\frac{e^{\omega\tau/2}}{e^{\omega\tau}-1}\right],$$
$$= \frac{\hbar\omega}{2}, \tag{5.1.7}$$

which is the correct zero-point energy for the quantum harmonic oscillator.

### 5.1.2 Feynman-Kac formula for the general quadratic Lagrangian

We now consider the case of general quadratic Hamiltonians, which take the form:

$$H_q(x, p) = ap^2 + bx^2 + 2cxp, \tag{5.1.8}$$

where $a, b, c \in \mathbb{R}$. These Hamiltonians are particularly significant because they're invariant under transformations generated by the group $SL(2, \mathbb{R})$ [38].

For this class of systems, the Wick-rotated star exponential has the following closed form [38]:

$$\text{Exp} \star \left(-\frac{\tau}{\hbar} H_q(x, p)\right) = \frac{1}{\cosh(\sqrt{ab - c^2}\tau)} e^{-\frac{iH_q(x,p)}{\hbar\sqrt{ab-c^2}} \tanh(\sqrt{ab-c^2}\tau)}. \tag{5.1.9}$$

To determine the ground state energy using the Feynman-Kac formula, we first integrate this star exponential over phase space, which yields:

$$\frac{1}{2\pi\hbar} \int_{\mathbb{R}^2} \text{Exp} \star \left(-\frac{\tau}{\hbar} H_q(x, p)\right) dx dp = \frac{1}{2}\text{csch}(\sqrt{ab - c^2}\tau). \tag{5.1.10}$$

Substituting this result into the Feynman-Kac formula gives the ground state energy:

$$E_0 = -\lim_{\tau \to \infty} \frac{\hbar}{\tau} \ln\left[\frac{1}{2}\text{csch}(\sqrt{ab - c^2}\tau)\right],$$
$$= \hbar\sqrt{ab - c^2}. \tag{5.1.11}$$

## 5.2 Feynman-Kac formula for fermions

Let us extend the formula for the fermionic case. As it can be guessed, some subtleties will arise, concerning the nature of the quantities.

We need to expand the Fourier-Dirichlet case for fermions. The eigenvalues of the Hamiltonian operator $\hat{H} = Q_W(H)$ will be the same ones as in the bosonic case; however the question of the diagonal Wigner functions MUST change, as well as the integration of Grassmann variables.

Consider the Wigner function of the projection operator:

$$\rho_n^f = \frac{1}{2\pi\hbar} Q_W^{-1}(P_{|n\rangle}^{\hat{f}}),$$
$$= \frac{1}{2\pi\hbar} Q_W^{-1}(|n_f\rangle \langle n_f|). \tag{5.2.1}$$

where $f$ denotes the fermionic case. In the Weyl-Wigner-Moyal integral formalism it implies:

$$\rho_W(\psi, \pi) = \hbar^{-n} \int D\lambda \exp\{-i\sum_{j=1}^{n} \pi_j \lambda_j\} \langle\psi + \frac{\hbar\lambda}{2}| \hat{\rho} |\psi - \frac{\hbar\lambda}{2}\rangle,$$
$$= \hbar^{-n} \int D\lambda \exp\{-i\sum_{j}^{n} \pi_j \lambda_j\} \phi_n(\psi - \frac{\hbar\lambda}{2}) \bar{\phi}(\psi + \frac{\hbar\lambda}{2}). \tag{5.2.2}$$

It is normalized:

$$\int D\pi D\psi \rho_W(\psi, \pi) = \hbar^{-n} \int D\pi D\psi \int D\lambda \exp\{-i \sum_j^n \pi_j \lambda_j\} \phi_n(\psi - \frac{\hbar\lambda}{2}) \bar{\phi}_n(\psi + \frac{\hbar\lambda}{2}),$$

$$= \hbar^{-n} \int D\lambda [\int D\pi \exp\{-i \sum_j^n \pi_j \lambda_j\}][\int D\psi \phi_n(\psi - \frac{\hbar\lambda}{2}) \bar{\phi}_n(\psi + \frac{\hbar\lambda}{2})],$$

$$= \hbar^{-n} \int D\lambda \delta(\lambda) \int D\psi \phi_n(\psi - \frac{\hbar\lambda}{2}) \bar{\phi}_n(\psi + \frac{\hbar\lambda}{2}),$$

$$= \hbar^{-n} \int D\psi \phi_n(\psi - \frac{\hbar\lambda}{2}) \bar{\phi}_n(\psi + \frac{\hbar\lambda}{2}),$$

$$= \hbar^{-n}. \tag{5.2.3}$$

$\implies$

$$\hat{\rho}_W(\psi, \pi) = \frac{1}{\hbar^n} \int D\lambda \exp\{-i \sum_j^n \pi_j \lambda_j\} \phi_n(\psi - \frac{\hbar\lambda}{2}) \bar{\phi}_n(\psi + \frac{\hbar\lambda}{2}). \tag{5.2.4}$$

Extending the procedure for bosons (available in detail in the appendix) we have the following:

$$\frac{1}{2\pi\hbar} \int \exp_\star \left\{-i \frac{tH(\psi, \pi)}{\hbar}\right\} D\pi D\psi = \int \sum_{n=0}^\infty e^{-i\frac{tE_n}{\hbar}} \rho_W^n(\psi, \pi) D\pi D\psi,$$

$$= \sum_{n=0}^\infty e^{-i\frac{tE_n}{\hbar}} \int \rho_W^n(\psi, \pi) D\pi D\psi \overset{1}{\longrightarrow}$$

$$= \sum_{n=0}^\infty e^{-i\frac{tE_n}{\hbar}}. \tag{5.2.5}$$

$$lim_{\tau \to \infty} \frac{e^{\frac{\tau E_0}{\hbar}}}{2\pi\hbar} \int \exp_\star\{-\frac{\tau H(\pi, \psi)}{\hbar}\} D\pi D\psi = 1, \tag{5.2.6}$$

$$E_0 = -\lim_{\tau \to \infty} \frac{\hbar}{\tau} ln \left(\frac{1}{2\pi\hbar} \int \exp_\star \left\{-\frac{\tau H(\Psi, \Pi)}{\hbar}\right\} D\Pi D\Psi\right), \tag{5.2.7}$$

where we used the *Wick rotation* method, the appropriate asymptotic expansion, and intermediate algebraic manipulations that are isomorphic to the bosonic case. We also renamed the integration variables using uppercase letters to avoid any potential confusion with the geometric factor in the denominator.

### 5.2.1 Naive approach for the Feynman-Kac formula

**Remediation (RMD)**

As we will see, a crucial subtlety arises regarding the imposition of **boundary conditions** in a specific model. This step is fundamental for correctly determining the system's spectrum. The argument can be sketched as follows. For a correct physical propagation of a solution, as well as for mathematical self-consistency, a set of boundary conditions must be imposed, given that we omitted the $\pi$ degrees of freedom within the propagator. If this is not the case, as will be considered in the following examples, a process denoted by us as **remediation (RMD)** must be realized. It consists of imposing a constraint on phase space that guarantees the integration along the $\pi_i = \psi_i$ axis. In Grassmann variables this admits a simple expression:

$$\delta(\pi, \psi) = \int d\eta \exp\{-\eta(\pi - \psi)\},$$
$$= (\pi - \psi). \tag{5.2.8}$$

This will alleviate the boundary conditions problem in a simple and pretty straightforward way.

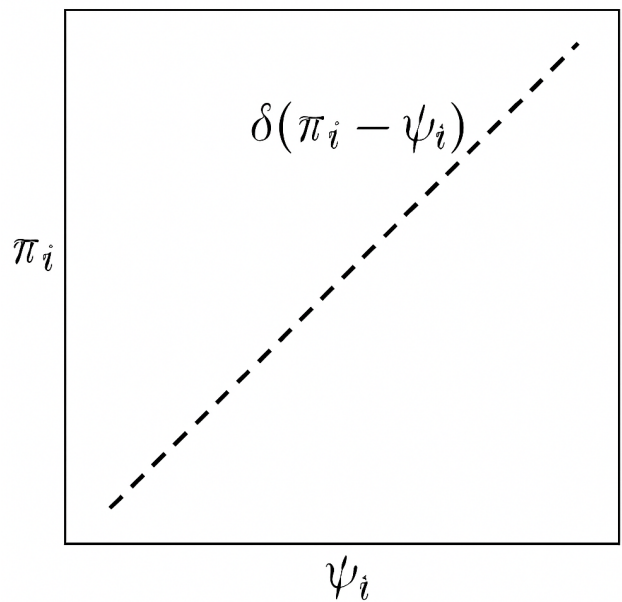

**Figure 5.1:** Phase space in 1D constrained with the delta function. Image generated using AI (ChatGPT, 2025).

Let us now proceed with the most important examples: **oscillators**.

### 5.2.2　Feynman-Kac formula for the fermionic harmonic oscillator

Using the previous results we find

$$
\begin{aligned}
E_0 &= -\lim_{\tau\to\infty}\frac{\hbar}{\tau}\ln\left|\frac{1}{2\pi\hbar}\int \exp_\star\left\{-\frac{\tau H(\Psi,\Pi)}{\hbar}\right\}\underbrace{(\Pi-\Psi)}_{\text{RMD}}D\Pi\,D\Psi\right|, \\
&= -\lim_{\tau\to\infty}\frac{\hbar}{\tau}\ln\left|\frac{1}{2\pi\hbar}\int \exp\left\{\frac{\tau\omega}{2}\right\}\left(-2\frac{i}{\hbar}\Pi+2\exp\{-\tau\omega\}\Psi\right)(\Pi-\Psi)\,D\Pi\,D\Psi\right|, \\
&= -\lim_{\tau\to\infty}\frac{\hbar}{\tau}\ln\left|\frac{1}{2\pi\hbar}\exp\left\{\frac{\tau\omega}{2}\right\}(2\frac{i}{\hbar}+2\exp\{-\tau\omega\})\right|, \\
&:= -\lim_{\tau\to\infty}\frac{\hbar}{\tau}\ln\left|a\exp\left\{\frac{\tau\omega}{2}\right\}+b\exp\left\{-\frac{\tau\omega}{2}\right\}\right|.
\end{aligned}
\tag{5.2.9}
$$

Now we will study the limit, showing that the nature of $a$, $b$ will be irrelevant.

$$
\begin{aligned}
ln|ae^{\frac{\tau\omega}{2}}+be^{\frac{-\tau\omega}{2}}| &= ln|e^{\frac{\tau\omega}{2}}(a+be^{-\tau\omega})|, \\
&= ln|e^{\frac{\tau\omega}{2}}|+ln|a+be^{-\tau\omega}|, \\
&= \frac{\tau\omega}{2}+ln|a+be^{-\tau\omega}|.
\end{aligned}
\tag{5.2.10}
$$

Returning to the previous calculation we find:

$$
\begin{aligned}
E_0 &= -\lim_{\tau\to\infty}\frac{\hbar}{\tau}(\frac{\tau\omega}{2}+\underbrace{ln|a+be^{-\tau\omega}|}_{\approx ln|a|\quad\text{as}\quad\tau\to\infty}), \\
&= -\frac{\hbar\omega}{2}.
\end{aligned}
\tag{5.2.11}
$$

This confirms the exact result.

### 5.2.3　Feynman-Kac formula for the fermionic driven harmonic oscillator

Let us develop, given the extensive character of the star exponential, first the case under the integral sign:

$$
\begin{aligned}
\frac{1}{2\pi\hbar}\int &[(2e^{-i\omega t})\Psi-\frac{i2}{\hbar}(1+f(\alpha,\omega,t))\pi-\frac{i2}{\hbar\omega}(1-e^{-i\omega t})(\alpha+\alpha^*)\pi\Psi, \\
&-\frac{1}{\omega}(1-e^{-i\omega t})(\alpha-\alpha^*)](\Pi-\Psi)D\Pi D\Psi, \\
&= \frac{1}{2\pi\hbar}[2e^{-i\omega t}+i\frac{2}{\hbar}(1+f(\alpha,\omega,t)], \\
&= \dots, \\
&= \left(\frac{1}{\pi\hbar}+i\frac{|\alpha|^2}{2\pi\omega\hbar^2}\right)e^{-\tau\omega}-\left(i\frac{|\alpha|^2\omega}{\hbar}\right)\tau-i\left(\frac{|\alpha|^2}{\hbar\omega}-\frac{2}{\hbar}\right),
\end{aligned}
\tag{5.2.12}
$$

where

$$f(\alpha, t, \omega) = -\frac{|\alpha|^2}{2\omega}(-i\omega t + e^{-i\omega t} - 1).$$

We must treat the complex nature of the constants, hence solving also the open case we assumed for the harmonic oscillator.

**Limit evaluation**

The following limit must be solved

$$L = \lim_{\tau \to \infty} \frac{1}{\tau} \ln(ae^{-\tau\omega} + b\tau + c),$$

given the following complex-valued constants:

$$a = \frac{1}{\pi\hbar} + \frac{i|\alpha|^2}{2\pi\omega\hbar^2},$$
$$b = -\frac{i|\alpha|^2}{\hbar}\omega,$$
$$c = -i\left(\frac{|\alpha|^2}{\hbar\omega} - \frac{2}{\hbar}\right).$$

The key insight is that the final value of the limit depends only on the parameter $\omega$, and not on the specific complex values of the constants $a$, $b$, and $c$.

**General Approach for Complex Constants**

Let $Z(\tau) = ae^{-\tau\omega} + b\tau + c$. Since $a, b, c$ are complex, $Z(\tau)$ is a complex number for any given $\tau$. The natural logarithm of a complex number $Z$ is defined as:

$$\ln(Z) = \ln|Z| + i \cdot \arg(Z),$$

where $|Z|$ is the magnitude of $Z$ and $\arg(Z)$ is its principal argument (phase).

Substituting this into our limit expression, we can split the limit into its real and imaginary parts:

$$L = \lim_{\tau \to \infty} \frac{1}{\tau} \left(\ln|Z(\tau)| + i \cdot \arg(Z(\tau))\right),$$
$$= \underbrace{\lim_{\tau \to \infty} \frac{1}{\tau} \ln|Z(\tau)|}_{\text{Real Part}} + i \underbrace{\lim_{\tau \to \infty} \frac{1}{\tau} \arg(Z(\tau))}_{\text{Imaginary Part}}.$$

To evaluate this, we must identify the term in $Z(\tau)$ that has the largest magnitude as $\tau \to \infty$. This is the **dominant term**. For very large $\tau$, the argument of $Z(\tau)$ will approach the constant argument of its dominant term. Let's say $\arg(Z(\tau)) \to \phi_{\text{dominant}}$. The imaginary part of the limit then becomes:

$$i \lim_{\tau \to \infty} \frac{1 \cdot \phi_{\text{dominant}}}{\tau} = 0.$$

Since the numerator is a constant while the denominator goes to infinity, the imaginary part of the limit is always zero. Therefore, the final result is purely real and is determined entirely by the limit of the magnitude term:

$$L = \lim_{\tau \to \infty} \frac{1}{\tau} \ln |Z(\tau)|.$$

This simplifies the problem to analyzing the magnitude of the dominant term in $Z(\tau)$, which depends on the sign of $\omega$.

### Evaluation Based on the Sign of $\omega$

We now analyze the three possible cases for the real parameter $\omega$.

### Case 1: $\omega > 0$

In this case, the term $e^{-\tau \omega}$ is a decaying exponential. As $\tau \to \infty$, the magnitudes of the terms in $Z(\tau)$ behave as follows:

- $|ae^{-\tau\omega}| = |a|e^{-\tau\omega} \to 0$,
- $|b\tau| = |b|\tau \to \infty$,
- $|c|$ is constant.

The dominant term is $b\tau$. Therefore, for large $\tau$, the magnitude $|Z(\tau)|$ is approximately $|b|\tau$. The limit becomes:

$$L = \lim_{\tau \to \infty} \frac{1}{\tau} \ln(|b|\tau),$$
$$= \lim_{\tau \to \infty} \frac{1(\ln |b| + \ln \tau)}{\tau}.$$

Since the linear function $\tau$ grows faster than the logarithmic function $\ln \tau$, this limit is **0**.

### Case 2: $\omega < 0$

In this case, it is helpful to write $\omega = -|\omega|$. The exponential term becomes $e^{-\tau\omega} = e^{\tau|\omega|}$, which is a growing exponential. The magnitudes behave as:

- $|ae^{\tau|\omega|}| \to \infty$ (exponential growth),
- $|b\tau| \to \infty$ (linear growth),

- $|c|$ is constant.

Exponential growth is much faster than linear growth, so the dominant term is $ae^{\tau|\omega|}$. The magnitude $|Z(\tau)|$ is approximately $|a|e^{\tau|\omega|}$. The limit becomes:

$$L = \lim_{\tau \to \infty} \frac{1}{\tau} \ln(|a|e^{\tau|\omega|}),$$

$$= \lim_{\tau \to \infty} \frac{1}{\tau}(\ln|a| + \ln(e^{\tau|\omega|})),$$

$$= \lim_{\tau \to \infty} \frac{1}{\tau}(\ln|a| + \tau|\omega|),$$

$$= \lim_{\tau \to \infty} \left(\frac{\ln|a|}{\tau} + \frac{\tau|\omega|}{\tau}\right).$$

The first term goes to zero, leaving:

$$L = |\omega|.$$

Since $\omega < 0$, we have $|\omega| = -\omega$. Thus, the limit is $\omega$.

#### Case 3: $\omega = 0$

If $\omega = 0$, the constant $b$ becomes zero:

$$b = -\frac{i|\alpha|^2}{\hbar}(0),$$

$$= 0.$$

The expression for $Z(\tau)$ simplifies to a constant, independent of $\tau$:

$$Z(\tau) = ae^0 + (0)\tau + c,$$

$$= a + c.$$

The limit is therefore:

$$L = \lim_{\tau \to \infty} \frac{\ln(a+c)}{\tau}.$$

Since $\ln(a+c)$ is just a complex constant, the limit is **0**.

#### Summary

The final value of the limit depends solely on the sign of $\omega$. The results are summarized below.

If we then recall that the Expression for $E_0$ has a global minus sign, modulo $\hbar$ we find:

$$E_0 = \omega. \tag{5.2.13}$$

**Table 5.1:** Final value of the limit $L$ based on the condition on $\omega$.

| Condition on $\omega$ | Final Value of the Limit ($L$) |
|---|---|
| $\omega > 0$ | 0 |
| $\omega < 0$ | $-\omega$ |
| $\omega = 0$ | 0 |

Now if we recall the result for the eigenenergies determined for the Hamiltonian we have:

$$\begin{aligned}\lambda &= \frac{\omega \pm \sqrt{\omega^2 + 4|\alpha|^2}}{2}, \\ &= \frac{\omega}{2} \pm \frac{1}{2}\sqrt{\omega^2 + 4|\alpha|^2}, \\ &\approx \begin{cases} \omega \\ 0 \end{cases} \iff |\alpha| \ll |\omega|. \end{aligned} \quad (5.2.14)$$

Let us study in greater detail some aspects of the approximation to understand why it is valid or not.

**Parameter space $(\omega, \alpha)$**

$$\lambda_\pm = \frac{\omega \pm \sqrt{\omega^2 + 4|\alpha|^2}}{2}, \quad \text{with} \quad g := |\alpha|.$$

**Table 5.2:** Approximate eigenvalues $\lambda_\pm$ under different physical conditions.

| Region | Condition | Approximate eigenvalues $\lambda_\pm$ |
|---|---|---|
| (1) Resonant | $\omega \to 0$, $g$ not weak | $\lambda_\pm \approx \pm g + \frac{\omega}{2} \pm \frac{\omega^2}{8g}$ $\;(O((\omega/g)^3))$ |
| (2) Dispersive ($\omega > 0$) | $|\omega| \gg g$ | $\lambda_+ \approx \omega + \frac{g^2}{\omega}, \quad \lambda_- \approx -\frac{g^2}{\omega}$ $\;(O(g^4/\omega^3))$ |
| (2) Dispersive ($\omega < 0$) | $|\omega| \gg g$ | $\lambda_- \approx \omega + \frac{g^2}{\omega}, \quad \lambda_+ \approx -\frac{g^2}{\omega}$ $\;(O(g^4/\omega^3))$ |
| (3) Strong coupling | $g \gg |\omega|$ | $\lambda_\pm \approx \pm g + \frac{\omega}{2} \pm \frac{\omega^2}{8g}$ $\;(O((\omega/g)^3))$ |
| (4) Weak coupling | $\alpha \to 0$ | $\lambda_+ \to \omega, \quad \lambda_- \to 0$ |

We thus confirm the approximate result within the weak coupling regime.

### 5.2.4 Meticulous approach for the Feynman-Kac formula

This proposal ensures Grassmann parity; therefore there is no need to apply a remediation process. The important formula is the following:

$$E_0 = -lim_{\tau \to \infty} \frac{\hbar}{\tau} ln(\frac{1}{2\pi\hbar} \int \exp\{-\frac{\tau H(\Psi, \Pi)}{\hbar}\} D\Pi D\Psi), \qquad (5.2.15)$$

with the Wick rotation $\tau = it$.

Let us now consider its applications with oscillators.

**Harmonic Oscillator**

Calculating the integral we find:

$$\int \exp_\star\left\{-\frac{itH(\Psi, \Pi)}{\hbar}\right\} D\Pi D\Psi$$

$$= \int -\left(\frac{\hbar^2}{5}\right)\left[\exp\left\{\frac{i}{\hbar}\Pi\Psi\right\}\left(\exp\left\{-\frac{i\omega t}{2}\right\} + i\frac{2}{\hbar}\exp\left\{i\frac{wt}{2}\right\}\right) + i\frac{2}{\hbar}\Pi\Psi \exp\left\{-\frac{i\omega t}{2}\right\}\right] D\Pi D\Psi,$$

$$= -\left(\frac{\hbar^2}{5}\right)\left[\left(\frac{i}{\hbar}\right)\left(\exp\left\{-\frac{i\omega t}{2}\right\} + i\frac{2}{\hbar}\exp\left\{i\frac{wt}{2}\right\}\right) + i\frac{2}{\hbar}(1)\exp\left\{-\frac{i\omega t}{2}\right\}\right],$$

$$= -\left(\frac{\hbar^2}{5}\right)\left[\frac{i}{\hbar}\exp\left\{-\frac{i\omega t}{2}\right\} - \frac{2}{\hbar^2}\exp\left\{i\frac{wt}{2}\right\} + \frac{2i}{\hbar}\exp\left\{-\frac{i\omega t}{2}\right\}\right],$$

$$= -\left(\frac{\hbar^2}{5}\right)\left[\frac{3i}{\hbar}\exp\left\{-\frac{i\omega t}{2}\right\} - \frac{2}{\hbar^2}\exp\left\{i\frac{wt}{2}\right\}\right],$$

$$= \frac{2}{5}\exp\left\{i\frac{wt}{2}\right\} - \frac{3i\hbar}{5}\exp\left\{-\frac{i\omega t}{2}\right\}. \qquad (5.2.16)$$

Provided we use the same Wick rotation as in the naive case we recover the structure

$$ln|ae^{\frac{\tau\omega}{2}} + be^{\frac{-\tau\omega}{2}}|, \qquad (5.2.17)$$

which corroborates the result:

$$E_0 = -\frac{\hbar\omega}{2}. \qquad (5.2.18)$$

**Driven harmonic oscillator**

Using (4.2.145) we get:

$$\int Exp_\star\left(-\frac{it}{\hbar}H(\Psi, \Pi)\right) D\Pi D\Psi =$$

$$-(\frac{\hbar^2}{5})\left[\{\frac{i}{\hbar}\}\left(\exp\{-i\omega t\} + i\frac{2}{\hbar}(1 - \frac{|\alpha|^2}{2\omega}(-i\omega t + \exp\{-i\omega t\} - 1))\right)\right.$$

$$\left. +i\frac{2}{\hbar}(\exp\{-i\omega t\}. \qquad (5.2.19)\right.$$

Again using the same Wick rotation one arrives at the foloowing structure:

$$ln|ae^{\frac{\tau\omega}{2}} + b\tau + c| \tag{5.2.20}$$

which as was proven for the naive method yields the two asymptotic results

$$\lambda \approx \begin{cases} \omega \\ 0 \end{cases} \quad \Longleftrightarrow \quad |\alpha| \ll |\omega|. \tag{5.2.21}$$

which correspond to the weak coupling region within the system's parameter space.

# Chapter 6
# Conclusions

Deformation quantization stands as an eclectic theoretical approach for merging classical and quantum mechanics. Inspired by the development of *phase-space quantum mechanics*, it is based on extending the algebra of observables into a more general structure that connects to classical phase-space quantities *asymptotically*. One of its most important elements is the existence of Star-Products. For decades, the only known example was the Moyal Star-Product. However, following his foundational paper [18], M. Kontsevich proved the existence of another kind of star product, later referred to as the *Kontsevich Star-Product*, which was inspired on perturbative methods from quantum field theory and string theory. This remarkable discovery brought attention back to the theory, inspiring new frameworks in which to apply its principles. However, the path has not been trivial. Star-exponentials—the symbols of the evolution operator and a building block of the theory—face an inevitable problem: convergence. In fact, proving or disproving the convergence of general star-exponentials remains an open problem in the literature.

This might be interpreted as a deep flaw within the formalism that nullifies its potential applications for quantizing physical systems. Nonetheless, this is not the case. J. Berra-Montiel, H. Garcia-Compean, and A. Molgado devised a mechanism that makes it possible to obtain this quantity without relying on its explicit formal series calculation. They found a closed-form integral relation between the star-exponential of a system and its propagator. This is a direct and useful application that alleviates the problem of convergence by using an a priori known physical quantity—the quantum propagator—to calculate its associated star-exponential. Its potential applications are multiple, one of the most important being its connection to the Feynman-Kac formula [38]. The original formalism, however, was developed only for the bosonic case, thus limiting its applications to a specific type of matter.

In this thesis, an extension of this formalism to the fermionic case was developed. To achieve this, a deep understanding of the mathematics associated with fermions (namely, Grassmann variables) was fundamental. A consistent formalism was obtained, and a direct application of it was deduced: the Feynman-Kac formula for fermions. With this tool, the ground state energy of a given fermionic system could be obtained. The two most elementary examples were developed: the harmonic oscillator and a "driven" generalization of it. Through these examples, several subtleties were identified and resolved. The first challenge was to obtain a closed-form expression for the fermionic oscillator propagator; although much information exists on path integrals for fermions, a concise formula suitable for this work was not readily available. For achieving a successful result, two alternative methods were developed: one intuitive and the other one more meticulous. The second consisted of understanding the nuances of Grassmann-Berezin integration, as it is non-trivial to obtain physically correct, non-zero results at least for the intuitive approach. Finally, a significant problem was the implementation of proper boundary conditions for determining the ground state energy via the Feynman-Kac formula. This was ultimately solved by constraining the integration to a specific subspace of the Grassmann phase space, though it appeared as an inexorable anomaly within that approach, not being present at all after more thoughtful considerations.

**Open Problems and Future Work**

First, it is important to independently verify that the formulas developed for fermions are valid. We showed that the approach works for the two most elementary examples available, but this is by no means sufficient. It is common knowledge in physics that if a theory cannot properly describe the harmonic oscillator, it must be rejected; however, its ability to model the oscillator does not obviously imply its general validity.

It is also important to merge bosonic and fermionic degrees of freedom as a theoretical advancement that could further elucidate the connection between them. To this end, a supersymmetric (SUSY) case must be developed. A useful guide for this would be the paper by Galaviz et al. [13], which includes comments concerning the SUSY case within the Weyl-Wigner-Moyal formalism, as well as relevant chapters in A. Das's book, *Field Theory: A Path Integral Approach* [22]. This will be the next step to be studied within our mathematical physics group, led by H. Garcia-Compean at Cinvestav. Finally, it is important to work in more general contexts, such as in quantum field theory and possibly string theory.

We have shown that this is an interesting and successful alternative quantization scheme that has multiple applications, as well as theoretically interesting intricacies; nonetheless, there is much more to be done.

# Appendix
# Appendix

## A  Deduction of the Commutator $[\hat{x}^n, \hat{p}^m]$

The following deduction establishes the general form for the commutator of integer powers of the position operator $\hat{x}$ and the momentum operator $\hat{p}$, based on the canonical commutation relation (CCR) $[\hat{x}, \hat{p}] = i\hbar$ [40].

**Lemma 1.** *For any positive integer n, the commutator of $\hat{x}^n$ and $\hat{p}$ is given by:*

$$[\hat{x}^n, \hat{p}] = i\hbar n \hat{x}^{n-1}$$

*Proof.* We proceed by mathematical induction on $n$.

**Base Case ($n = 1$):** The statement is $[\hat{x}, \hat{p}] = i\hbar(1)\hat{x}^0 = i\hbar$, which is the canonical commutation relation. Thus, the base case holds.

**Inductive Step:** Assume the formula holds for some integer $k \geq 1$, i.e., $[\hat{x}^k, \hat{p}] = i\hbar k \hat{x}^{k-1}$ (the inductive hypothesis). Consider the commutator $[\hat{x}^{k+1}, \hat{p}]$:

$$\begin{aligned}
[\hat{x}^{k+1}, \hat{p}] &= [\hat{x}\hat{x}^k, \hat{p}] \\
&= \hat{x}[\hat{x}^k, \hat{p}] + [\hat{x}, \hat{p}]\hat{x}^k & \text{(Leibniz Rule)} \\
&= \hat{x}(i\hbar k \hat{x}^{k-1}) + (i\hbar)\hat{x}^k & \text{(Inductive Hypothesis and CCR)} \\
&= i\hbar k \hat{x}^k + i\hbar \hat{x}^k \\
&= i\hbar(k+1)\hat{x}^k
\end{aligned}$$

By the principle of induction, the lemma is proven. □

**Lemma 2.** *For any positive integer m, the commutator of $\hat{x}$ and $\hat{p}^m$ is given by:*

$$[\hat{x}, \hat{p}^m] = i\hbar m \hat{p}^{m-1}$$

*Proof.* The proof is symmetric to that of Lemma 1, proceeding by induction on $m$ and using the Leibniz rule in the form $[\hat{A}, \hat{B}\hat{C}] = [\hat{A}, \hat{B}]\hat{C} + \hat{B}[\hat{A}, \hat{C}]$. □

**Theorem 1.** *For any positive integers n and m, the commutator $[\hat{x}^n, \hat{p}^m]$ is given by the expression:*

$$[\hat{x}^n, \hat{p}^m] = \sum_{k=0}^{n-1} i\hbar m \, \hat{x}^k \hat{p}^{m-1} \hat{x}^{n-1-k}$$

*Proof.* Let $C_{n,m} = [\hat{x}^n, \hat{p}^m]$. We will establish a recurrence relation for $C_{n,m}$ in the index $n$ and solve it by iteration.

Using the Leibniz identity $[\hat{A}\hat{B}, \hat{C}] = \hat{A}[\hat{B}, \hat{C}] + [\hat{A}, \hat{C}]\hat{B}$ with $\hat{A} = \hat{x}$, $\hat{B} = \hat{x}^{n-1}$, and $\hat{C} = \hat{p}^m$:

$$C_{n,m} = [\hat{x}\hat{x}^{n-1}, \hat{p}^m] = \hat{x}[\hat{x}^{n-1}, \hat{p}^m] + [\hat{x}, \hat{p}^m]\hat{x}^{n-1} \tag{A.1}$$

In our notation, this is $C_{n,m} = \hat{x} C_{n-1,m} + [\hat{x}, \hat{p}^m]\hat{x}^{n-1}$. Applying Lemma 2, we obtain the recurrence relation:

$$C_{n,m} = \hat{x} C_{n-1,m} + (i\hbar m \hat{p}^{m-1})\hat{x}^{n-1}$$

The recursion terminates with the base case $C_{0,m} = [\hat{x}^0, \hat{p}^m] = [\hat{I}, \hat{p}^m] = 0$.

We solve this recurrence by expanding it iteratively:

$$\begin{aligned}
C_{n,m} &= \hat{x} C_{n-1,m} + i\hbar m \hat{p}^{m-1} \hat{x}^{n-1} \\
&= \hat{x}\left(\hat{x} C_{n-2,m} + i\hbar m \hat{p}^{m-1} \hat{x}^{n-2}\right) + i\hbar m \hat{p}^{m-1} \hat{x}^{n-1} \\
&= \hat{x}^2 C_{n-2,m} + i\hbar m \hat{x} \hat{p}^{m-1} \hat{x}^{n-2} + i\hbar m \hat{p}^{m-1} \hat{x}^{n-1} \\
&= \hat{x}^2 \left(\hat{x} C_{n-3,m} + i\hbar m \hat{p}^{m-1} \hat{x}^{n-3}\right) + i\hbar m \hat{x} \hat{p}^{m-1} \hat{x}^{n-2} + i\hbar m \hat{p}^{m-1} \hat{x}^{n-1} \\
&= \hat{x}^3 C_{n-3,m} + i\hbar m \hat{x}^2 \hat{p}^{m-1} \hat{x}^{n-3} + i\hbar m \hat{x} \hat{p}^{m-1} \hat{x}^{n-2} + i\hbar m \hat{p}^{m-1} \hat{x}^{n-1} \\
&= \ldots
\end{aligned}$$

Observing the emerging pattern, if we continue this expansion $n$ times until we reach $C_{0,m} = 0$, the terms form a sum. Reversing the order of summation for clarity gives:

$$C_{n,m} = i\hbar m \left(\hat{p}^{m-1} \hat{x}^{n-1} + \hat{x} \hat{p}^{m-1} \hat{x}^{n-2} + \cdots + \hat{x}^{n-1} \hat{p}^{m-1}\right)$$

This can be written compactly as a sum:

$$[\hat{x}^n, \hat{p}^m] = \sum_{k=0}^{n-1} i\hbar m \, \hat{x}^k \hat{p}^{m-1} \hat{x}^{n-1-k}$$

□

# B   The Symplectic to Poisson Correspondence

**Definition B.1** (Symplectic Manifold)**.** *A **symplectic manifold** is a pair $(M, \omega)$, where $M$ is a smooth manifold and $\omega$ is a 2-form on $M$ that is [41]:*

1. **Closed:** *Its exterior derivative is zero, $d\omega = 0$.*

2. **Non-degenerate:** *For every point $p \in M$, the map $v \mapsto \omega_p(v, \cdot)$ from the tangent space $T_pM$ to the cotangent space $T_p^*M$ is a linear isomorphism.*

**Definition B.2** (Poisson Manifold). *A **Poisson manifold** is a pair $(M, \{\cdot, \cdot\})$, where $M$ is a smooth manifold and $\{\cdot, \cdot\}$ is an $\mathbb{R}$-bilinear map on the algebra of smooth functions $C^\infty(M)$, called the Poisson bracket, satisfying for all $f, g, h \in C^\infty(M)$ [41]:*

1. **Antisymmetry:** $\{f, g\} = -\{g, f\}$.
2. **Leibniz's Rule (Derivation):** $\{fg, h\} = f\{g, h\} + g\{f, h\}$.
3. **Jacobi Identity:** $\{\{f, g\}, h\} + \{\{g, h\}, f\} + \{\{h, f\}, g\} = 0$.

**Theorem 2.** *Every symplectic manifold $(M, \omega)$ is a Poisson manifold.*

The proof can be sketched in [41]; however the key ingredient lies in proving the Jacobi identity. It can be useful to employ Cartan's magic/homotopy formula:

**Definition B.3** (Cartan's magic formula).

$$\mathcal{L}_X = i_X d + d i_X$$

This equation should be understood as an identity between operators acting on the space of differential forms on a manifold.

### The Ingredients

- **The Lie Derivative ($\mathcal{L}_X$):** This operator measures the change of a tensor field (such as a differential form) along the flow of a given **vector field** $X$. It quantifies how the form deforms as it is dragged along the trajectories defined by $X$.
- **The Exterior Derivative ($d$):** This operator takes a $k$-form and produces a $(k+1)$-form, generalizing the concepts of gradient, curl, and divergence. It captures the intrinsic local change of the form and satisfies the crucial property $d^2 = 0$.
- **The Interior Product ($i_X$):** Also known as contraction, this operator takes a $k$-form and a **vector field** $X$ and produces a $(k-1)$-form. It works by "plugging" the vector field $X$ into the first argument of the differential form.

### The "Magic" Connection

The "magic" of the formula is that it **elegantly connects the geometric and intuitive concept of the Lie derivative (change along a flow) to the purely algebraic and more fundamental operations of the exterior derivative and the interior product**.

# C  Formal definitions and results within Deformation Quantization

**Definition C.1.** *A* Poisson algebra *[24] is a real vector space A equipped with a commutative associative algebra structure*

$$(f, g) \longrightarrow fg$$

*and a Lie algebra structure*

$$(f, g) \longrightarrow \{f, g\}$$

*which satisfy the compatibility condition*

$$\{fg, h\} = f\{g, h\} + \{f, h\}g$$

**Definition C.2.** *A* Poisson manifold *is a manifold M whose function space $C^\infty(M)$ is a Poisson algebra with respect to the usual pointwise multiplication of functions and a prescribed Lie algebra structure.*

**Definition C.3.** *A* formal deformation *of the algebra $A = C^\infty(M)$, or equivalently a* star-product $*$*, is defined to be a map*

$$* : A \times A \longrightarrow A[\hbar]$$

$$(f, g) \longmapsto \sum_{k=0}^{\infty} c_k(f, g) \hbar^k$$

*satisfying*

(i) *formal associativity, i.e. for all $p \geq 0$*

$$\sum_{k+l=p} [c_k(c_l(f, g), h)) - c_k(f, c_l(g, h))] = 0$$

(ii) $c_0(f, g) = fg$

(iii) $(1/2)(c_1(f, g) - c_1(g, f)) = \{f, g\}$ *where* $\{\ ,\ \}$ *is the Poisson bracket.*

(iv) *Each map $c_k : A \times A \longrightarrow A$ should be a bidifferential operator.*

**Definition C.4.** *A* formal deformation *of the Poisson bracket is a skew-symmetric map*

$$[\ ,\ ] : A \times A \longrightarrow A[\hbar]$$

$$(f, g) \longmapsto \sum_{k=0}^{\infty} T_k(f, g) \hbar^k$$

*satisfying:*

(i) *the formal Jacobi identity, i.e. for all $p \geq 0$*

$$\sum \big( \sum_{k+l=p} T_k(T_l(f, g), h) \big) = 0$$

*where the outer sum is taken over the cyclic permutations of the set $\{f, g, h\}$.*

(ii) $T_0(f, g) = \{f, g\}$ *where* $\{\ ,\ \}$ *is the Poisson bracket.*

(iii) *Each map $T_k : A \times A \longrightarrow A$ should be a bidifferential operator.*

## C.1 Cohomological perspective

Let A be an associative algebra (over some commutative ring K) and for simplicity assume it is a module over itself with the adjoint action (i.e. algebra multiplication).

**Definition C.5.** *A p-cochain is a p-linear map C from $A^p$ into the module A, and its coboundary $\partial C$ is given by:*

$$\partial C(u_0, u_1, ..., u_p) = u_0 C(u_1, u_2, .., u_p) - C(u_0 u_1, u_2, .., u_p)$$
$$+ .. + (-1)^p C(u_0, u_1, .., u_{p-1} u_p) + C(u_0, u_1, ..., u_{p-1}) u_p$$

*This is a complex, i.e. $\partial^2 = 0$.*

**Definition C.6.** *A p-cochain C is a p-cocycle if $\partial C = 0$.*

**Definition C.7.** *Let $Z^p(A, A)$ be the space of all p-cocyles and $B^p(A, A)$ the space of those p-cocycles that are coboundaries (of a (p-1)-cochain). The pth Hochschild cohomology space (of A valued in A) is defined as*

$$H^p(A, A) = Z^p(A, A)/B^p(A, A)$$

Let A be a Lie algebra, with bracket { , }. The p-cochains here are skew-symmetric, i.e. they are linear maps $B : \Lambda^p A \longrightarrow A$ and the Chevalley coboundary operator $\partial_c$ is defined on a p-cochain B by:

$$\partial_c C(u_0, u_1, ..u_p) = \sum_{j=0}^{p}(-1)^j \{u_j, C(u_0, .., \hat{u}_j, ...., u_p)\}$$
$$+ \sum_{i<j}(-1)^{i+j} C(\{u_i, u_j\}, u_0, .., \hat{u}_i, .., \hat{u}_j, ..., u_p)$$

(where $\hat{u}$ means that u has to be omitted.)

**Definition C.8.** *A differential graded Lie algebra (DGLA) is a $\mathbb{Z}$-graded Lie superalgebra*

$$L = \bigoplus_{i \geq 0} L^i$$

*with a map $d : L^i \longrightarrow L^{i+1}$ such that*

$$d[a, b] = [da, b] + (-1)^i [a, db] \text{ for } a \in L^i, b \in L$$

*(where the bracket is denoted by [ , ]).*

Let $A = C^\infty(M)$ be the algebra of smooth functions on a smooth real manifold M. Let $C^\circ(A, A)$ be the (local) Hochschild complex of the algebra A over M, i.e. for any n,

$$C^n(A, A) = \{\phi \in Hom(A^{\otimes n}, A) | \phi(f_1, f_2, .., f_n)$$

is a differential operator in each entry $f_i$}

Then the corresponding Hochschild cohomology satisfies:

$$H^n(A, A) = \Lambda^n TM = \text{smooth multivector fields on M}$$

Both the Hochschild complex and the cohomology are differential graded Lie algebras (DGLAs).

**Definition C.9.** *Two DGLAs L, L' are* quasi-isomorphic *if there is a chain*

$$L \longrightarrow L_1 \longleftarrow L_2 \longrightarrow \ldots \longleftarrow L_n \longrightarrow L'$$

*of DGLA homomorphisms all of which induce isomorphisms of cohomology.*

## C.2 Formality Conjecture

**Conjecture 1** (Kontsevich's Formality Conjecture)**.** *The Hochschild complex $C^n$ is quasi-isomorphic as a DGLA to its (Hochschild) cohomology $H^n$.*

Maxim Kontsevich proved that 1, for a manifold M, implies deformation quantization of any Poisson structure on M.

**Theorem 3.** *Let M be a smooth manifold and $A = C^\infty(M)$. Then there is a natural isomorphism between equivalence classes of deformations of the null Poisson structure on M and equivalence classes of smooth deformations of the associative algebra A. In particular any Poisson bracket on M comes from a canonically defined (modulo equivalence) star product.*

The proof is available in [18], which implies **any Poisson structure can be deformed**, solving a 20 years old open problem.

# D Detailed calculations for the fermionic star exponential

**Theorem 4.** *Let $|\psi(t)\rangle$ be the state of a quantum system evolving under Schrödinger's equation:*

$$i\hbar \frac{d}{dt}|\psi(t)\rangle = \hat{H}|\psi(t)\rangle, \tag{D.1}$$

*where $\hat{H}$ is the Hamiltonian of the system. The form of the Schrödinger equation remains invariant under a global phase transformation:*

$$|\psi(t)\rangle \to |\psi'(t)\rangle = e^{i\alpha}|\psi(t)\rangle, \tag{D.2}$$

*with $\alpha \in \mathbb{R}$*

## Proof

We apply the transformation to the state vector:

$$|\psi'(t)\rangle = e^{i\alpha} |\psi(t)\rangle. \tag{D.3}$$

Taking the time derivative:

$$\frac{d}{dt} |\psi'(t)\rangle = \frac{d}{dt} \left( e^{i\alpha} |\psi(t)\rangle \right) \tag{D.4}$$

$$= e^{i\alpha} \frac{d}{dt} |\psi(t)\rangle, \tag{D.5}$$

since $e^{i\alpha}$ is constant.

Multiplying both sides by $i\hbar$:

$$i\hbar \frac{d}{dt} |\psi'(t)\rangle = i\hbar e^{i\alpha} \frac{d}{dt} |\psi(t)\rangle \tag{D.6}$$

$$= e^{i\alpha} \left( i\hbar \frac{d}{dt} |\psi(t)\rangle \right). \tag{D.7}$$

Using Schrödinger's equation for $|\psi(t)\rangle$:

$$i\hbar \frac{d}{dt} |\psi'(t)\rangle = e^{i\alpha} \hat{H} |\psi(t)\rangle = \hat{H} \left( e^{i\alpha} |\psi(t)\rangle \right) = \hat{H} |\psi'(t)\rangle. \tag{D.8}$$

Therefore, $|\psi'(t)\rangle$ also satisfies Schrödinger's equation:

$$i\hbar \frac{d}{dt} |\psi'(t)\rangle = \hat{H} |\psi'(t)\rangle. \tag{D.9}$$

Physical predictions, which depend on $\langle \psi | \psi \rangle$ and expectation values like $\langle \psi | \hat{A} | \psi \rangle$, remain unchanged. This shows that the Hamiltonian $\hat{H}$ is invariant under global phase shifts.

Now consider the calculation of the harmonic oscillator propagator.

## D.1 Propagator of the Harmonic Oscillator: deduction

Consider first the following equation:

$$K(\psi_f, t; \psi_i, 0) = \langle \psi_f | \exp\{-i\hat{H}(t)\} |\psi_i\rangle, \tag{D.10}$$

where the Hamiltonian has the following form:

$$\hat{H} = \omega \hat{\bar{\psi}} \hat{\psi}, \tag{D.11}$$

with $\psi, \bar{\psi}$ being Grassmann variables. The following completeness relations are useful:

$$\int d\bar{\psi} d\psi \, e^{-\bar{\psi}\psi} |\psi\rangle \langle \psi| = 1, \tag{D.12}$$

$$\int d\bar{\psi} d\psi \, \bar{\psi}\psi = 1. \tag{D.13}$$

Inserting multiple resolutions of the identity between infinitesimal time slices turns the transition amplitude into a path integral over Grassmann trajectories $\{\psi(t'), \bar{\psi}(t')\}$.

**Discretizing time and setting up the integral**

1. Consider firstly the time slicing:
   Divide the interval $[0, t]$ into N equal slices of size $\omega = \frac{t}{N}$. Label the intermediate time $t_k = k\omega \ \forall \ \ k = 0, \cdots, N..$
   At each time slice insert $\int d\bar{\psi}_k d\psi_k e^{-\bar{\psi}_k \psi_k} |\psi_k\rangle \langle \psi_k|$.
2. The boundary conditions are $\psi_o = \psi_i$, $\psi_N = \psi_f$ and $\bar{\psi}_N, \bar{\psi}_o$ are left free since they are in bra-kets.
3. The short-time transition amplitude between $\psi_k$ and $\psi_{k+1}$ for a time step $\omega$ is approximately

$$\langle \psi_{k+1}| e^{-i\hat{H}\omega} |\psi_k\rangle = \exp\{\bar{\psi}_{k+1}\psi_k - i\omega \underbrace{H(\bar{\psi}_{k+1}, \psi_k)}_{w\bar{\psi}_{k+1}\psi_k}\} + O(\omega). \tag{D.14}$$

4. Collecting all slices yields the discretized path integral:

$$K \approx \int [\prod_{k=1}^{N-1} d\bar{\psi}_k d\psi_k e^{-\bar{\psi}_k \psi_k}] \exp\{\sum_{k=0}^{N-1} (\bar{\psi}_{k+1}\psi_k - i\omega\omega\bar{\psi}_{k+1}\psi_k)\}, \tag{D.15}$$

where $\psi_0 = \psi_i$, $\psi_N = \psi_f$ and $\bar{\psi}_N$ remains unconstrained in the bra.

**Continuum Limit**

Now consider the following quantities in the continuum limit

$$\sum_{k=0}^{N-1} \bar{\psi}_{k+1}\psi_k = \sum_{k=0}^{N-1} \bar{\psi}_{k+1} \frac{\psi_{k+1} - \psi_k}{\omega} \omega, \tag{D.16}$$

$$\overset{\omega \to \infty}{\approx} \int_0^t dt' \bar{\psi}(t') \dot{\psi}(t'). \tag{D.17}$$

$$\sum_{k=0}^{N-1} (-i\omega\omega\bar{\psi}_{k+1}\psi_k) \overset{\omega \to \infty}{\approx} \int_0^t (-i\omega dt' \bar{\psi}(t')\psi(t')). \tag{D.18}$$

It yields the following:

$$K(\psi_f, t; \psi_o, 0) = \int D\bar{\psi} D\psi \exp\{i \int_0^t \underbrace{(\bar{\psi}\dot{\psi} - \omega\bar{\psi}\psi) dt'}_{L_2}\} \tag{D.19}$$

which confirms the validity of the **path integral formula.** Let us now proceed further.

**Classical Action**

Recall the equations of motion of $L_2$:

$$i\dot{\bar{\psi}} + \omega\bar{\psi} = 0; \quad i\dot{\psi} + \omega\psi = 0 \tag{D.20}$$

Consider now the Ansatz:

$$\psi_{cl}(t') = \psi_i e^{-i\omega t'}, \quad \bar{\psi}_{cl}(t') = \bar{\psi}_f e^{-i\omega(t-t')}. \tag{D.21}$$

Plugging it into the classical action $S_{cl}$ yields:

$$S_{cl} = \int_0^t dt' [\bar{\psi}_{cl}(t')\dot{\psi}_{cl}(t) - \omega\bar{\psi}_{cl}(t')\psi_{cl}(t')]. \tag{D.22}$$

Let us solve each term independently, separating them between Kinetic and Potential terms, respectively.

**Kinetic Term:**

$$\psi_{cl}(t') = \psi_i e^{-i\omega t'} \quad \Longrightarrow \quad \dot{\psi}_{cl}(t') = -i\omega\psi_{cl}(t') \tag{D.23}$$

$$\bar{\psi}_{cl}(t') = \bar{\psi}_f e^{-i\omega t} e^{i\omega t'}. \quad \Longrightarrow \quad \begin{aligned}\bar{\psi}_{cl}(t')\dot{\psi}_{cl}(t') &= (\bar{\psi}_f e^{-i\omega t} e^{i\omega t'})(-i\omega\psi_{cl}(t')), \\ &= -i\omega(\bar{\psi}_f e^{-i\omega t}\psi_i).\end{aligned} \tag{D.24}$$

This yields the following for the first integral:

$$I_1 = \int_0^t dt' \bar{\psi}_{cl}\dot{\psi}_{cl}, \tag{D.25}$$

$$= -i\omega(\bar{\psi}_f e^{-i\omega t}\psi_i) \int_0^t dt', \tag{D.26}$$

$$= -i\omega t(\bar{\psi}_f e^{-i\omega t}\psi_i). \tag{D.27}$$

**Potential Term**

$$\begin{aligned}\bar{\psi}_{cl}(t')\psi_{cl}(t') &= (\bar{\psi}_f e^{-i\omega t})e^{i\omega t'} \times \psi_i e^{-i\omega t'}, \\ &= \bar{\psi}_f e^{-i\omega t}\psi_i.\end{aligned} \tag{D.28}$$

$$\Longrightarrow$$

$$\begin{aligned}I_2 &= ..., \\ &= \omega t(\bar{\psi}_f e^{-i\omega t}\psi_i).\end{aligned} \tag{D.29}$$

Finally calculating $I_1 - I_2$, absorbing in the measures normalization the $\omega$ factor and performing the continuum limit in the time variable yields:

$$S_{cl} = \bar{\psi}_f e^{-i\omega t}\psi_i \tag{D.30}$$

# E   Proof of the Gaussian Fermionic Integral

It is important to note that the formula $\det(M) \exp(-b^T M^{-1} a)$ can sometimes be found in the literature, but it is not the most general result. The standard derivation, which we present here, yields a different ordering of the source terms in the exponent. The two expressions are only equivalent if the matrix $M$ is anti-symmetric. We will prove the general form.

## Proposition

Let $\mathbf{u} = (u_1, \ldots, u_n)^T$ and $\mathbf{v} = (v_1, \ldots, v_n)^T$ be column vectors whose components are independent generators of a Grassmann algebra. Let $\mathbf{a} = (a_1, \ldots, a_n)^T$ and $\mathbf{b} = (b_1, \ldots, b_n)^T$ be column vectors of external, constant Grassmann numbers. Let $M$ be an $n \times n$ invertible matrix whose entries are commuting complex numbers (c-numbers). The Berezin integral is then given by:

$$\int D\mathbf{u}\, D\mathbf{v}\, \exp(\mathbf{v}^T M \mathbf{u} + \mathbf{u}^T \mathbf{a} + \mathbf{v}^T \mathbf{b}) = \det(M)\, \exp(-\mathbf{a}^T M^{-1} \mathbf{b})$$

where the integration measure is defined as $D\mathbf{u}\, D\mathbf{v} = du_n \ldots du_1\, dv_n \ldots dv_1$.

## Proof

The proof proceeds by completing the square in the exponent through a suitable linear shift of the integration variables. A key property of Berezin integration is that the Jacobian for a constant shift of variables is unity.

*Proof.*   1. **Define the Integral:** Let the integral be denoted by $I$:

$$I = \int D\mathbf{u}\, D\mathbf{v}\, \exp(\mathbf{v}^T M \mathbf{u} + \mathbf{u}^T \mathbf{a} + \mathbf{v}^T \mathbf{b})$$

2. **Change of Variables:** We introduce new integration variables $\mathbf{u}'$ and $\mathbf{v}'$ via the shift:

$$\mathbf{u} = \mathbf{u}' + \boldsymbol{\alpha} \quad \text{and} \quad \mathbf{v} = \mathbf{v}' + \boldsymbol{\beta}$$

where $\boldsymbol{\alpha}$ and $\boldsymbol{\beta}$ are vectors of constant Grassmann numbers to be determined. The Berezinian (Jacobian) of this transformation is 1, so the measure is invariant: $D\mathbf{u}\, D\mathbf{v} = D\mathbf{u}'\, D\mathbf{v}'$.
Substituting these shifts into the exponent gives:

$$\begin{aligned} \text{Exponent} &= (\mathbf{v}' + \boldsymbol{\beta})^T M (\mathbf{u}' + \boldsymbol{\alpha}) + (\mathbf{u}' + \boldsymbol{\alpha})^T \mathbf{a} + (\mathbf{v}' + \boldsymbol{\beta})^T \mathbf{b} \\ &= \mathbf{v}'^T M \mathbf{u}' + \mathbf{v}'^T M \boldsymbol{\alpha} + \boldsymbol{\beta}^T M \mathbf{u}' + \boldsymbol{\beta}^T M \boldsymbol{\alpha} + \mathbf{u}'^T \mathbf{a} + \boldsymbol{\alpha}^T \mathbf{a} + \mathbf{v}'^T \mathbf{b} + \boldsymbol{\beta}^T \mathbf{b} \end{aligned}$$

3. **Completing the Square:** We choose $\boldsymbol{\alpha}$ and $\boldsymbol{\beta}$ to eliminate the terms linear in $\mathbf{u}'$ and $\mathbf{v}'$. The linear terms are:

$$(\mathbf{v}'^T M \boldsymbol{\alpha} + \mathbf{v}'^T \mathbf{b}) + (\boldsymbol{\beta}^T M \mathbf{u}' + \mathbf{u}'^T \mathbf{a}) = \mathbf{v}'^T (M\boldsymbol{\alpha} + \mathbf{b}) + (\boldsymbol{\beta}^T M + \mathbf{a}^T)\mathbf{u}'$$

Setting the coefficients to zero yields:

$$M\boldsymbol{\alpha} + \mathbf{b} = 0 \quad \Longrightarrow \quad \boldsymbol{\alpha} = -M^{-1}\mathbf{b}$$
$$\boldsymbol{\beta}^T M + \mathbf{a}^T = 0 \quad \Longrightarrow \quad \boldsymbol{\beta} = -(M^T)^{-1}\mathbf{a}$$

4. **Calculating the Resulting Exponent:** With the linear terms removed, the exponent simplifies to $\mathbf{v}'^T M \mathbf{u}' + C$, where the constant term $C$ is:

$$C = \boldsymbol{\beta}^T M \boldsymbol{\alpha} + \boldsymbol{\alpha}^T \mathbf{a} + \boldsymbol{\beta}^T \mathbf{b}$$

Using the conditions $b = -M\boldsymbol{\alpha}$ and $a = -M^T \boldsymbol{\beta}$, we substitute for $\mathbf{a}$ and $\mathbf{b}$:

$$C = \boldsymbol{\beta}^T M \boldsymbol{\alpha} + \boldsymbol{\alpha}^T(-M^T \boldsymbol{\beta}) + \boldsymbol{\beta}^T(-M\boldsymbol{\alpha})$$
$$= \boldsymbol{\beta}^T M \boldsymbol{\alpha} - \boldsymbol{\alpha}^T M^T \boldsymbol{\beta} - \boldsymbol{\beta}^T M \boldsymbol{\alpha}$$

Since $\boldsymbol{\alpha}^T M^T \boldsymbol{\beta} = (\boldsymbol{\beta}^T M \boldsymbol{\alpha})^T = \boldsymbol{\beta}^T M \boldsymbol{\alpha}$ for a scalar term, this simplifies to:

$$C = \boldsymbol{\beta}^T M \boldsymbol{\alpha} - \boldsymbol{\beta}^T M \boldsymbol{\alpha} - \boldsymbol{\beta}^T M \boldsymbol{\alpha} = -\boldsymbol{\beta}^T M \boldsymbol{\alpha}$$

Substituting the explicit expressions for $\boldsymbol{\alpha}$ and $\boldsymbol{\beta}$:

$$C = -\left(-(M^T)^{-1}\mathbf{a}\right)^T M \left(-M^{-1}\mathbf{b}\right)$$
$$= -\left(-\mathbf{a}^T M^{-1}\right) M \left(-M^{-1}\mathbf{b}\right)$$
$$= -\mathbf{a}^T (M^{-1} M M^{-1})\mathbf{b} = -\mathbf{a}^T M^{-1}\mathbf{b}$$

5. **Final Integration:** The integral $I$ is now:

$$I = \int D\mathbf{u}' \, D\mathbf{v}' \, \exp(\mathbf{v}'^T M \mathbf{u}' - \mathbf{a}^T M^{-1}\mathbf{b})$$

The exponential of the constant term factors out:

$$I = \exp(-\mathbf{a}^T M^{-1}\mathbf{b}) \int D\mathbf{u}' \, D\mathbf{v}' \, \exp(\mathbf{v}'^T M \mathbf{u}')$$

The remaining integral is the standard definition of the Berezinian, which is the determinant of the matrix:

$$\int D\mathbf{u}' \, D\mathbf{v}' \, \exp(\mathbf{v}'^T M \mathbf{u}') = \det(M)$$

Combining these gives the final result:

$$I = \det(M) \exp(-\mathbf{a}^T M^{-1}\mathbf{b})$$

□

# F Detailed calculations for the bosonic Feynman-Kac formula

Let us calculate the case for a single bosonic degree of freedom.

Consider the following

$$\frac{1}{2\pi\hbar}\int_{\mathbb{R}^2}\exp_\star\{-\frac{itH}{\hbar}\}dxdp = \int_{\mathbb{R}^2}\sum_{n=0}^{\infty}e^{-\frac{itE_n}{\hbar}}\rho_n dxdp \tag{F.1}$$

$$= \sum_{n=0}^{\infty}e^{-\frac{itE_n}{\hbar}}\int_{\mathbb{R}^2}\rho_n dxdp \tag{F.2}$$

$$= \sum_{n=0}^{\infty}e^{-\frac{itE_n}{\hbar}}\int_{\mathbb{R}^2}Q_W^{-1}(|n\rangle\langle n|)dxdp \tag{F.3}$$

$$= \sum_{n=0}^{\infty}e^{-\frac{itE_n}{\hbar}}\underbrace{\int_{\mathbb{R}^2}\rho(x,p)dxdp}_{1} \tag{F.4}$$

$$= \sum_{n=0}^{\infty}e^{-\frac{itE_n}{\hbar}} \tag{F.5}$$

where we used the normalization of the Wigner function

$$\int_{\mathbb{R}^2}\rho_n(x,p)dxdp = \int_{\mathbb{R}^2}\frac{1}{(2\pi\hbar)}\psi(x+\frac{y}{2})\bar{\psi}(x-\frac{y}{2})e^{-\frac{i}{\hbar}yp}dydxdp \tag{F.6}$$

$$= \frac{1}{2\pi\hbar}(-\hbar)\int\psi(x+\frac{y}{2})\bar{\psi}(x-\frac{y}{2})e^{i(\frac{-p}{\hbar})y}d(-\frac{1}{\hbar}p)dxdy \tag{F.7}$$

$$= \int\psi(x+\frac{y}{2})\bar{\psi}(x-\frac{y}{2})\delta(-y)dxdy \tag{F.8}$$

$$= \int\psi(x)\bar{\psi}(x)dx \tag{F.9}$$

$$= 1 \tag{F.10}$$

Let us now perform an analytic continuation of the time variable via a Wick Rotation: $\tau = it$ :

$$\sum_{n=0}^{\infty}e^{-\frac{i}{\hbar}tE_n} = \sum_{n=0}^{\infty}e^{-\frac{\tau}{\hbar}E_n} \tag{F.11}$$

$$= e^{-\frac{\tau}{\hbar}E_0} + O(E_1) \tag{F.12}$$

This implies:

$$\lim_{\tau\to\infty}\frac{e^{\frac{\tau E_0}{\hbar}}}{2\pi\hbar}\int_{\mathbb{R}^2}\exp\{-\frac{\tau H}{\hbar}\}dxdp = 1 \tag{F.13}$$

where $E_0 \to non-degenerate\ ground\ state\ energy$. Equivalently:

$$e^{\frac{\tau E_0}{\hbar}} \lim_{\tau \to \infty} \frac{1}{2\pi\hbar} \int_{\mathbb{R}^2} \exp_\star\{-\frac{\tau H}{\hbar}\} dx dp = 1 \tag{F.14}$$

$\iff$

$$\lim_{\tau \to \infty} \frac{1}{2\pi\hbar} \int_{\mathbb{R}^2} \exp_\star\{-\frac{\tau H}{\hbar}\} dx dp = e^{\frac{-\tau E_0}{\hbar}} \tag{F.15}$$

$\iff$

$$\ln\left|\lim_{\tau \to \infty} \frac{1}{2\pi\hbar} \int_{\mathbb{R}^2} \exp_\star\left\{-\frac{\tau H}{\hbar}\right\} dx\, dp\right| = -\frac{\tau E_0}{\hbar} \tag{F.16}$$

$\iff$

$$E_0 = -\frac{\hbar}{\tau} \ln\left|\lim_{\tau \to \infty} \frac{1}{2\pi\hbar} \int_{\mathbb{R}^2} \exp_\star\left\{-\frac{\tau H}{\hbar}\right\} dx\, dp\right| \tag{F.17}$$

$$= -\frac{\hbar}{\tau} \lim_{\tau \to \infty} \ln\left|\frac{1}{2\pi\hbar} \int_{\mathbb{R}^2} \exp_\star\left\{-\frac{\tau H}{\hbar}\right\} dx\, dp\right| \tag{F.18}$$

$$= -\lim_{\tau \to \infty} \frac{\hbar}{\tau} \ln\left|\frac{1}{2\pi\hbar} \int_{\mathbb{R}^2} \exp_\star\{-\frac{\tau H}{\hbar}\} dx dp\right| \tag{F.19}$$

# G  Analysis of approximation regions: parameter space

We will now analyze each region presented in the table.

## G.1  Case 1: Resonant & Strong Coupling ($|\omega/g| \ll 1$)

This condition applies to both the **Resonant** region ($\omega \to 0$, $g$ not weak) and the **Strong coupling** region ($g \gg |\omega|$), as both are mathematically controlled by the dominance of $g$ over $\omega$.

**Step-by-Step Derivation**

1. We start by factoring out the dominant term, $2g$, from the square root in the eigenvalue formula:

$$\lambda_\pm = \frac{\omega \pm \sqrt{4g^2 + \omega^2}}{2}$$

$$= \frac{\omega \pm \sqrt{4g^2\left(1 + \frac{\omega^2}{4g^2}\right)}}{2}$$

$$= \frac{\omega \pm 2g\sqrt{1 + \left(\frac{\omega}{2g}\right)^2}}{2}$$

2. Define $x = (\omega/2g)^2$. Given the condition $|\omega/g| \ll 1$, it follows that $|x| \ll 1$. We can therefore apply the Taylor series expansion for $\sqrt{1+x}$:

$$\sqrt{1 + \left(\frac{\omega}{2g}\right)^2} \approx 1 + \frac{1}{2}\left(\frac{\omega}{2g}\right)^2 - \frac{1}{8}\left(\frac{\omega}{2g}\right)^4 + \cdots = 1 + \frac{\omega^2}{8g^2} - \frac{\omega^4}{128g^4} + \cdots$$

3. Substitute this expansion back into the expression for $\lambda_\pm$:

$$\lambda_\pm \approx \frac{\omega \pm 2g\left(1 + \frac{\omega^2}{8g^2}\right)}{2}$$

$$= \frac{\omega}{2} \pm \left(g + \frac{\omega^2}{4g}\right)$$

4. Finally, rearranging the terms gives the desired approximation:

$$\lambda_\pm \approx \pm g + \frac{\omega}{2} \pm \frac{\omega^2}{8g}$$

−

## G.2 Case 2: Dispersive ($\omega > 0$ and $|\omega| \gg g$)

This condition corresponds to the limit where $|g/\omega| \ll 1$.

**Step-by-Step Derivation**

1. Since $\omega$ is the dominant term and is positive, we factor it out from the square root. For $\omega > 0$, $\sqrt{\omega^2} = \omega$:

$$\lambda_\pm = \frac{\omega \pm \sqrt{\omega^2\left(1 + \frac{4g^2}{\omega^2}\right)}}{2}$$

$$= \frac{\omega \pm \omega\sqrt{1 + \frac{4g^2}{\omega^2}}}{2}$$

2. Let $x = 4g^2/\omega^2$. Since $|g/\omega| \ll 1$, we have $|x| \ll 1$. Applying the Taylor expansion:

$$\sqrt{1 + \frac{4g^2}{\omega^2}} \approx 1 + \frac{1}{2}\left(\frac{4g^2}{\omega^2}\right) - \frac{1}{8}\left(\frac{4g^2}{\omega^2}\right)^2 + \cdots = 1 + \frac{2g^2}{\omega^2} - \frac{2g^4}{\omega^4} + \cdots$$

3. Substitute this expansion back to find the approximations for $\lambda_+$ and $\lambda_-$:

   - For $\lambda_+$:

$$\lambda_+ \approx \frac{\omega + \omega\left(1 + \frac{2g^2}{\omega^2}\right)}{2} = \frac{2\omega + \frac{2g^2}{\omega}}{2} = \boldsymbol{\omega + \frac{g^2}{\omega}}$$

- For $\lambda_-$:

$$\lambda_- \approx \frac{\omega - \omega\left(1 + \frac{2g^2}{\omega^2}\right)}{2} = \frac{-\frac{2g^2}{\omega}}{2} = -\frac{g^2}{\omega}$$

—

## G.3   Case 3: Dispersive ($\omega < 0$ and $|\omega| \gg g$)

The condition is still $|g/\omega| \ll 1$, but now $\omega$ is negative.

**Step-by-Step Derivation**

1. We factor out $|\omega|$ from the square root. It is critical to recognize that since $\omega < 0$, we have $\sqrt{\omega^2} = |\omega| = -\omega$.

$$\lambda_\pm = \frac{\omega \pm \sqrt{\omega^2\left(1 + \frac{4g^2}{\omega^2}\right)}}{2}$$
$$= \frac{\omega \pm (-\omega)\sqrt{1 + \frac{4g^2}{\omega^2}}}{2}$$

2. We use the same Taylor expansion for the square root as in the previous case: $\sqrt{1 + 4g^2/\omega^2} \approx 1 + 2g^2/\omega^2$.
3. Substitute this expansion back to find the eigenvalues:

    - For $\lambda_+$:

$$\lambda_+ \approx \frac{\omega - \omega\left(1 + \frac{2g^2}{\omega^2}\right)}{2} = \frac{\omega - \omega - \frac{2g^2}{\omega}}{2} = -\frac{g^2}{\omega}$$

    - For $\lambda_-$:

$$\lambda_- \approx \frac{\omega + \omega\left(1 + \frac{2g^2}{\omega^2}\right)}{2} = \frac{2\omega + \frac{2g^2}{\omega}}{2} = \omega + \frac{g^2}{\omega}$$

—

## G.4   Case 4: Weak Coupling ($\alpha \to 0$)

This condition implies that $g = |\alpha| \to 0$.

**Step-by-Step Derivation**

1. We take the limit of the exact formula as $g \to 0$:

$$\lim_{g \to 0} \lambda_{\pm} = \lim_{g \to 0} \frac{\omega \pm \sqrt{\omega^2 + 4g^2}}{2}$$

$$= \frac{\omega \pm \sqrt{\omega^2}}{2} = \frac{\omega \pm |\omega|}{2}$$

2. The result depends on the sign of $\omega$:

    - If $\omega > 0$, then $|\omega| = \omega$. The limits are $\lambda_+ \to \frac{\omega+\omega}{2} = \omega$ and $\lambda_- \to \frac{\omega-\omega}{2} = 0$.
    - If $\omega < 0$, then $|\omega| = -\omega$. The limits are $\lambda_+ \to \frac{\omega-\omega}{2} = 0$ and $\lambda_- \to \frac{\omega+\omega}{2} = \omega$.

# Bibliography

[1] W. Heisenberg, "Über quantentheoretische umdeutung kinematischer und mechanischer beziehungen," *Zeitschrift für Physik*, vol. 33, pp. 879–893, 1925.

[2] D. Kaiser, "History: Shut up and calculate!," *Nature*, vol. 505, no. 7482, pp. 153–155, 2014.

[3] Y. Ge, "The weyl quantization: a brief introduction," 2021.

[4] G. F. T. del Castillo, *An Introduction to Hamiltonian Mechanics*. Modern Birkhäuser Classics, Boston: Birkhäuser, 2010.

[5] C. Cohen-Tannoudji, B. Diu, and F. Laloe, "Quantum mechanics, volume 1," *Quantum Mechanics*, vol. 1, p. 898, 1986.

[6] H. J. Groenewold, "On the principles of elementary quantum mechanics," *Physica*, vol. 12, no. 7, pp. 405–460, 1946. Available at https://www.rug.nl/research/vsi/events/groenewold/groenewold-article.pdf.

[7] C. Zachos, "On groenewold's theorem and classical and quantum hamiltonians." Physics Stack Exchange, 2018. https://physics.stackexchange.com/questions/389944/on-groenewolds-theorem-and-classical-and-quantum-hamiltonians (accessed June 2025).

[8] J. Berra-Montiel, H. García-Compeán, and A. Molgado, "Star exponentials from propagators and path integrals," *Annals of Physics*, vol. 468, p. 169744, 2024.

[9] I. Todorov, "Quantization is a mystery," *Bulg. J. Phys.*, vol. 39, pp. 107–149, 2012.

[10] T. L. Curtright, D. B. Fairlie, and C. K. Zachos, *A concise treatise on quantum mechanics in phase space*. World Scientific Publishing Company, 2013.

[11] J. Gleick, *Genius: The Life and Science of Richard Feynman*. New York: Pantheon Books, 1992.

[12] Stanford Encyclopedia of Philosophy, "Copenhagen interpretation of quantum mechanics." https://plato.stanford.edu/archives/win2016/entries/qm-copenhagen/. First published May 3, 2002; substantive revision July 31, 2024; accessed July 2025.

[13] I. Galaviz, H. Garcia-Compean, M. Przanowski, and F. Turrubiates, "Weyl–wigner–moyal formalism for fermi classical systems," *Annals of Physics*, vol. 323, no. 2, pp. 267–290, 2008.

[14] W. C. Myrvold, "Philosophical issues in quantum theory." Stanford Encyclopedia of Philosophy. First published July 25, 2016; substantive revision March 23, 2022; accessed July 2025.

[15] M. Rédei, "Uniqueness of canonical commutation relation representations revisited," *nLab / private communication*, 2024. Available at https://ncatlab.org/nlab/files/RedeiCCRRepUniqueness.pdf (accessed July 2025).

[16] F. Bayen, M. Flato, C. Fronsdal, A. Lichnerowicz, and D. Sternheimer, "Deformation theory and quantization. i. deformations of symplectic structures," *Annals of Physics*, vol. 111, no. 1, pp. 61–110, 1978.

[17] F. Bayen, M. Flato, C. Fronsdal, A. Lichnerowicz, and D. Sternheimer, "Deformation theory and quantization. ii. physical applications," *Annals of Physics*, vol. 111, no. 1, pp. 111–151, 1978.

[18] M. Kontsevich, "Deformation quantization of poisson manifolds," *Letters in Mathematical Physics*, vol. 66, pp. 157–216, 2003.

[19] P. Bressler and Y. Soibelman, "Mirror symmetry and deformation quantization," *arXiv preprint hep-th/0202128*, 2002.

[20] T. Asakawa and I. Kishimoto, "Noncommutative gauge theories from deformation quantization," *Nuclear Physics B*, vol. 591, no. 3, pp. 611–635, 2000.

[21] R. P. Feynman, A. R. Hibbs, and D. F. Styer, *Quantum mechanics and path integrals: Emended edition*. Dover Publications, 2005.

[22] A. Das, *Field theory: a path integral approach*, vol. 83. World Scientific, 2019.

[23] A. Hirshfeld, "Current aspects of deformation quantization," in *Proceedings of the Third International Conference on Geometry, Integrability and Quantization*, vol. 3, pp. 290–304, Bulgarian Academy of Sciences, Institute for Nuclear Research and Nuclear Energy, 2002.

[24] G. Karaali, "Deformation quantization – a brief survey," 1999. Unpublished manuscript, UC Berkeley.

[25] M. Gerstenhaber, "On the deformation of rings and algebras," *Annals of Mathematics*, vol. 79, no. 1, pp. 59–103, 1964.

[26] M. Reed and B. Simon, *Methods of modern mathematical physics: Functional analysis*, vol. 1. Gulf Professional Publishing, 1980.

[27] S. Waldmann, "Convergence of star products: From examples to a general framework," *EMS Surveys in Mathematical Sciences*, vol. 6, no. 1, pp. 1–31, 2019.

[28] K. Fujikawa, "Spin-statistics theorem in path integral formulation," *International Journal of Modern Physics A*, vol. 16, no. 24, pp. 4025–4044, 2001.

[29] N. Lang, "Lecture 4 (free fermions → the integer quantum hall effect)." Institute for Theoretical Physics III, University of Stuttgart, talk notes (PDF), 2025. Institute for Theoretical Physics III Lecture 4 PDF.

[30] S. Weinberg, *The quantum theory of fields*, vol. 2. Cambridge university press, 1995.

[31] L. A. Takhtajan, *Quantum Mechanics for Mathematicians*, vol. 95 of *Graduate Studies in Mathematics*. Providence, RI: American Mathematical Society, 2008. Hardcover ISBN; also available as eBook.

[32] J. Zinn-Justin, *Quantum field theory and critical phenomena*, vol. 171. Oxford university press, 2021.

[33] P. A. Horváthy, "The maslov correction in the semiclassical feynman integral," *Central European Journal of Physics*, vol. 9, no. 1, pp. 1–12, 2011.

[34] J. Berra-Montiel, "Star product representation of coherent state path integrals," *The European Physical Journal Plus*, vol. 135, no. 11, p. 906, 2020.

[35] M. E. Peskin, *An Introduction to quantum field theory*. CRC press, 2018.

[36] J. Glimm and A. Jaffe, *Quantum physics: a functional integral point of view*. Springer Science & Business Media, 2012.

[37] P. Sharan, "Star-product representation of path integrals," *Physical Review D*, vol. 20, no. 2, p. 414, 1979.

[38] A. Molgado, J. Berra-Montiel, and H. García-Compeán, "The feynman-kac formula in deformation quantization," *Available at SSRN 5142294*.

[39] J. Dito, "Star-product approach to quantum field theory: the free scalar field," *letters in mathematical physics*, vol. 20, no. 2, pp. 125–134, 1990.

[40] Don'tKnowMuch, "How do you compute the commutator $[p^m, x^n]$?." Physics Forums, Apr. 2012. https://www.physicsforums.com/threads/how-do-you-compute-the-commutator-p-m-x-n.598902/ (accessed June 2025).

[41] I. Terek, "A guide to symplectic geometry (symplectic geometry crash course)." Lecture notes, The Ohio State University, 2021. Updated May 6, 2022; available at https://web.williams.edu/Mathematics/it3/texts/symp_geo.pdf (accessed July 2025).